\documentclass[a4paper,10pt]{article}
\usepackage[utf8]{inputenc}
\pdfoutput=1 

\usepackage{jcappub} 

\usepackage[T1]{fontenc} 
\usepackage{graphicx}
\usepackage[makeroom]{cancel}
\usepackage{makecell}
\usepackage{bm}
\usepackage{amssymb}
\usepackage{amsmath}
\usepackage{epsfig}
\usepackage{hyperref}
\usepackage{mathtools}
\usepackage{hhline}
\usepackage{multirow}
\usepackage[normalem]{ulem}
\usepackage{subfig}
\usepackage{tikz}
\usepackage{epstopdf}
\usetikzlibrary{snakes}
\usepackage{slashed}
\usepackage{float}
\allowdisplaybreaks

\usepackage{xspace}

\definecolor{jd}{rgb}{0.858, 0.188, 0.478}

\def\lapp{\mathrel{\rlap{\raise.5ex\hbox{$<$}}
                    {\lower.5ex\hbox{$\sim$}}}}
\def\gapp{\mathrel{\rlap{\raise.5ex\hbox{$>$}}
                    {\lower.5ex\hbox{$\sim$}}}}

{
{

\newcommand{\bmt}{\begin{pmatrix}}
\newcommand{\emt}{\end{pmatrix}}
\newcommand{\ba}{\begin{array}{c}}
\newcommand{\ea}{\end{array}}
\newcommand{\be}{\begin{equation}}
\newcommand{\ee}{\end{equation}}
\newcommand{\bea}{\begin{eqnarray}}
\newcommand{\eea}{\end{eqnarray}}

\newcommand{\bi}{\begin{itemize}}
\newcommand{\ei}{\end{itemize}}

\newcommand{\baz}{\begin{array}{cc}}
\newcommand{\besub}{\begin{subequations}}
\newcommand{\eesub}{\end{subequations}}

\newcommand{\mathsym}[1]{{}}

\newcommand{\bt}{\begin{tabular}}
\newcommand{\et}{\end{tabular}}

\newcommand{\benu}{\begin{enumerate}}
\newcommand{\eenu}{\end{enumerate}}
\newcommand{\fsl}[1]{\ensuremath{\mathrlap{\!\not{\phantom{#1}}}#1}}

\def\a{\alpha}

\def\b{\beta}
\def\g{\gamma}

\def\l{\lambda}
\def\m{\mu}

\def\q2 {q^2}

\def\bt{\begin{table}}
\def\et{\end{table}}
\def\mZ{\mathbb{Z}}

\newcommand{\bav}{\begin{array}{cccc}}

\title{Multicomponent dark matter in extended $U(1)_{B-L}$: neutrino mass and high scale validity}
\author[a]{Subhaditya Bhattacharya,}
\emailAdd{subhab@iitg.ac.in}
\author[b,c]{Nabarun Chakrabarty,}
\emailAdd{chakran@iisc.ac.in}
\author[a]{Rishav Roshan,}
\emailAdd{rishav.roshan@iitg.ac.in}
\author[a]{Arunansu Sil}
\emailAdd{asil@iitg.ac.in}
\affiliation[a]{Department of Physics, Indian Institute of Technology Guwahati, North Guwahati, Assam-781039, India}
\affiliation[b]{Physics Division, National Center for Theoretical Sciences, Hsinchu, Taiwan 30013, R.O.C.}
\affiliation[c]{Centre for High Energy Physics, Indian Institute of Science, C.V. Raman Avenue, Bangalore 560012, India}

\abstract{Standard Model with right handed neutrinos charged under additional $U(1)_{B-L}$ gauge symmetry offer solutions to both dark matter (DM) problem 
and neutrino mass generation, although constrained severely from relic density, direct search and Higgs vacuum stability.
We therefore investigate a multicomponent DM scenario augmented by 
an extra inert scalar doublet, that is neutral under  $U(1)_{B-L}$, which aids 
to enlarge parameter space allowed by DM constraints and Higgs vacuum stability. 
The lightest right-handed neutrino and the $CP$-even inert scalar are taken as the dark matter candidates and
constitute a two component dark matter framework as they are rendered stable by an unbroken 
$\mathbb{Z}_2 \times \mathbb{Z}_2^\prime$ symmetry. DM-DM conversion processes   
 turn out crucial to render requisite relic abundance in mass regions of the RH neutrino that do not appear in the stand-alone $U(1)_{B-L}$ scenario. 
In addition, the one-loop renormalisation 
group (RG) equations in this model demonstrate that the electroweak (EW) vacuum can be stabilised till $\sim 10^{9}$ GeV in a parameter region compatible with 
the observed relic, the direct detection bound and other relevant constraints. We finally comment on the possibility of including the freeze-in mechanism in the same set-up.}

\DeclareUnicodeCharacter{2212}{-}

\begin{document}

\maketitle
\flushbottom

\setcounter{footnote}{0}
\renewcommand*{\thefootnote}{\arabic{footnote}}
\section{Introduction}
\label{sec:intro}

The Higgs boson of mass around 125 GeV discovered at the Large Hadron Collider (LHC)~\cite{Chatrchyan:2012xdj,Aad:2012tfa} 
completes the particle spectrum of the Standard Model (SM). Moreover, the couplings of this particle to the other SM particles are progressively 
getting closer to the corresponding SM values. However, certain pressing experimental evidences of phenomena ranging from dark matter in the 
universe to non-zero neutrino mass continue to advocate dynamics beyond the SM (BSM). And on the theoretical side, a rather pertinent
question is to ask whether the SM by itself can ensure a stable electroweak (EW) vacuum~
\cite{Degrassi:2012ry,Buttazzo:2013uya,Zoller:2014cka,EliasMiro:2011aa,Isidori:2001bm} at scales above that of electroweak symmetry breaking (EWSB). 
That is, the SM quartic coupling turns negative during renormalisation group (RG) evolution thereby destabilising the vacuum and the energy scale where 
that happens can vary several orders of magnitude depending upon the $t$-quark mass chosen. However, additional bosonic degrees of freedom over and 
above the SM ones can help the Higgs quartic coupling overcome the destabilising effect coming dominantly from the $t$-quark. This motivates
to look for extensions of the SM scalar sector.

Observation of galactic rotation curves, gravitational lensing and anisotropies in cosmic microwave background collectively hint towards the 
existence of cosmologically stable dark matter (DM) in the present universe~\cite{Spergel:2006hy,Aghanim:2018eyx}. 
Assuming DM has an elementary particle character, no such particle candidate(s) can be accommodated
in the Standard Model alone. Hence physics beyond the SM is inevitable. Hitherto the only information known about DM is its relic abundance and is precisely
determined by experiments studying anisotropies in cosmic microwave background radiation (CMBR) like 
Wilkinson Microwave Anisotropy Probe (WMAP)\cite{Bennett:2012zja} and PLANCK \cite{Aghanim:2018eyx}. Apart from this, we do not have any other
information about DM, such as its mass, spin, interaction strength etc. As a result, the nature of DM being
a scalar, a fermion, or a vector boson or an admixture of them, cannot be inferred. In addition to
gravity, if the DM interacts to the visible sector \emph{weakly}, it can thermalise in the early universe
at a temperature above its mass scale. As the universe cools down due to Hubble expansion, the
DM freezes-out from the thermal plasma at a temperature below its mass scale and gets red-
shifted since then. It is miraculous that the observed DM abundance implies to thermal freeze-out
cross-section of DM: of typical weak interaction strength and therefore it is largely believed that
the DM is a weakly interacting massive particle (WIMP)~\cite{Roszkowski:2017nbc}. 
Alternatively, DM can also be produced non-thermally
from decays or annihilation of particles present in early universe and freezes in as the temperature drops below DM mass. 
As the required interaction strength between DM-SM is substantially small for obtaining the correct relic density, such a framework is 
often referred as feebly interacting massive particle (FIMP) \cite{Hall:2009bx,Biswas:2016bfo,Bernal:2017kxu}.

The lack of precise information on dark matter quantum numbers opens up the possibility that DM consists of more than one type of particle. Multiparticle DM frameworks are interesting since they open up the possibility of DM-DM interaction (see~\cite{Biswas:2013nn,Bhattacharya:2013hva,Bian:2013wna,Esch:2014jpa,Bhattacharya:2016ysw,Ahmed:2017dbb,Herrero-Garcia:2017vrl,Herrero-Garcia:2018qnz,Poulin:2018kap,Aoki:2018gjf,Aoki:2017eqn,Elahi:2019jeo,Borah:2019aeq,Bhattacharya:2019fgs,Biswas:2019ygr} 
for a partial list of some recent studies). While such processes can contribute to the thermal relic, they do not
have a role in the direct detection rates. A multipartite DM model therefore can evade the ever
tightening bound on the direct detection (DD) rates while enlarging relic density allowed parameter space. We
have considered such a framework in this paper. The model is a hybrid of the two following single component DM models.

The minimal $U(1)_{B-L}$ framework~\cite{Davidson:1978pm,Mohapatra:1980qe,Marshak:1979fm,Davidson:1987mh} necessitates the introduction of additional fermions in order to be free of triangle anomalies. One possibility in that direction (a partial list is~\cite{Hasegawa:2019amx,Okada:2010wd,Basak:2013cga,El-Zant:2013nta,Okada:2016gsh,DelleRose:2017ukx,Escudero:2018fwn,Barman:2019aku,Basso:2010jm,Banerjee:2015hoa,Iso:2010mv,Okada:2012fs}) is to add 3 right-handed (RH)
neutrinos $N_{1,2,3}$ and make them couple to a scalar $S$ appropriately charged under $U(1)_{B-L}$.
Masses for the RH neutrinos are generated when $S$ receives a vacuum expectation value (VEV) and spontaneously break $U(1)_{B-L}$.
Annihilation of the lightest RH neutrino, say $N_1$ (rendered stable by an additional unbroken $\mathbb{Z}_2$ symmetry), 
via the exchange of scalars and the $U(1)_{B-L}$ gauge boson $Z_{BL}$ to the SM
particles can give rise to the observed DM thermal relic. It turns out that the relic density can only be satisfied in the resonance region(s). 
$N_1$ being the DM, the two other heavier right handed neutrinos can generate light neutrino masses 
through the so-called type-I seesaw mechanism \cite{Antusch:2011nz}. Hence this framework allows addressing DM and neutrino mass generation under the same umbrella. 
The allowed parameter space is also severely constrained by EW vacuum stability as additional fermions drag the quartic coupling $\beta$ functions 
to negative direction.\\
 Inert doublet model (IDM), with an extra $SU(2)_L$ scalar doublet charged negatively under a $\mathbb{Z}_2$ symmetry, and thus rendered stable against decays to purely SM fields provides a potential dark matter candidate in terms of the lightest among the $CP$-even and $CP$-odd components (\cite{LopezHonorez:2006gr,Honorez:2010re,Belyaev:2016lok,Choubey:2017hsq,LopezHonorez:2010tb,Ilnicka:2015jba,Arhrib:2013ela,Cao:2007rm,
Lundstrom:2008ai,Gustafsson:2012aj,Kalinowski:2018ylg,Bhardwaj:2019mts} and the references therein). The annihilation cross section for such a DM is often too large and renders a large intermediate region ($M_W$ - 500 GeV), for which the dark matter remains under abundant. 
This very feature plays a key role in embedding inert doublet DM into a multipartite 
framework, where under abundance of individual components naturally becomes legitimate. This model also can accommodate a
non-zero neutrino mass generated at the one-loop level when RH neutrinos are further added.

We have combined the two aforementioned models into a hybrid scenario in this work keeping the intermediate range of dark matter 
masses, between $M_W$ to 500 GeV in focus. Apart from three RH neutrinos and a scalar $S$ having appropriate $U(1)_{B-L}$ charges, a second
scalar doublet $\phi_2$ neutral under the same is introduced. The ensuing interactions are governed
by a $\mathbb{Z}_2 \times \mathbb{Z}_2^\prime$  discrete symmetry. The inert doublet and two RH neutrinos ($N_{2,3}$) carry negative
$\mathbb{Z}_2$ charges thereby opening up the possibility of a radiatively generated non zero neutrino mass
mimicking the scotogenic mechanism~\cite{Ma:2006km}. On the other hand, $N_1$ is non-trivially charged under $\mathbb{Z}_2^\prime$ 
and hence segregated from the rest of the RH neutrinos. Such an assortment of the discrete charges gives rise to a two-component DM 
scenario comprising $N_1$ and the lightest neutral scalar component of $\phi_2$ as the DM candidates.  It is worthy noting at this point, 
that analyses with same field content, $i.e.$ RH neutrinos and IDM, but transforming under a single $\mathbb{Z}_2$ symmetry \cite{Kanemura:2011vm,Borah:2018smz} have been addressed before. 
This evidently renders the lightest under $\mathbb{Z}_2$ stable and provide a single component DM framework. In such circumstances, the DM can only enjoy 
co-annihilation with the heavier component (at the expense of having small mass difference between them) \cite{Borah:2018smz} on top of usual annihilation to 
SM to help it evade direct search bounds. In our case however, with two DM components present, DM-DM conversion
plays a crucial role to yield necessary depletion of heavier DM component in obtaining correct relic density and evade direct search bound 
in a larger parameter space. In particular, the Yukawa coupling required in our case turns out much smaller to respect DM constraints than in \cite{Borah:2018smz}, 
making our model more viable in terms of high-scale validity.

In this paper, we study the DM phenomenology of the two-component model in detail and emphasize the role of DM-DM conversion as mentioned above.  
The behaviour of the set-up at high energy scales is also looked at using one-loop RG equations. In other words, we explore the enticing 
possibility of correlating the DM-allowed parameter space (or, more specifically, the `conversion' region) with high scale validity under RG. 
We also comment on the possibility to  accommodate non-thermal production of $N_1$ through \emph{freeze-in}.

The paper is organized as follows. The model is introduced in section~\ref{model} and the various
theoretical and experimental constraints deemed relevant here are detailed in section \ref{constraints}. Sections \ref{dm_pheno}
and \ref{vacstab} shed slight on the DM phenomenology and the RG-behaviour of the model respectively. In
section \ref{dm+hsv}, we combine the constraints coming from DM and high scale behaviour and in section \ref{summary}, we conclude. 
 Various important formulae
are relegated to the Appendix.
\section{The scenario}\label{model}

Augmenting the SM gauge group by an $U(1)_{B-L}$ symmetry,
we extend the minimal $U(1)_{B-L}$ framework, that comprises three RH neutrinos $N_1,N_2, N_3$ and a complex scalar $S$, 
with an inert scalar Higgs $\phi_2$. The quarks and leptons respectively carry $U(1)_{B-L}$ charges $\frac{1}{3}$ and -1. An additional $\mathbb{Z}_2 \times {\mathbb{Z}_2}^{\prime}$ symmetry is invoked. 
The charges of the additional fields under the gauge $\mathcal{G} = SU(2)_L \times U(1)_Y \times U(1)_{B-L}$ and discrete symmetries are shown in Table~\ref{qno}.

\begin{table}[h]
\centering
\begin{tabular}{|c c c c c|}
\hline
Field   & ~~~~$SU(2)_L \times U(1)_Y$ & ~~~~$Y_{BL}$ & ~~~~$\mathbb{Z}_2$ & ~~~~${\mathbb{Z}_2}^{\prime}$\\ \hline \hline
$\phi_2$ & (2,$~\frac{1}{2}$) & 0 & - & -\\ \hline
$N_1$ & (1,~0) & -1 & - & +\\ \hline
$N_2,N_3$ & (1,~0) & -1 & - & -\\ \hline
$S$  & (1,~0) & 2 & + & +\\ \hline
\end{tabular}
\caption{The additional fields and their quantum numbers under 
$\mathcal{G} \times \mathbb{Z}_2 \times {\mathbb{Z}_2}^{\prime}$. Here, $Y_{BL}$ refers to the $U(1)_{B-L}$ charge.}
\label{qno}
\end{table}
This particular assignment of the $B-L$ charges eliminates the triangular $B-L$ gauge anomalies. 
It is important to note $\mZ_2, \mZ^{'}_2$ charges of $N_1, N_{2,3}, \phi_2$. $N_{2,3}, \phi_2$ having same charge under $\mZ_2 \times \mZ^{'}_2$ offers the lightest 
amongst them to be stable. We will assume $\phi_2$ to be lighter and constitute one of the DM components of the model. Absence of any other particle with $[-,+]$ charge under 
$\mZ_2 \times \mZ^{'}_2$, $N_1$ is always stable and contributes as the second DM component in our model. The other motivation(s) for segregating 
$N_{1}$ and $N_{2,3}$ charges will be spelled after introducing the Yukawa interactions allowed in the model.


The kinetic terms for the additional fields are
\besub
\bea
\mathcal{L}_{KE} &=& |D_{\mu}S|^2+\sum_{i=1,2,3}\bar{N_i}i\gamma^{\mu}D_{\mu}N_i-\frac{1}{4}Z_{\mu\nu}Z^{\mu\nu}, \\
\rm{where}~~Z^{\mu\nu}&=&\partial^{\mu}Z_{BL}^{\nu}-\partial^{\nu}Z_{BL}^{\mu},\\
D_{\mu}&=& \partial_{\mu}+i[Yg^{\prime}+Y_{BL}g_{BL}](Z_{BL})_{\m}.
\eea
\eesub

We will consider the pure $U(1)_{B-L}$ model here, that is defined by $g^{\prime} = 0$. This forbids $Z$-$Z_{BL}$ mixing at the tree level. It is obvious that 
$g_{BL}$ refers to $U(1)_{B-L}$ coupling, and serves as a key parameter for the model.\\
The Yukawa Lagrangian in this set up has the form

\bea
-\mathcal{L}_{Y} \supset 
\zeta_{i\a} \bar{L}_{Li} \phi_2 N_\a
 + y_{11} \bar{N_1^c} N_1 S + y_{\a \b} \bar{N_\a^c} N_\b S, 
\label{eq1a}
\eea

All parameters in the above are taken to be real. \\
In addition, the most general scalar potential complying with $\mathcal{G} \times \mathbb{Z}_2 \times {\mathbb{Z}_2}^{\prime}$ is given by 

\bea
V(\phi_1,\phi_2,S) &=& -\mu_1^2 \phi_1^{\dagger} \phi_1
 + \mu_2^2 \phi_2^{\dagger} \phi_2 - \mu_{S}^2 |S|^2
+ \frac{\l_1}{2} (\phi_1^{\dagger} \phi_1)^2
+ \frac{\l_2}{2} (\phi_2^{\dagger} \phi_2)^2 \nonumber \\
&&
 + \l_3 (\phi_1^{\dagger} \phi_1) (\phi_2^{\dagger} \phi_2)
 + \l_4 (\phi_1^{\dagger} \phi_2) (\phi_2^{\dagger} \phi_1)
 + \frac{\l_5}{2} \Big[(\phi_1^{\dagger} \phi_2)^2 + 
 (\phi_2^{\dagger} \phi_1)^2 \Big] \nonumber \\
&&
+ \l_6 (\phi_1^{\dagger} \phi_1) |S|^2 + 
\l_7 (\phi_2^{\dagger} \phi_2) |S|^2 + \l_8 |S|^4.
\eea

Electroweak symmetry breaking (EWSB) is triggered for $\mu_1^2, \mu_S^2 > 0$. The $CP$-even components of $\phi_1$ and $S$ then receive VEVs $v$ and $v_{BL}$ respectively through the tadpole conditions below:
\besub
\bea
\mu_1^2 &=& \frac{\l_1}{2} v^2 + \frac{\l_6}{2} v_{BL}^2, \\
\mu_S^2 &=& \frac{\l_6}{2} v_1^2 + \l_8 v_{BL}^2.
\eea
\eesub
One must demand $\mu_2^2 >0$ so that $\phi_2$ does not develop a VEV and a spontaneous breakdown of $\mathbb{Z}_2$ is avoided. 
Following EWSB, the scalar multiplets can then be parametrised as
\besub
\bea
\phi_1 &=&
  \begin{pmatrix}
    G^+\\
 \frac{1}{\sqrt 2}(v + \phi_h + i G^0)
  \end{pmatrix},~~S = \frac{1}{\sqrt 2}(v_{BL} + \phi_S ), \\
\phi_2 &=&
  \begin{pmatrix}
    H^+\\
 \frac{1}{\sqrt 2}(H + i A).
  \end{pmatrix}
\eea
\eesub

The component scalars $H,A,H^+$ of the inert doublet do not mix with 
$\phi_1$ and $S$ and therefore have the masses
\besub
\bea
M_H^2 &=& 
\mu_2^2 + \frac{1}{2}(\l_3 + \l_4 + \l_5)v^2 + \frac{1}{2}\l_7 v_{BL}^2, \\
M_A^2 &=& 
\mu_2^2 + \frac{1}{2}(\l_3 + \l_4 - \l_5)v^2 + \frac{1}{2}\l_7 v_{BL}^2, \\
M_{H^+}^2 &=& 
\mu_2^2 + \frac{1}{2} \l_3 v^2 + \frac{1}{2}\l_7 v_{BL}^2 .
\eea
\eesub
One defines a $\l_L = \l_3 + \l_4 + \l_5$ which is physically relatable since the interaction strength of the $H-H-h$ coupling in the pure IDM is given by 
-$\l_L v$. On the other hand, a non-zero $\phi_h - \phi_S$ mixing leads to the following mass terms
\bea
V \supset \frac{1}{2} \begin{pmatrix}
    \phi_h & \phi_S
  \end{pmatrix} \begin{pmatrix}
    \l_1 v^2 & \l_6 v v_{BL} \\
    \l_6 v v_{BL} & 2 \l_8 v_{BL}^2 
  \end{pmatrix} 
 \begin{pmatrix}
    \phi_h \\
    \phi_S 
  \end{pmatrix}.
\eea

The mass matrix is diagonalised using
\bea
\begin{pmatrix}
    \phi_h \\
    \phi_s 
  \end{pmatrix} = \begin{pmatrix}
    c_\theta & s_\theta \\
    -s_\theta & c_\theta 
  \end{pmatrix}
  \begin{pmatrix}
    h \\
    s 
  \end{pmatrix}
\eea
with 
\bea
\text{tan} 2\theta &=& \frac{-2 \l_6 v v_{BL}}{\l_1 v^2-2 \l_8 v^2_{BL} }.
\eea

The mass eigenstates ($h,s$) then have masses
\besub
\bea
M^2_{h,s} &=& \frac{1}{2} \Big[\big(\l_1 v^2 + 2 \l_8 v^2_{BL}\big) \pm \sqrt{(\l_1 v^2 - 2 \l_8 v^2_{BL}\big)^2 + 4 \l_6^2 v^2 v_{BL}^2}\Big]. 
\eea
\eesub
We choose the masses and the mixing angle $\theta$ as the independent variables. With that choice, 
the independent parameters in the scalar sector are:  
\bea
\nonumber
\{M_h,M_s,s_\theta,M_H, M_A,M_{H^+},\l_L,\l_2,\l_7\}. 
\eea
The various model parameters are expressible in terms of the physical quantities as follows:
\besub
\bea
\mu_2^2 &=& M_H^2 - \frac{1}{2}\l_L v^2 - \frac{1}{2}\l_7 v_{BL}^2,  \\
\l_1 &=& \frac{(M^2_h c^2_\theta + M^2_s s^2_\theta)}{v^2}, \\
\l_3 &=& \l_L + \frac{2(M_{H^+}^2 - M_H^2)}{v^2}, \\
\l_4 &=& \frac{M_H^2 + M_A^2 - 2 M_{H^+}^2}{v^2}, \\
\l_5 &=& \frac{(M^2_H - M^2_A)}{v^2}, \\
\l_6 &=& \frac{(M^2_s - M^2_h)s_\theta c_\theta}{vv_{BL}}, \\
\l_8 &=& \frac{(M^2_h s^2_\theta + M^2_s c^2_\theta)}{2 v_{BL}^2}.
\eea
\label{quartic}
\eesub

\noindent where $\a, ~\b = 2,~3$ and summation over repeated indices is implied. The motivation behind imposing the additional $\mathbb{Z}_2^{\prime}$ symmetry is to distinguish $N_1$ from $N_2,N_3$. In that case, $N_1$ does not enter the one-loop diagrams that generate $m_\nu$, and, it also does not participate in leptogenesis. In such a case, it is expected to be free of constraints that stem from the two aforementioned issues.

In addition, EWSB gives rise to the following mass matrix for $N_{1,2,3}$.
\bea
M_N = \sqrt{2}~v_{BL}\begin{pmatrix}
    y_{11} & 0 & 0 \\
    0 & y_{22} & y_{23} \\
    0 & y_{23} & y_{33}     
  \end{pmatrix}.
\eea
 We take $y_{23} = 0$ for simplicity for the rest of the analysis,
in which case $M_N$ is diagonal with entries $M_i =  \sqrt{2}~y_{ii} v_{BL}$.

\section{Theoretical and experimental constraints}\label{constraints}
The scenario introduced here faces various constraints both from theory and experiments. We discuss these in this section.
\subsection{Theory constraints}
The scalar potential remains bounded from below in various directions in the field space once the following conditions are met:
\besub
\bea
\text{vsc1}: \l_1 > 0, \\
\text{vsc2}: \l_2 > 0, \\
\text{vsc3}: \l_8 > 0, \\
\text{vsc4}: \l_3 + \sqrt{\l_1 \l_2} > 0, \\
\text{vsc5}: \l_3 + \l_4 - |\l_5| + \sqrt{\l_1 \l_2} > 0, \\
\text{vsc6}: \l_6 + \sqrt{2\l_1 \l_8} > 0, \\
\text{vsc7}: \l_7 + \sqrt{2\l_2 \l_8} > 0. 
\eea
\eesub\label{vsc}

\noindent In addition, a perturbative theory demands that the model parameters
obey
\bea
|\l_i|< 4\pi,~ |g_i|< \sqrt{4\pi} ,~ |y_i|< \sqrt{4\pi}.
\eea

\subsection{Experimental constraints}
The main experimental constraints stem from oblique parameters, collider search, neutrino mass and dark matter as detailed below.
\subsubsection{Oblique parameters}
Amongst the oblique parameters $S,T,U$~\cite{Peskin:1991sw}, the strongest constraint on a multi-Higgs scenario is
in fact imposed by the $T$-parameter. More precisely, this restricts the mass splitting between the
scalars belonging to an $SU(2)_L$ multiplet. The scalar $s$ contributes negligibly in the small $s_\theta$ limit and contribution coming from the IDM is expressed as
follows~\cite{Grimus:2008nb}:
\bea
\Delta T &=& \frac{g^2}{64\pi^2 m^2_{W}\alpha}[F(M^2_{H^+},M^2_{H})+F(M^2_{H^+},M^2_{A})-F(M^2_{H},M^2_{A})].
\eea
where $F(x,y)=\frac{1}{2}(x+y)-\frac{xy}{x-y}\log(\frac{x}{y})$ for $x \neq y$. We use the latest bound~\cite{PhysRevD.98.030001}
\bea
\Delta T &=& 0.07\pm 0.12.
\eea
\subsubsection{Collider constraints}
Non-observation of neutral and charged scalars at the LEP have put lower limits on their masses. In Ref. \cite{Lundstrom:2008ai}, it is shown that the points 
satisfying the intersection of the following conditions\\
$$M_H<80~ \text{GeV},~ M_A<100~ \text{GeV} ~\text{and}~ M_A-M_H>8~ \rm{GeV},$$
\\
are excluded by the LEP II data as they would lead to a di-lepton/di-jet signature along with missing energy. We have adopted the more conservative 
$M_{H,A,H^+} > 200$ GeV in this work that easily bypasses the aforementioned constraints.\\

In the absence of any mixing between $h$ and the $\mathbb{Z}_2$ odd scalars, the tree level couplings of $h$ with the fermions and gauge bosons get scaled by a factor of $c_\theta$ \emph{w.r.t} the SM values. This implies that the $g g \to h$ production cross section is accordingly scaled by $c^2_\theta$. The signal strength in the diphoton channel then becomes $\mu_{\g \g} = c^2_\theta 
\frac{BR_{h \to \g \g}}{BR_{h \to \g \g}^{\text{SM}}} \simeq 
c^2_\theta \frac{\Gamma_{h \to \g \g}}{\Gamma_{h \to \g \g}^{\text{SM}}}$. 
The charged Higgs $H^+$ coming from the inert doublet leads to an additional one-loop term in the $h \to \g \g $ amplitude~\cite{Arhrib:2012ia,Swiezewska:2012eh}. 
That is,
\bea
\mathcal{M}_{h \to \g \g} &=& 
\frac{4}{3} c_\theta A_f\Big(\frac{M^2_h}{4 M^2_t}\Big)
 + c_\theta A_V\Big(\frac{M^2_h}{4 M^2_W}\Big)
 + \frac{\l_{h H^+ H^-} v}{2 M^2_{H^+}} A_S\Big(\frac{M^2_h}{4 M^2_{H^+}} \Big), \nonumber \\
\Gamma_{h \to \g \g} &=& \frac{G_F \a^2 M_h^3}{128 \sqrt{2} \pi^3} |\mathcal{M}_{h \to \g \g}|^2.
\eea
In the above, $G_F$ and $\a$ denote respectively the Fermi constant and the QED fine-structure constant. The expression for $\l_{h H^+ H^-}$ can be seen in the Appendix. The loop functions are listed below~\cite{Djouadi:2005gj}.
\bea
A_f(x) &=& \frac{2}{x^2}\big((x + (x -1)f(x)\big), \nonumber \\
A_V(x) &=& -\frac{1}{x^2}\big((2 x^2 + 3 x + 3(2 x -1)f(x)\big),\nonumber \\
A_S(x) &=& -\frac{1}{x^2}\big(x - f(x)\big), \nonumber \\
\text{with} ~~f(x) &=& \big(\text{sin}^{-1}\sqrt{x}\big)^2.
\eea

\noindent where $A_f(x), A_V(x)$ and $A_S(x)$ are the respective amplitudes for the spin-$\frac{1}{2}$, spin-1 and spin-0 particles in the loop and $x = m_h^2/4 m_{f/V/S}^2$. 
The latest $\mu_{\gamma \gamma}$ values from 13 TeV LHC read~\cite{Aaboud:2018xdt,Sirunyan:2018ouh} 
\besub
\bea
\mu_{\gamma \gamma} &=& 0.99^{+0.14}_{-0.14} ~(\text{ATLAS}),\\
&=& 1.18^{+0.17}_{-0.14} ~(\text{CMS}).
\eea
\eesub

\noindent Upon using the standard combination of signal strengths and uncertainties\footnote{
The signal strength data from the ATLAS and CMS for a given channel can be combined to yield a resultant central value $\mu$ and a resultant 1-sigma uncertainty $\sigma$ as 
$\frac{1}{\sigma^2} = 
\frac{1}{\sigma^2_{\text{ATLAS}}} + \frac{1}{\sigma^2_{\text{CMS}}}$ and \\
$\frac{\mu}{\sigma^2} = 
\frac{\mu_{\text{ATLAS}}}{\sigma^2_{\text{ATLAS}}} + \frac{\mu_{\text{CMS}}}{\sigma^2_{\text{CMS}}}$.}, we obtain $\mu_{\g\g} \simeq 1.06 \pm 0.1$.

One should also note that the observed signal strength of the 125 GeV Higgs boson at the LHC provides a limit on $\sin{\theta}$ as 
$|\sin{\theta}|\leq 0.36$ \cite{Robens:2016xkb}. Additionally, we obey the $\frac{M_{Z_{BL}}}{g_BL}\geq 7.1~\rm{TeV}$ exclusion limit from LEP-II \cite{Carena:2004xs,Cacciapaglia:2006pk} and lastly, we also obey the  constraints from opposite sign dilepton searches at LHC which mostly exclude the model for 150 GeV $<M_{Z_{BL}}<$ 3 TeV \cite{Escudero:2018fwn}.
\subsection{Neutrino Mass}
\begin{figure}[H]
\centering
\includegraphics[scale=0.5]{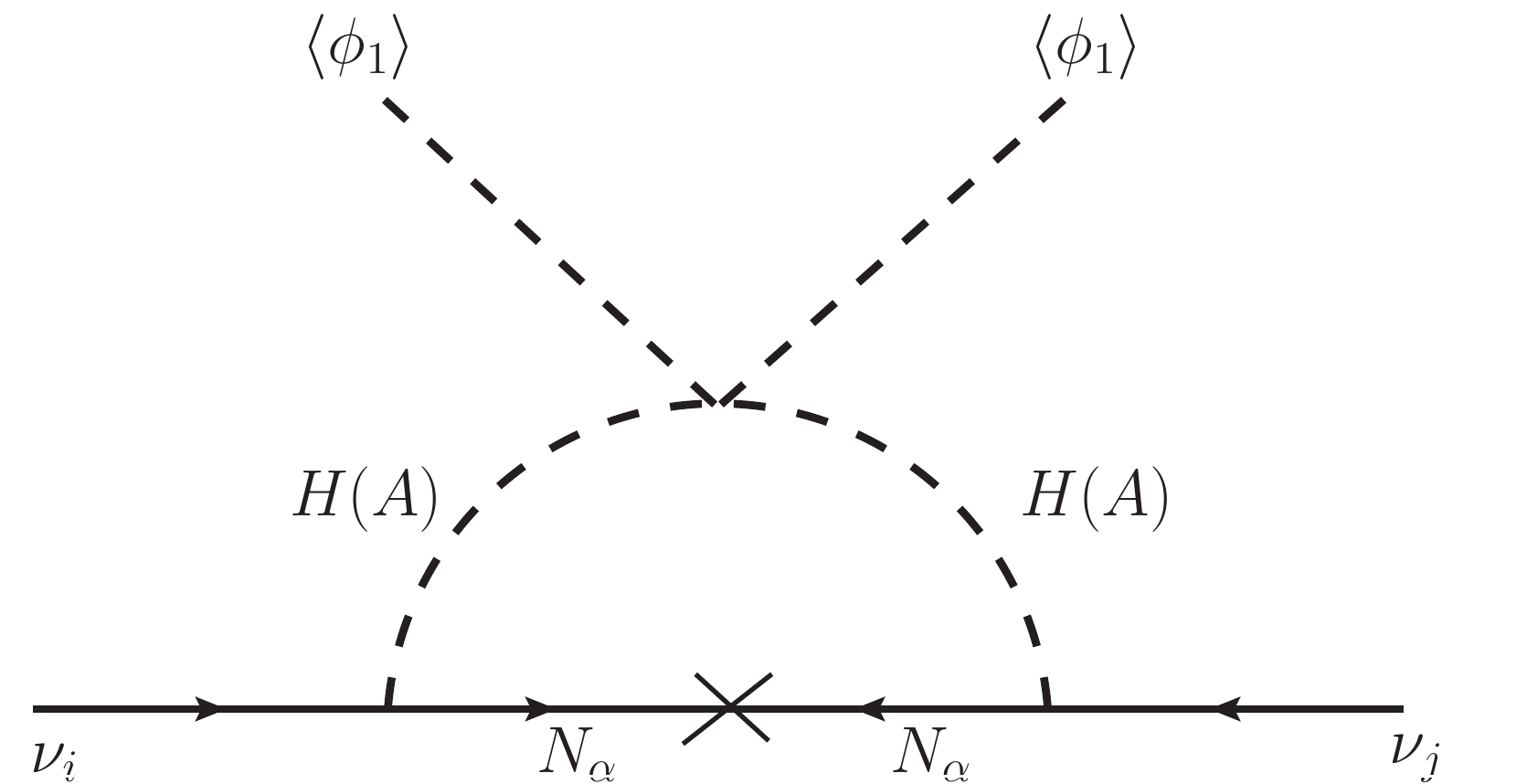}
\caption{Radiative generation of light neutrino mass.}\label{nu_loop}
\end{figure}
In any scotogenic scenario, the SM neutrinos acquire a non-zero Majorana mass at one-loop with the RH neutrinos and the inert scalars circulating in the loop~\cite{Ma:2006km}. The circulating particles are $H,A$ and $N_{2,3}$ 
for this model as shown in Fig.\ref{nu_loop}. The neutrino mass elements $(M_\nu)_{ij}$ are given by
\bea
(M_\nu)_{ij} &=& \sum_{\a = 2,3} \frac{M_a \zeta_{i \a} \zeta_{j \a}}{32 \pi^2}
\Big[\frac{M_H^2}{M^2_H - M^2_\a}\text{log}\Big(\frac{M_H^2}{M_\a^2}\Big) - \frac{M_A^2}{M^2_A - M^2_\a}\text{log}\Big(\frac{M_A^2}{M_\a^2}\Big)\Big].\label{mnu}
\eea
Eqn.(~\ref{mnu}) is recasted using matrices as
\bea
M_\nu = \zeta^* \Lambda \zeta^\dagger. \label{mnu_CI} 
\eea
Here, $M_\nu = [(M_\nu)_{ij}], \zeta = [\zeta_{i\a}]$ and $\Lambda = [\Lambda_{\a\b}]$ are $3\times3$, $3\times2$ and $2\times2$ matrices respectively. One notes
\bea
\Lambda_{\a \b} &=& \frac{M_\a}{32 \pi^2}
\Big[\frac{M_H^2}{M^2_H - M^2_\a}\text{log}\Big(\frac{M_H^2}{M_\a^2}\Big) - \frac{M_A^2}{M^2_A - M^2_\a}\text{log}\Big(\frac{M_A^2}{M_\a^2}\Big)\Big] \delta_{\a \b}.
\eea

The complex symmetric $M_\nu$ is diagonalized 
by the Pontecorvo Maki-Nakagawa-Sakata (PMNS) leptonic mixing matrix $U$ as $M^d_\nu = U M_\nu U^T$, where $M^d_\nu$ = diag($0,m_2,m_3$) is the diagonal neutrino mass matrix\footnote{A scotogenic model with only 2 RH neutrinos predicts one massless SM neutrino.}. Further, parametrisation introduced in \cite{Casas:2001sr} enables to express $\zeta$ as
\bea
\zeta_{i \a} &=& \Big(U ({M^d_\nu})^{\frac{1}{2}} 
R^\dagger ({\Lambda^d})^{-\frac{1}{2}}\Big)_{i\a}.\label{CI}
\eea
where $\Lambda^d$ denotes the  diagonalised $\Lambda$ and the arbitrary complex matrix $R$ satisfies $R^T R$ = $\mathcal{I}$. Note that due to the involvement of masses of the inert Higgs doublet components in $\Lambda^d$, which plays a significant role in DM phenomenology, a correlation between neutrino mass and DM is expected in the set-up. 
Taking, for instance, $M_{H/A} \simeq 500$ GeV, $M_A - M_H$ = 10 GeV and $M_{2,3} \simeq 1$ TeV, and assuming a typical $M_\nu ~\text{element}$ in the  [0.01, 0.1] eV range, one gets $\zeta_{i\a} \sim \mathcal{O}(10^{-5})$. This tiny coupling \footnote{It is possible to have large Yukawa $\sim\mathcal{O}(0.1)$ along with $M_{2,3}\sim 1$ TeV which can explain neutrino mass $\sim0.1$ eV through CI parametrisation \cite{Ghosh:2017fmr} with the introduction of a complex orthogonal matrix R. However, this choice is nonetheless fine-tuned and we will not consider this possibility in the ensuing analysis.} does not have any impact on the RG running of quartic coupling of $\phi_2$.
\subsection{Lepton flavour violation} 
Loop-induced lepton flavor violating decays of the $l_i \to l_j \gamma$ type are turned on in presence of the inert doublet and the RH neutrinos 
(with $N_i$ and $H^+$ running in the loop). The most restrictive amongst these is the $\mu \to e \gamma$ mode that carries the bound 
$\text{BR}_{\mu \to e \gamma} < 4.2 \times 10^{-13}$~\cite{MEGexp:2016wtm}. However, for 
$\zeta_{i\a} \sim 10^{-5}$, $M_{H^+} \simeq 500$ GeV and RH neutrinos of mass $\sim$ 1 TeV, one obtains 
$\text{BR}_{\mu \to e \gamma} \sim 10^{-27}$~\cite{Toma:2013zsa,Vicente:2014wga} which is well below the current limit.
\subsection{DM constraints} 
The observed amount of relic abundance of the dark matter  is provided the Planck experiment\cite{Aghanim:2018eyx}
\bea
0.1166\leq\Omega_{DM}h^2\leq0.1206.
\eea
Furthermore, the dark matter parameter space is constrained significantly by the direct detection experiments such as  LUX~\cite{Akerib:2016vxi}, PandaX-II~\cite{Zhang:2018xdp} and Xenon-1T~\cite{Aprile:2018dbl}. The detailed discussions on the dark matter phenomenology are presented in section \ref{dm_pheno}.

\section{DM phenomenology for thermal production}\label{dm_pheno}

In this section, we elaborate on thermal relic density of the two-component 
DM set up in this model with an emphasis on DM-DM conversion.

\subsection{Relic Density}

\begin{figure}[H]
$$
\includegraphics[height=4 cm, width=5 cm,angle=0]{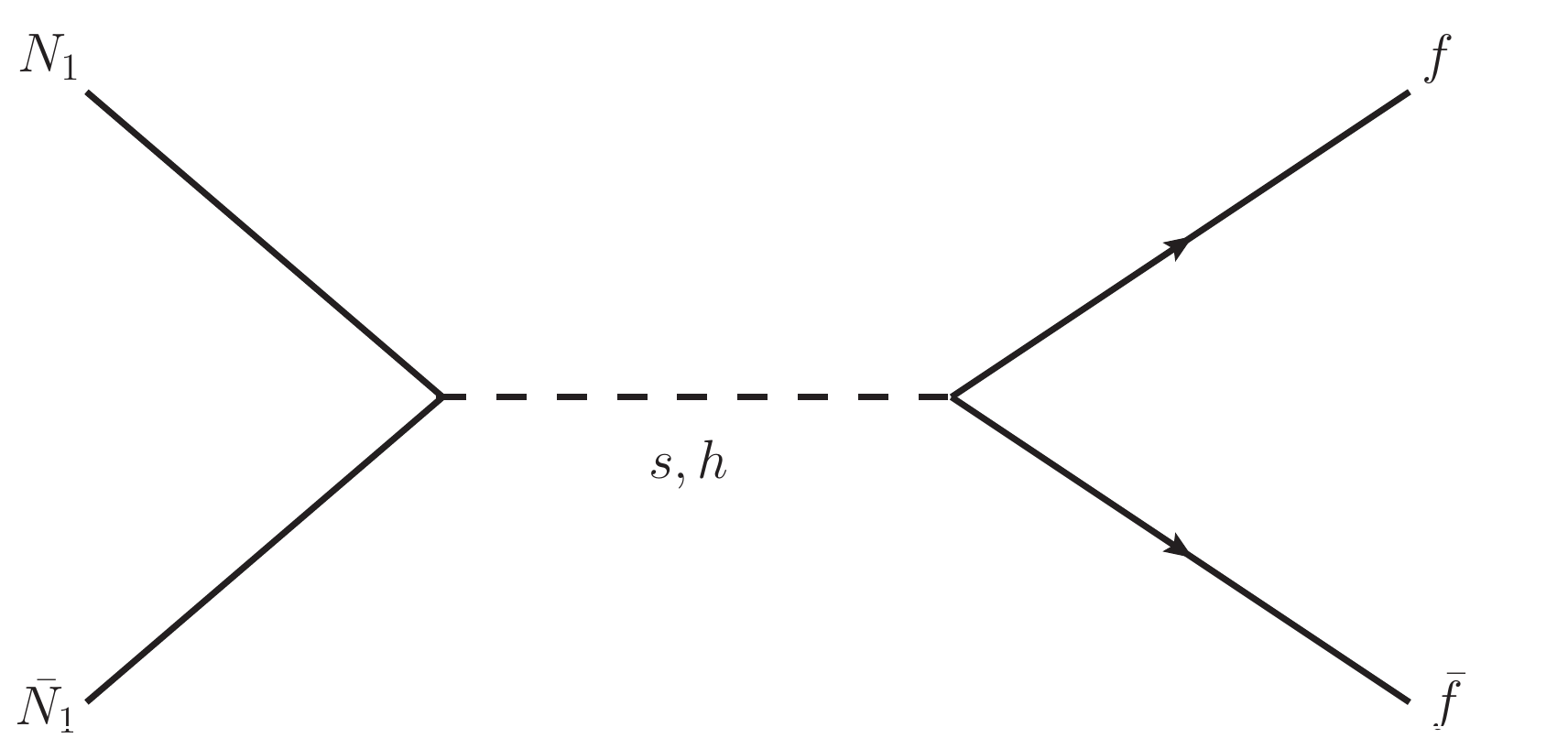}~~~
\includegraphics[height=4 cm, width=5 cm,angle=0]{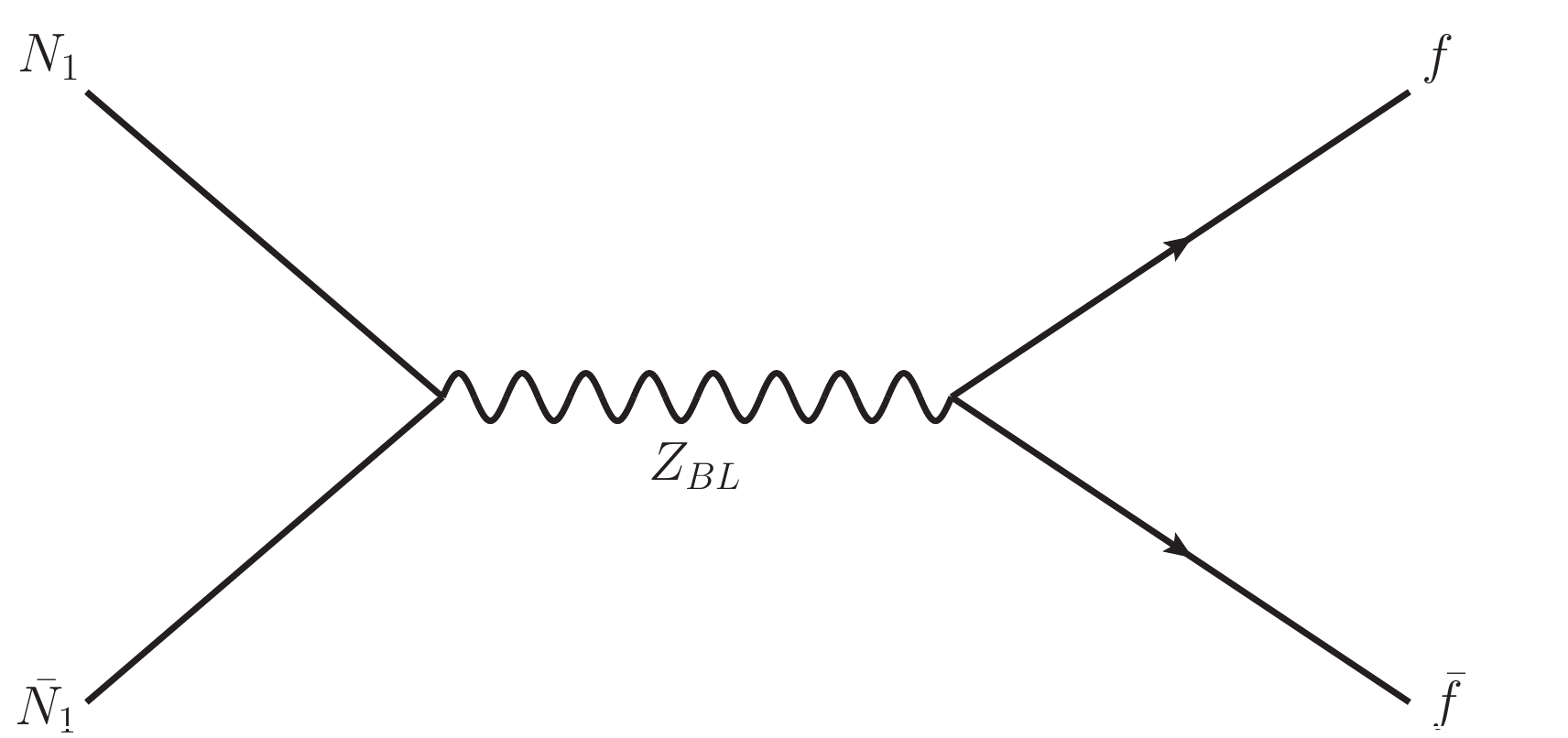}~~~
\includegraphics[height=4 cm, width=5 cm,angle=0]{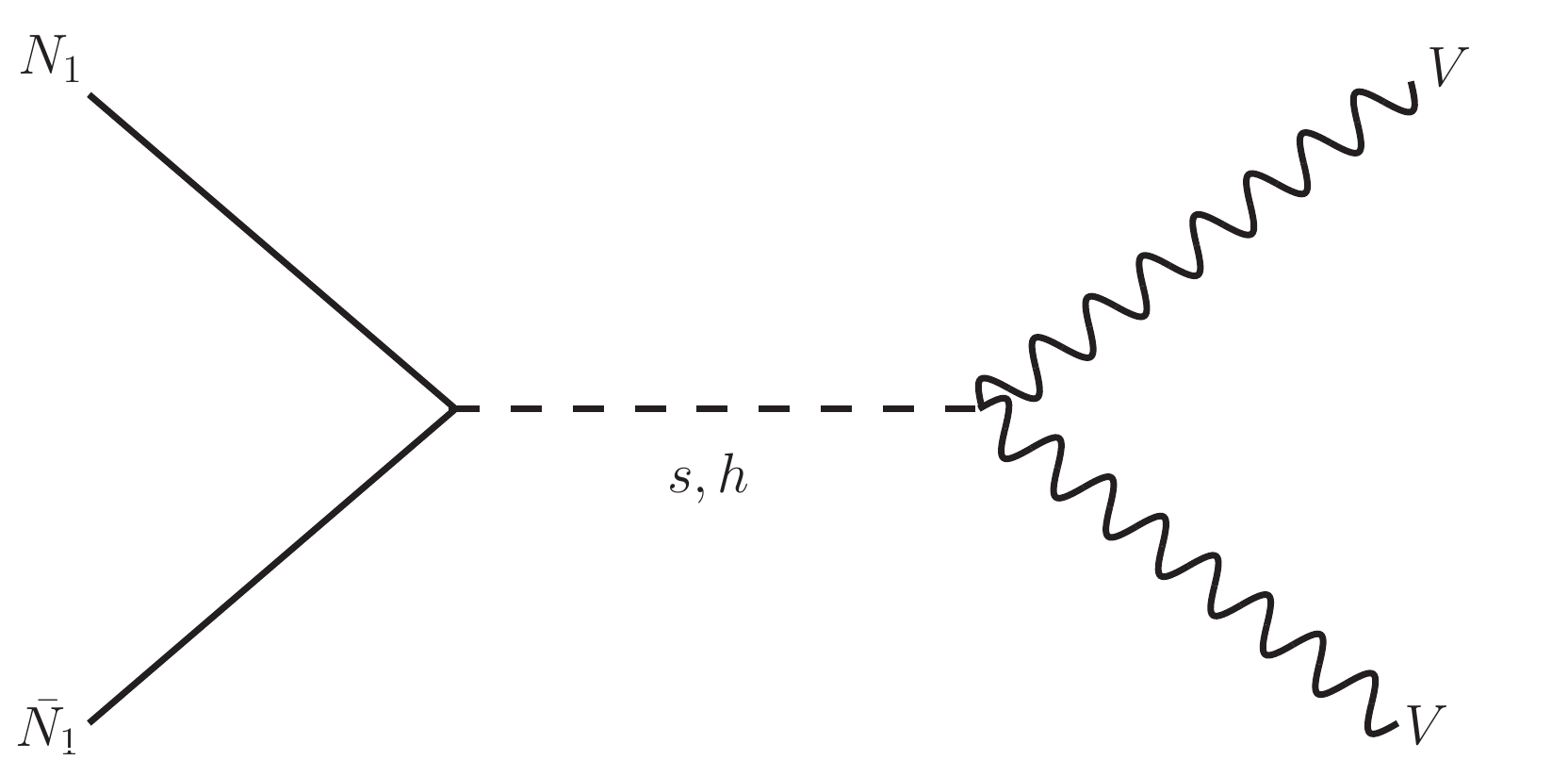}
$$
$$
\includegraphics[height=4 cm, width=5 cm,angle=0]{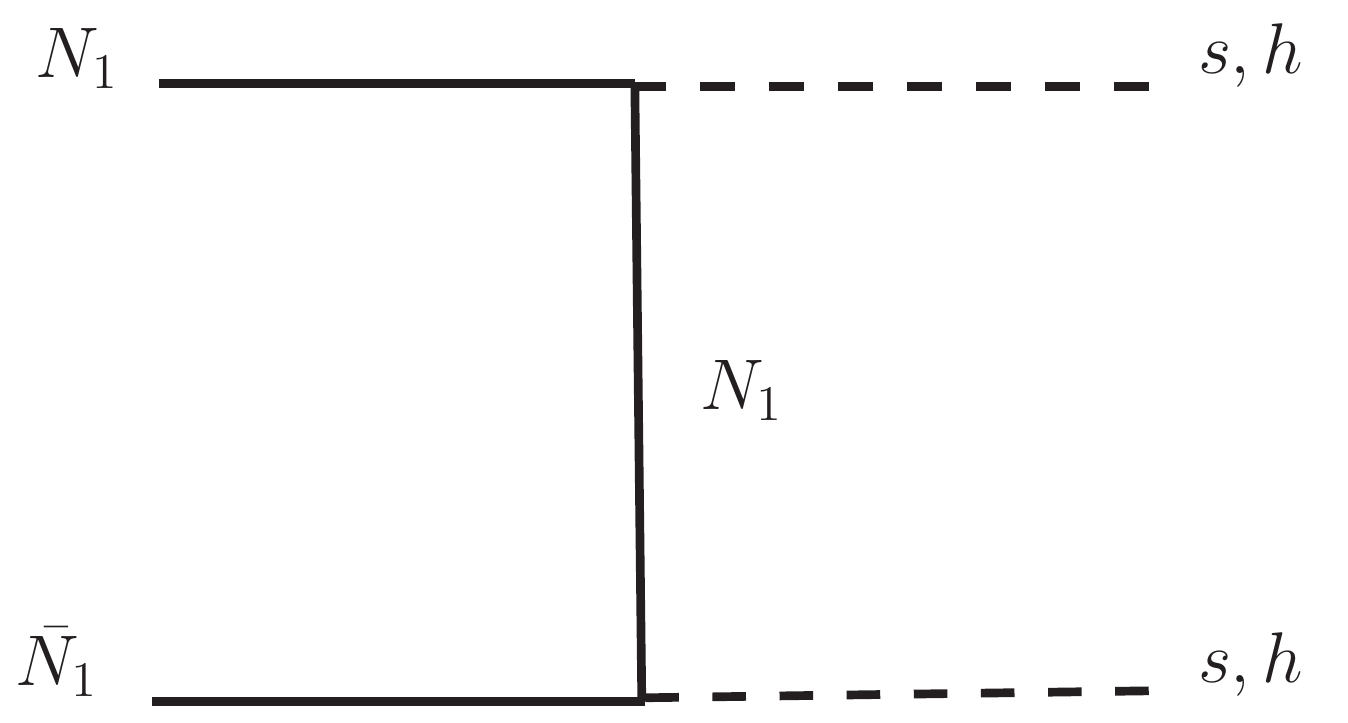}
\includegraphics[height=4 cm, width=5 cm,angle=0]{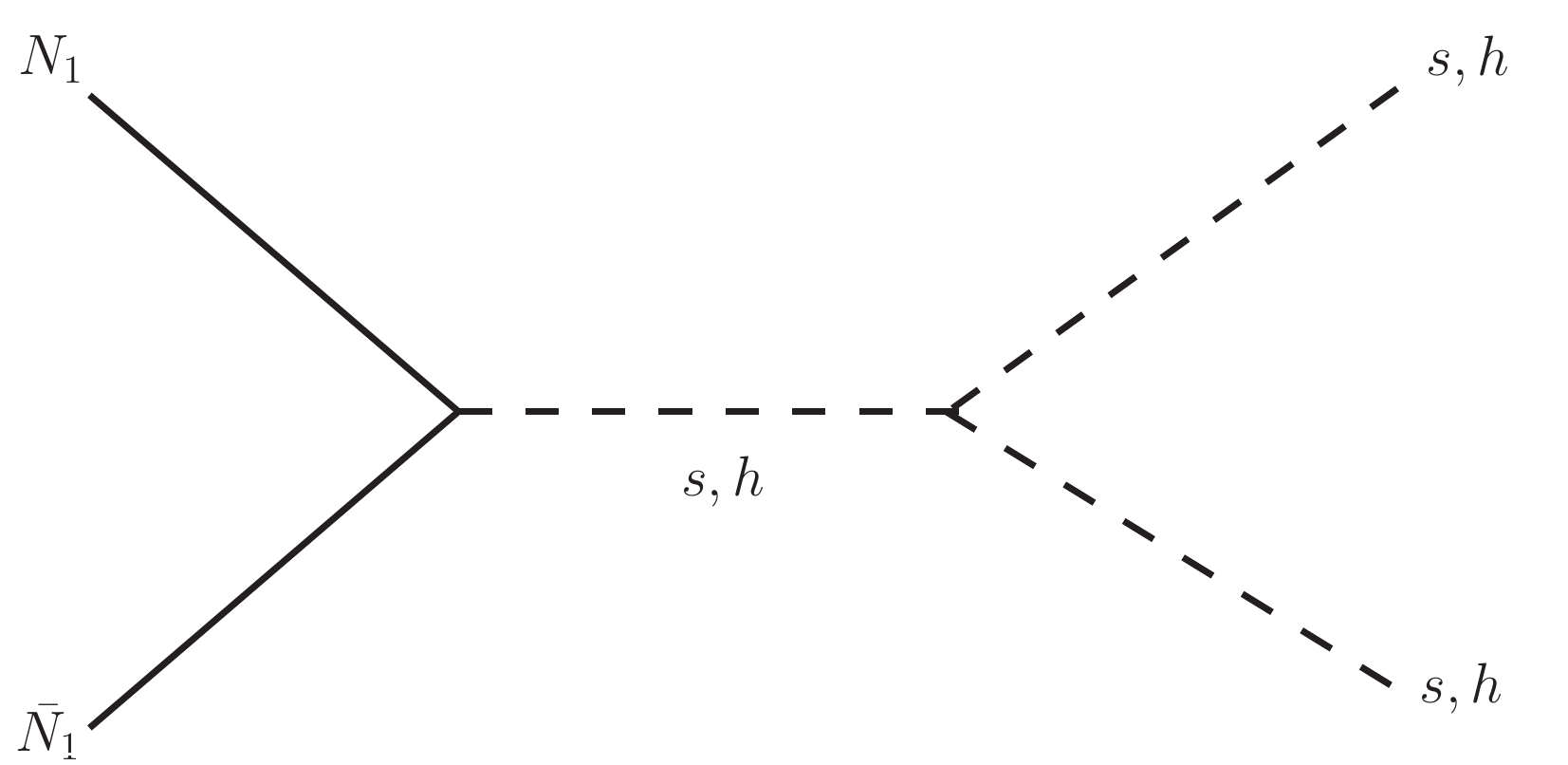}
\includegraphics[height=4 cm, width=5 cm,angle=0]{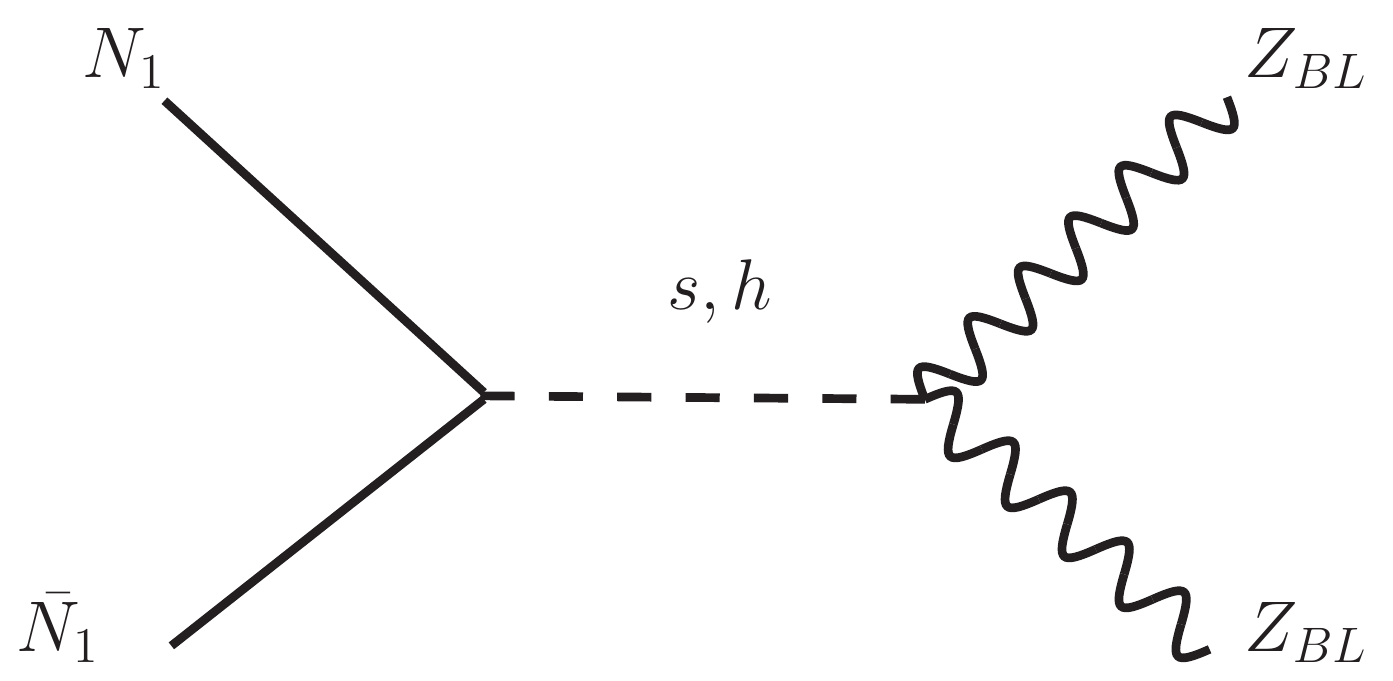}
$$
$$
\includegraphics[height=4 cm, width=5 cm,angle=0]{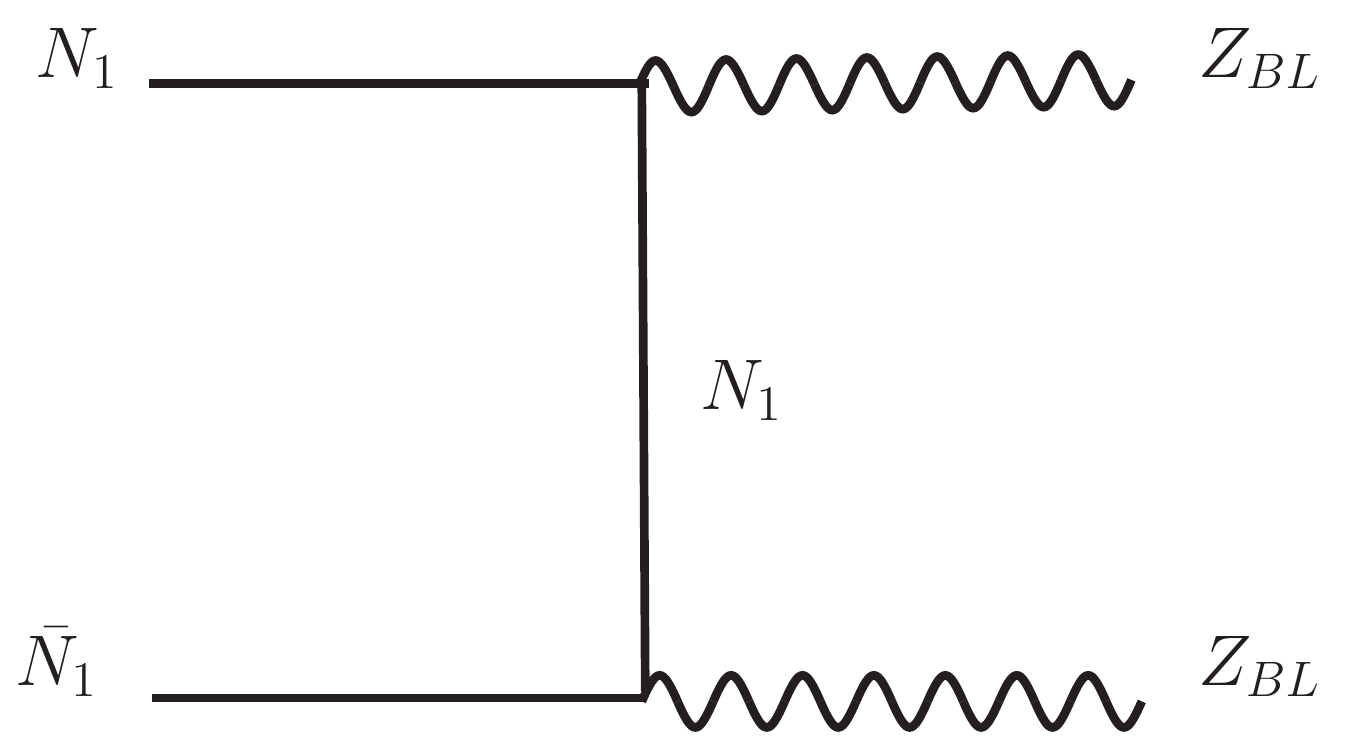}
$$
\caption{Annihilation processes for $N_1$ to SM.}
\label{anni_N}
\end{figure}

\begin{figure}[H]
$$
\includegraphics[height=4 cm, width=5 cm,angle=0]{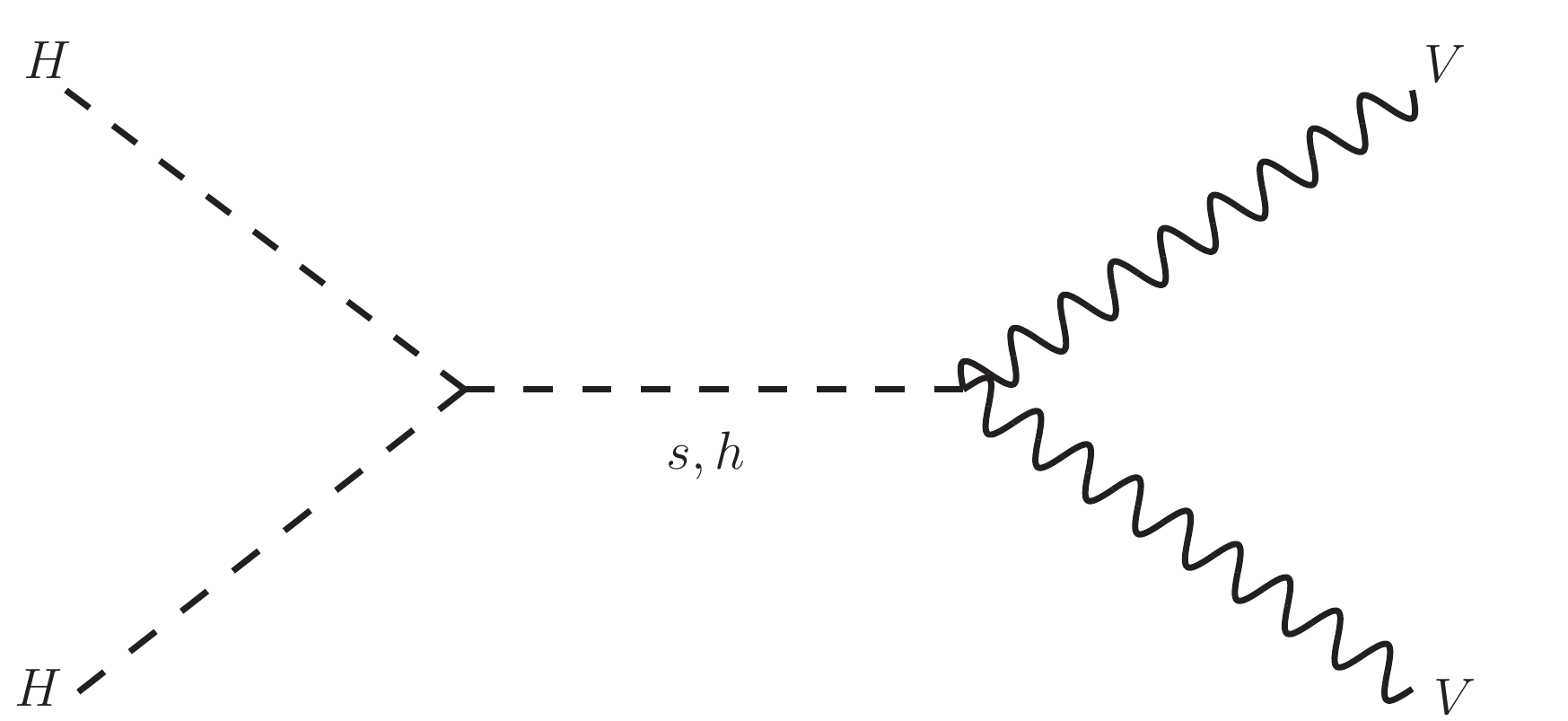}~~~
\includegraphics[height=4 cm, width=5 cm,angle=0]{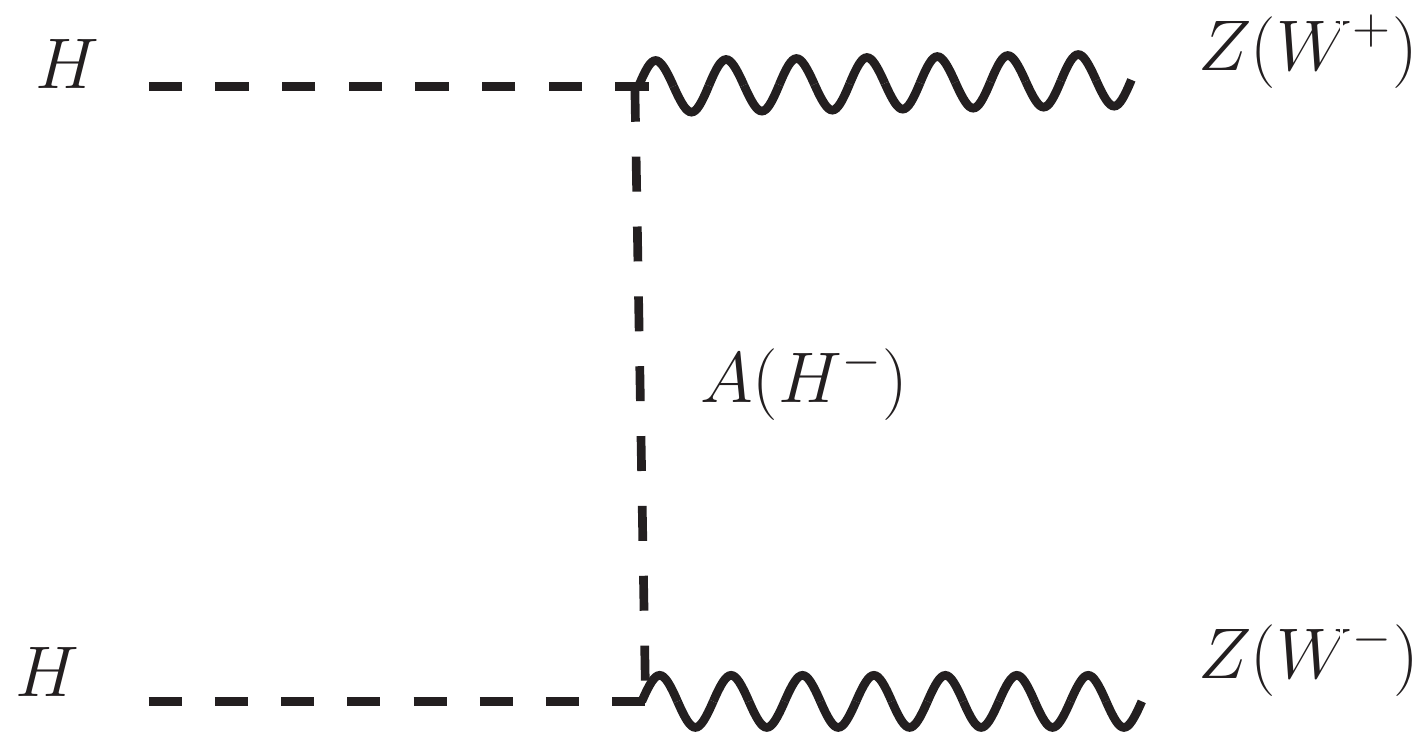}~~~
\includegraphics[height=4 cm, width=5 cm,angle=0]{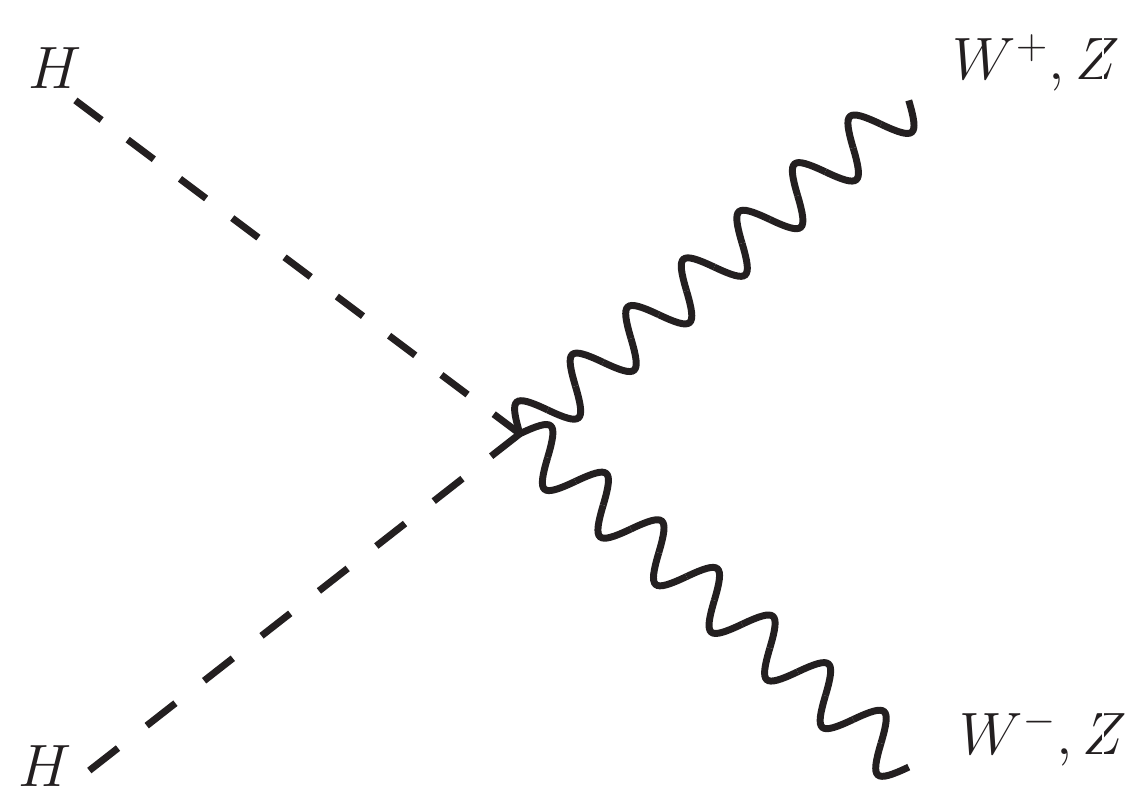}
$$
$$
\includegraphics[height=4 cm, width=5 cm,angle=0]{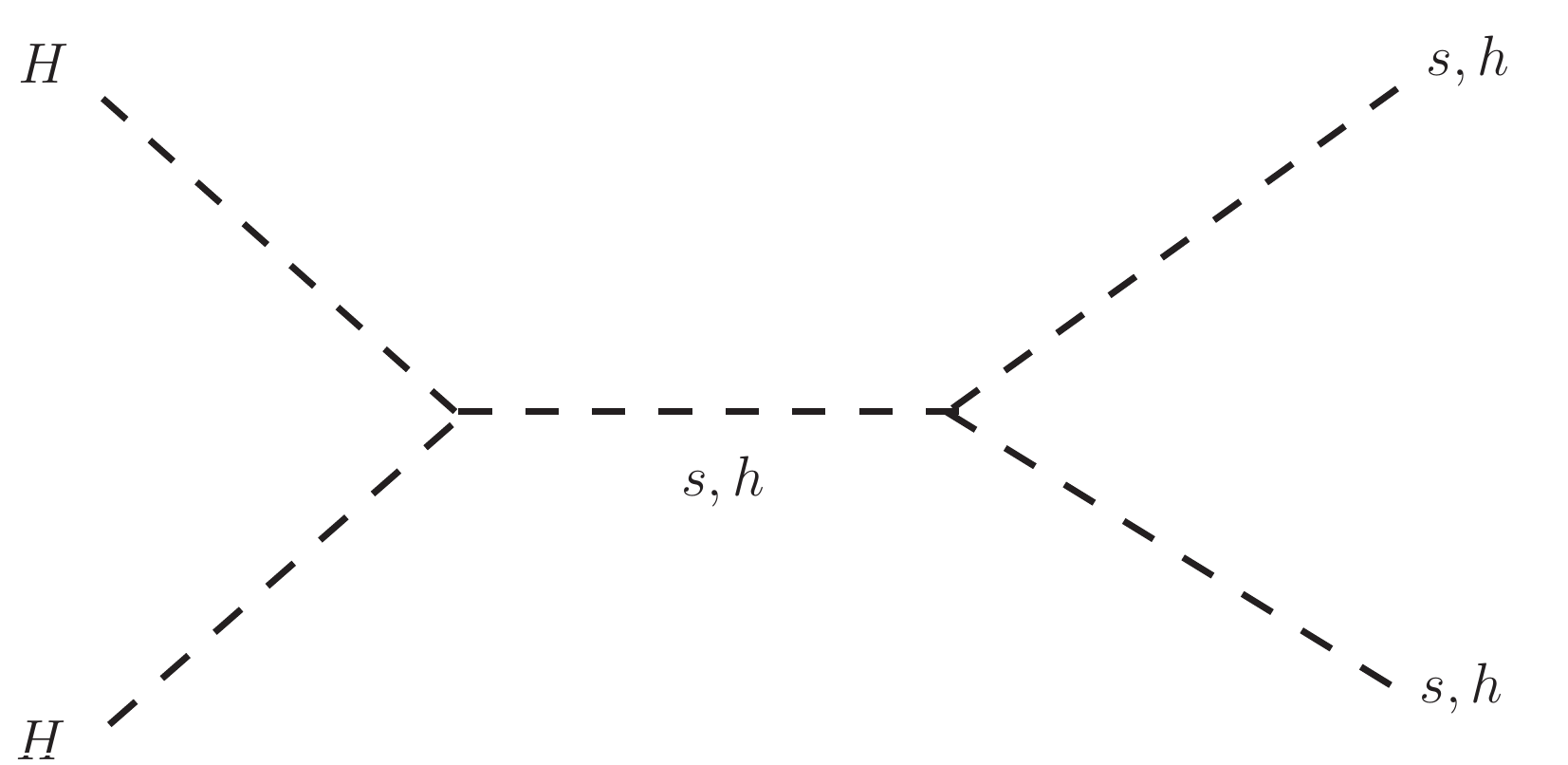}~~~
\includegraphics[height=4 cm, width=5 cm,angle=0]{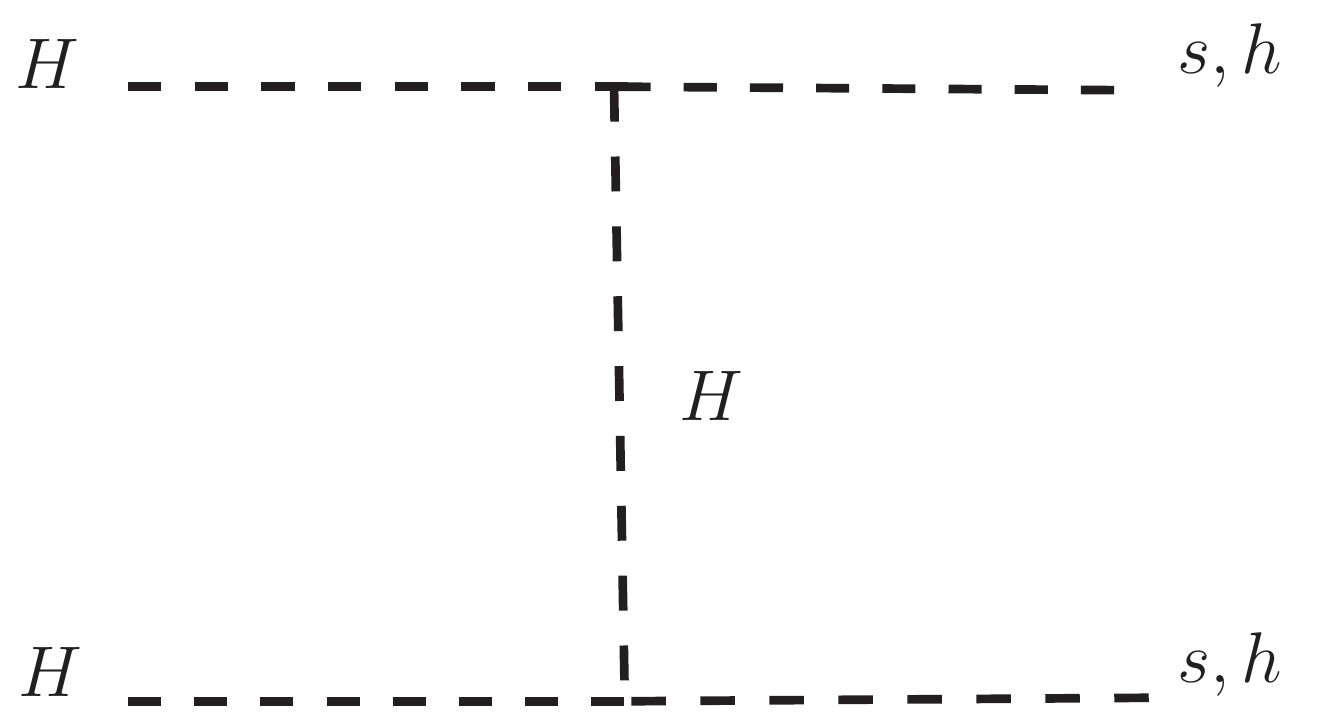}~~~
\includegraphics[height=4 cm, width=5 cm,angle=0]{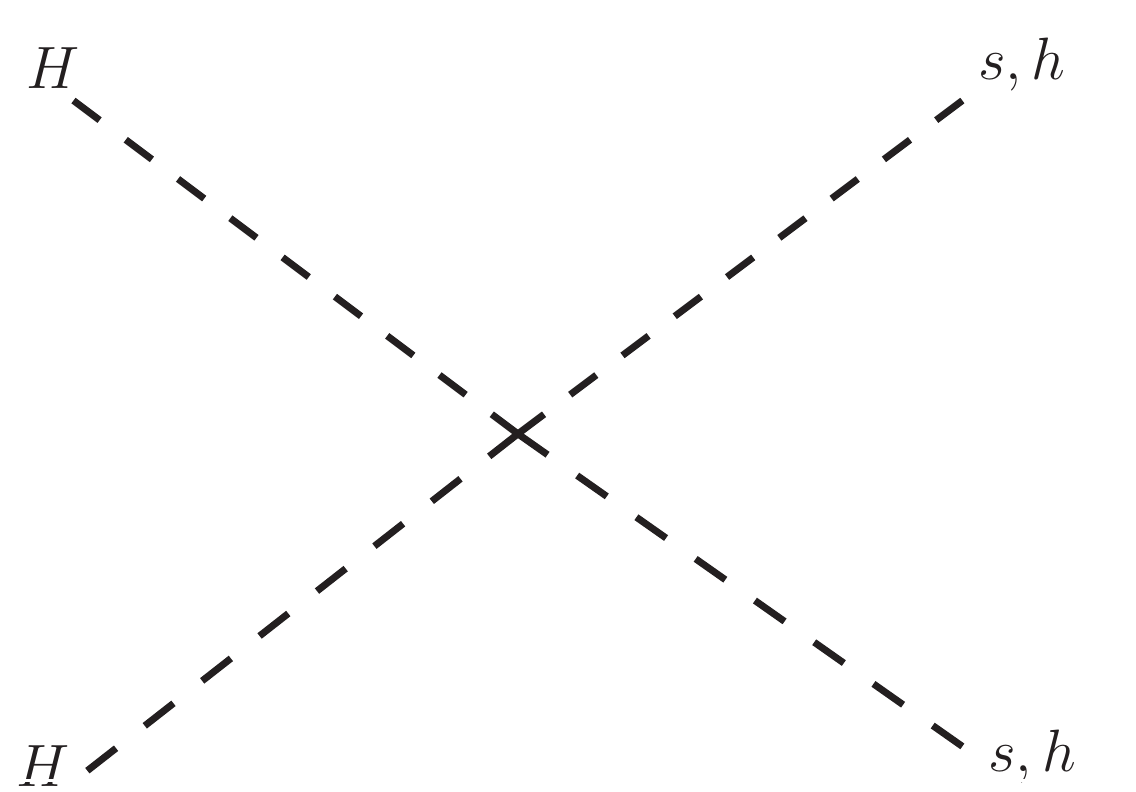}
$$
$$
\includegraphics[height=4 cm, width=5 cm,angle=0]{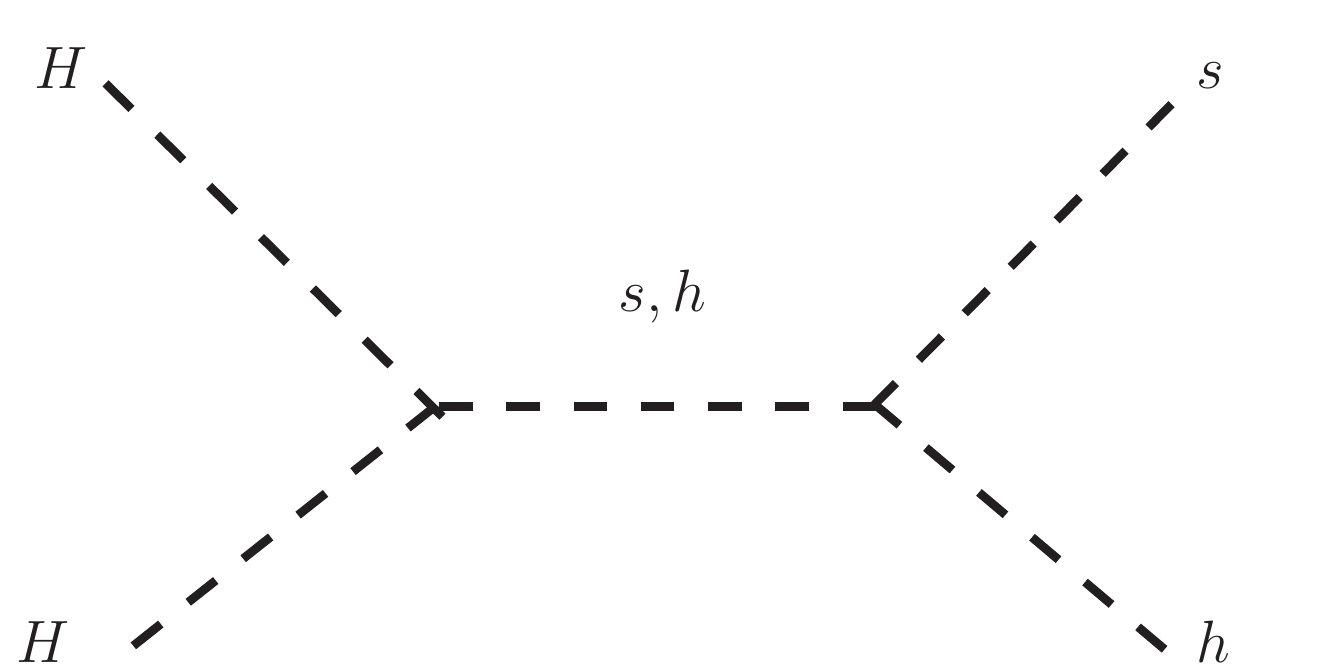}~~~
\includegraphics[height=4 cm, width=5 cm,angle=0]{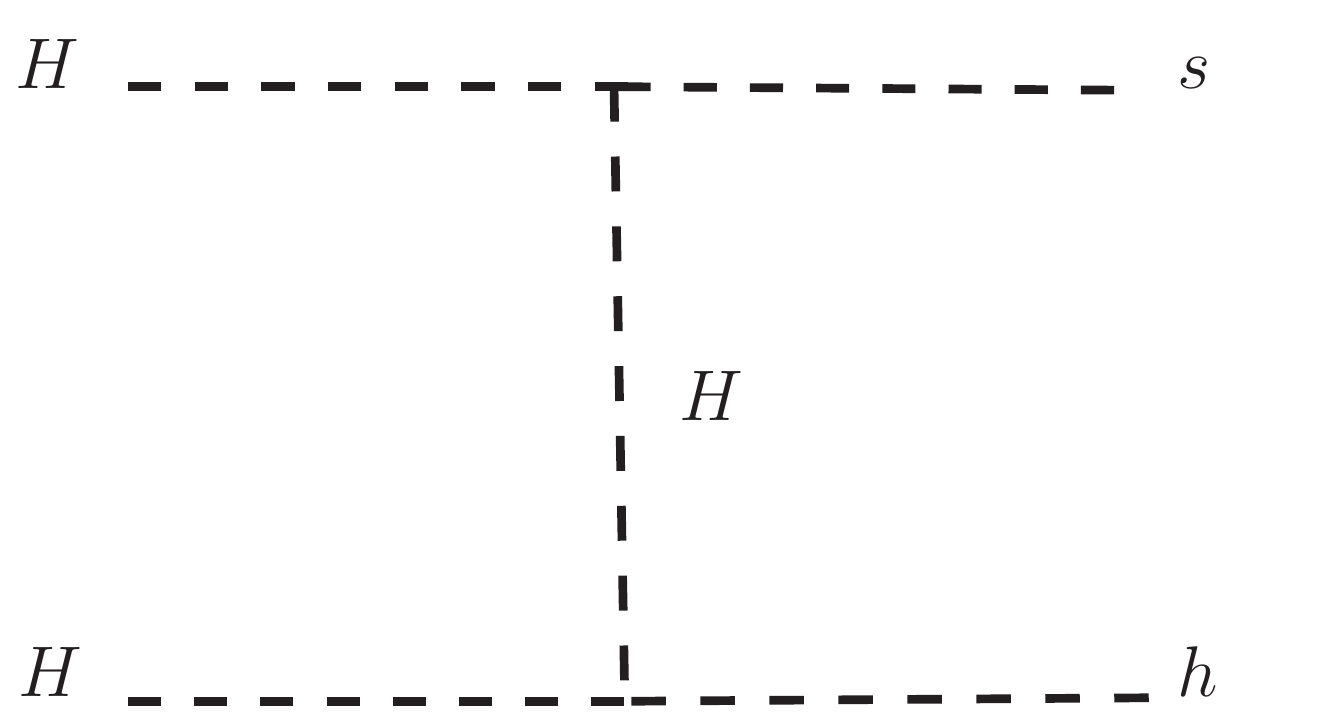}~~~
\includegraphics[height=4 cm, width=5 cm,angle=0]{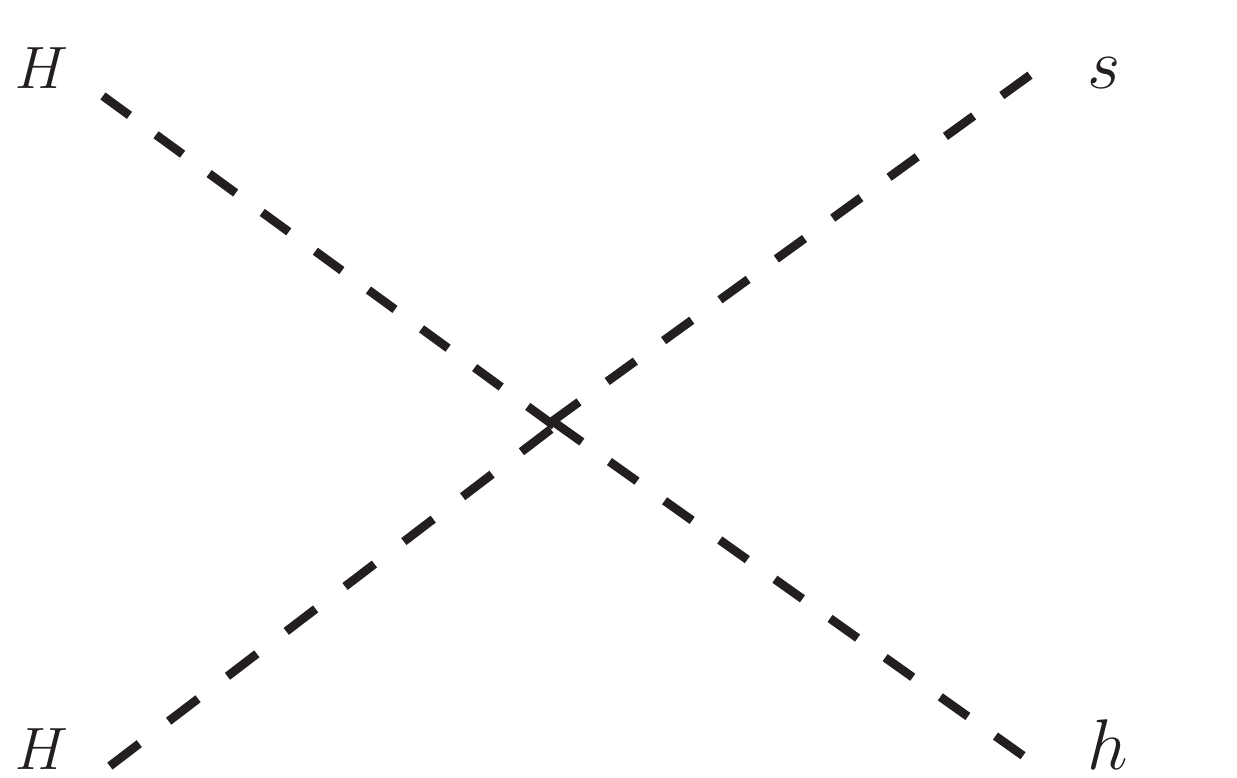}
$$
$$
\includegraphics[height=4 cm, width=5 cm,angle=0]{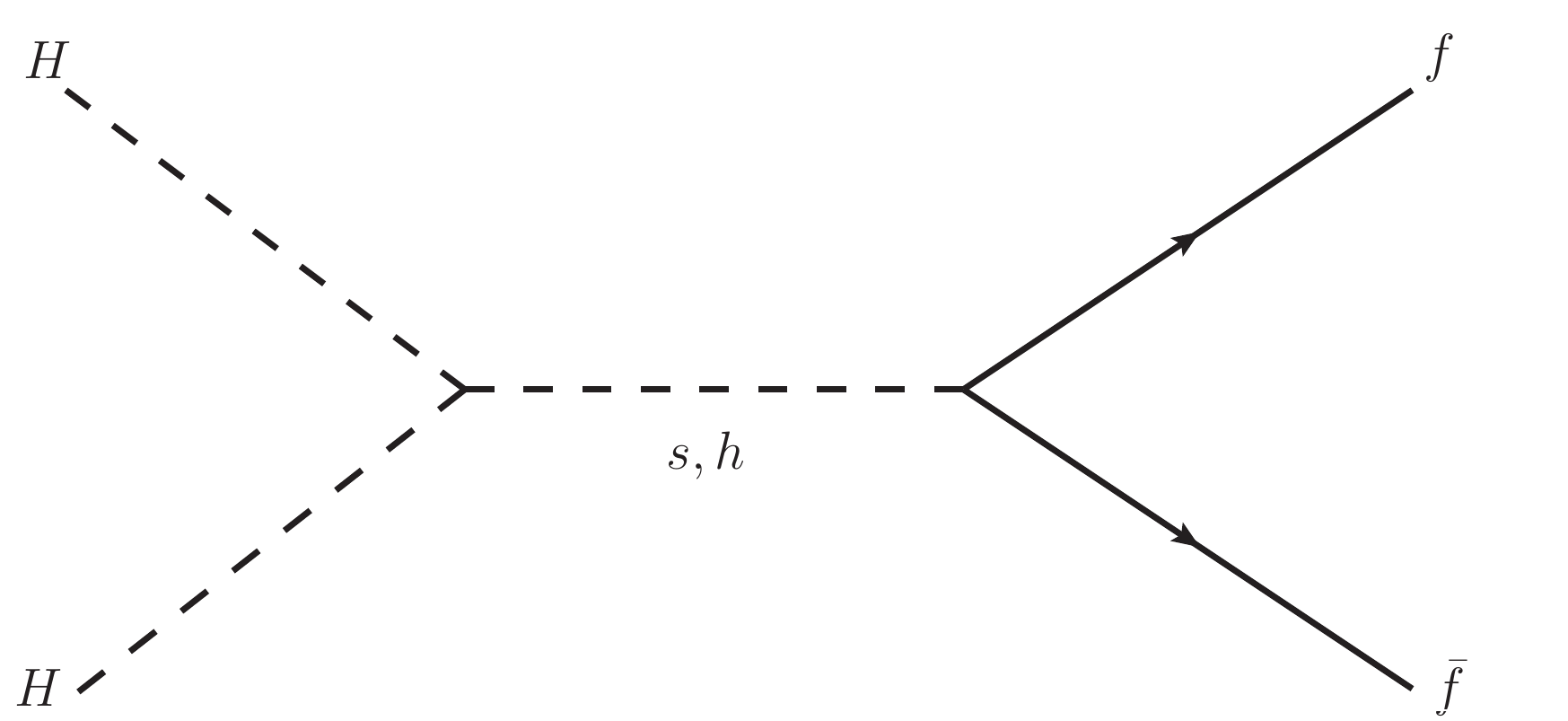}
$$
\caption{Annihilation processes for $H$}
\label{anni_H}
\end{figure}

\begin{figure}[H]
$$
\includegraphics[height=4 cm, width=5 cm,angle=0]{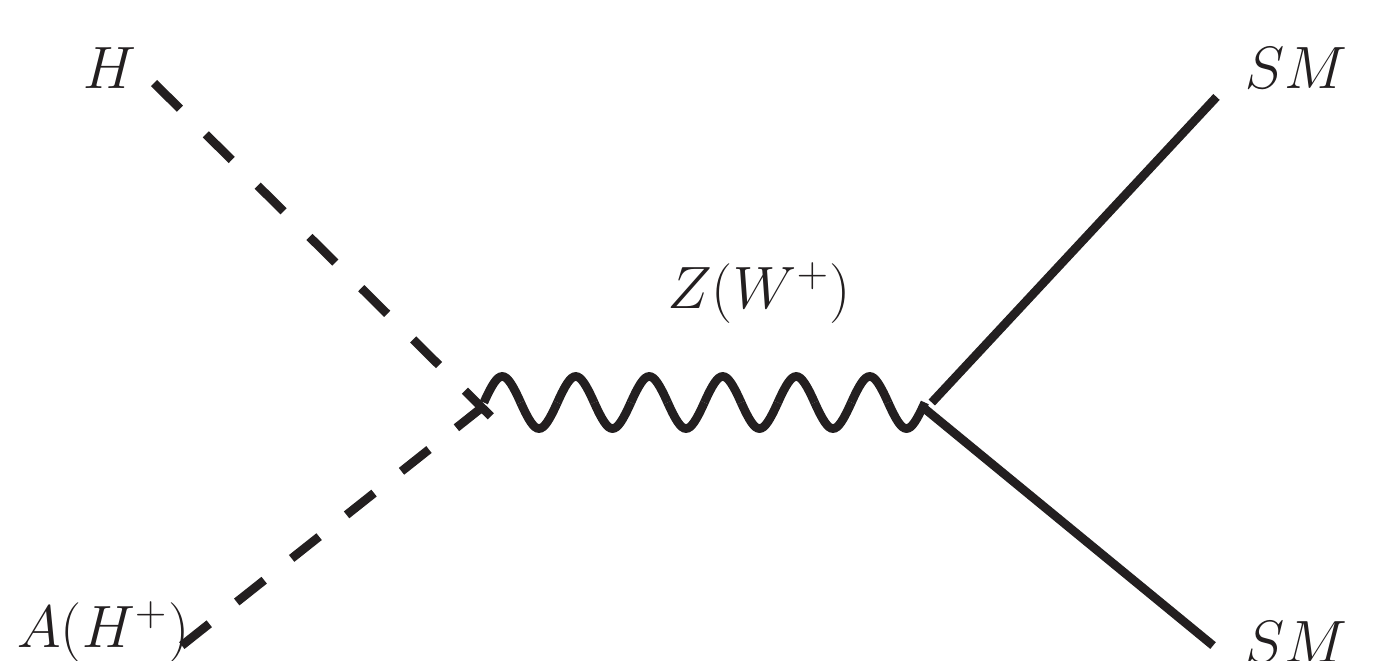}~~~
\includegraphics[height=4 cm, width=5 cm,angle=0]{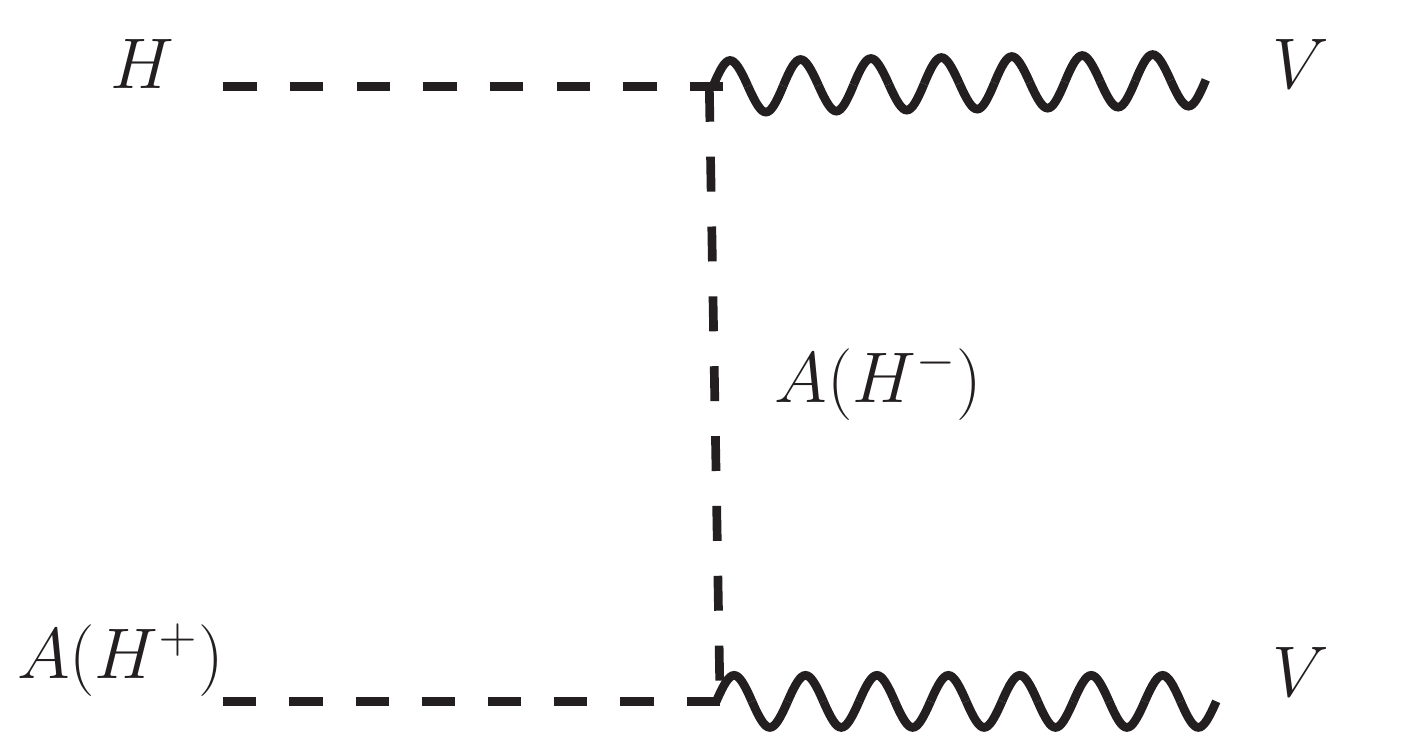}~~~
\includegraphics[height=4 cm, width=5 cm,angle=0]{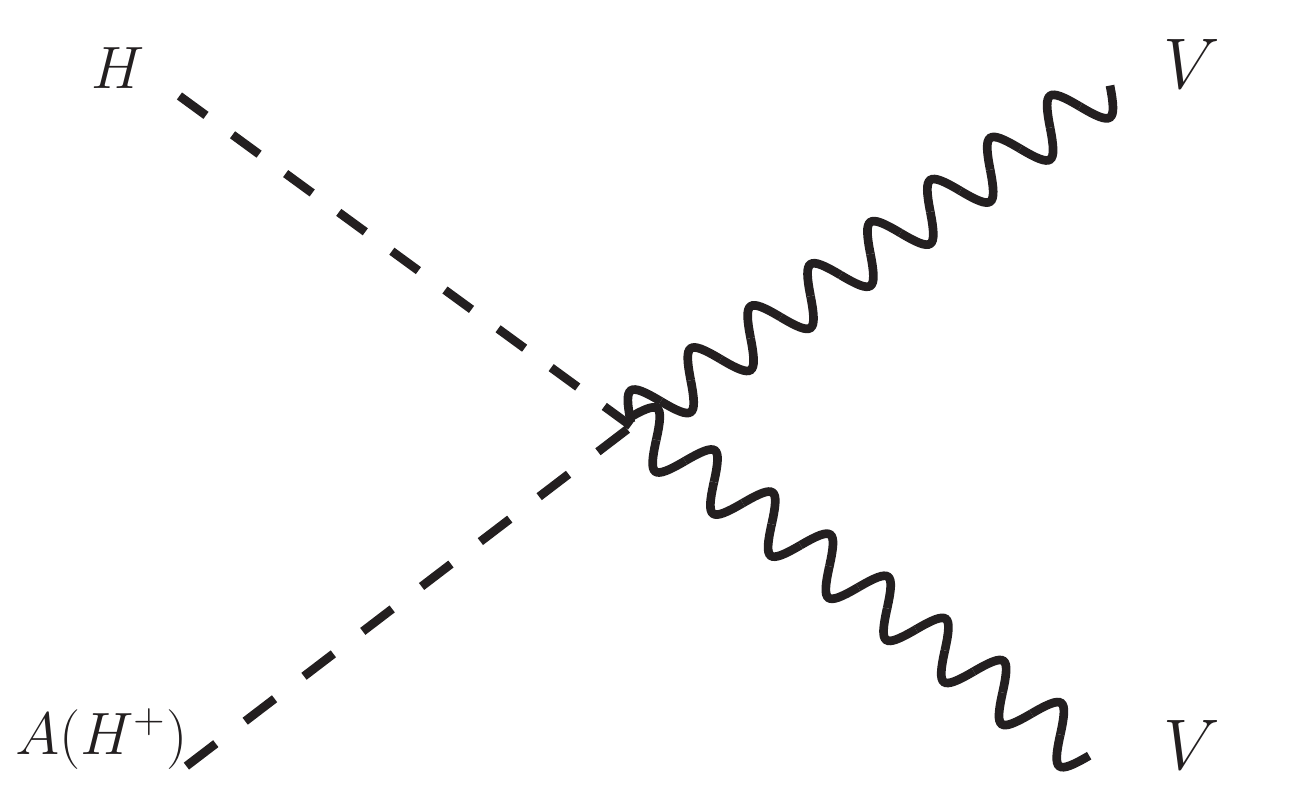}
$$
$$
\includegraphics[height=4 cm, width=5 cm,angle=0]{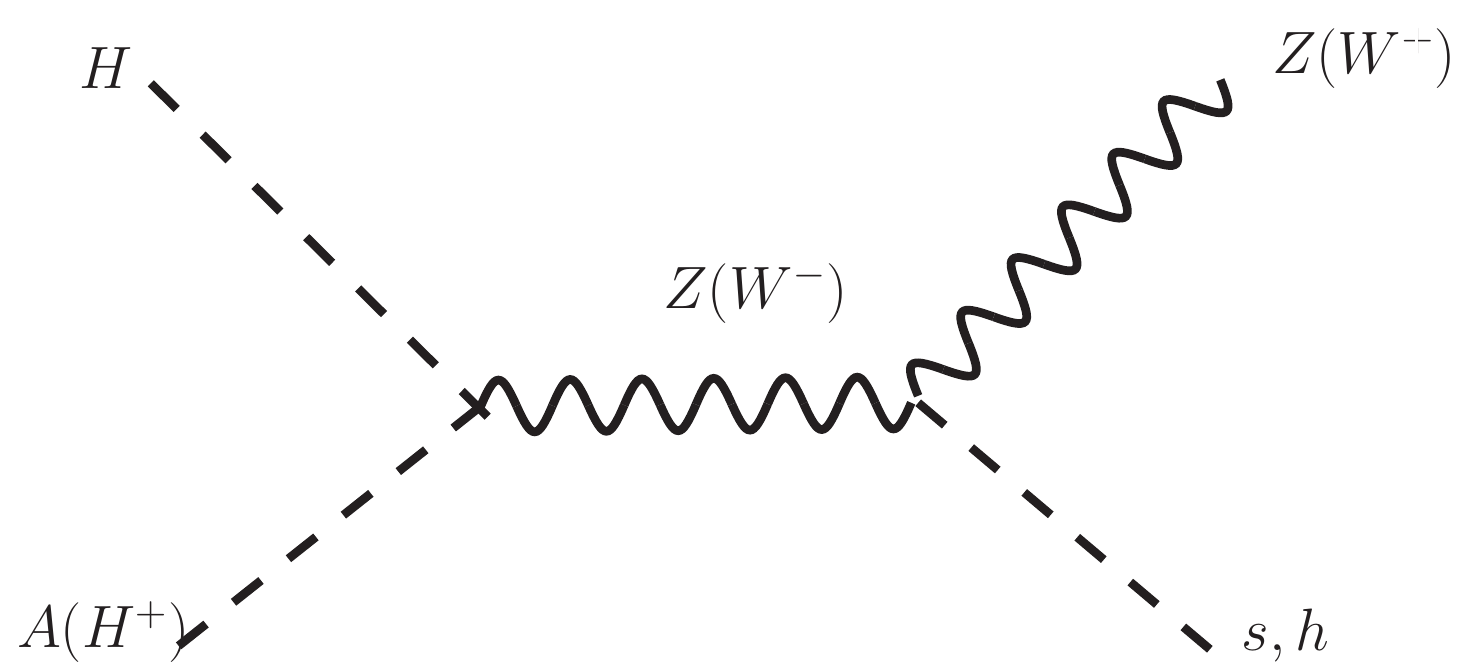}~~~
\includegraphics[height=4 cm, width=5 cm,angle=0]{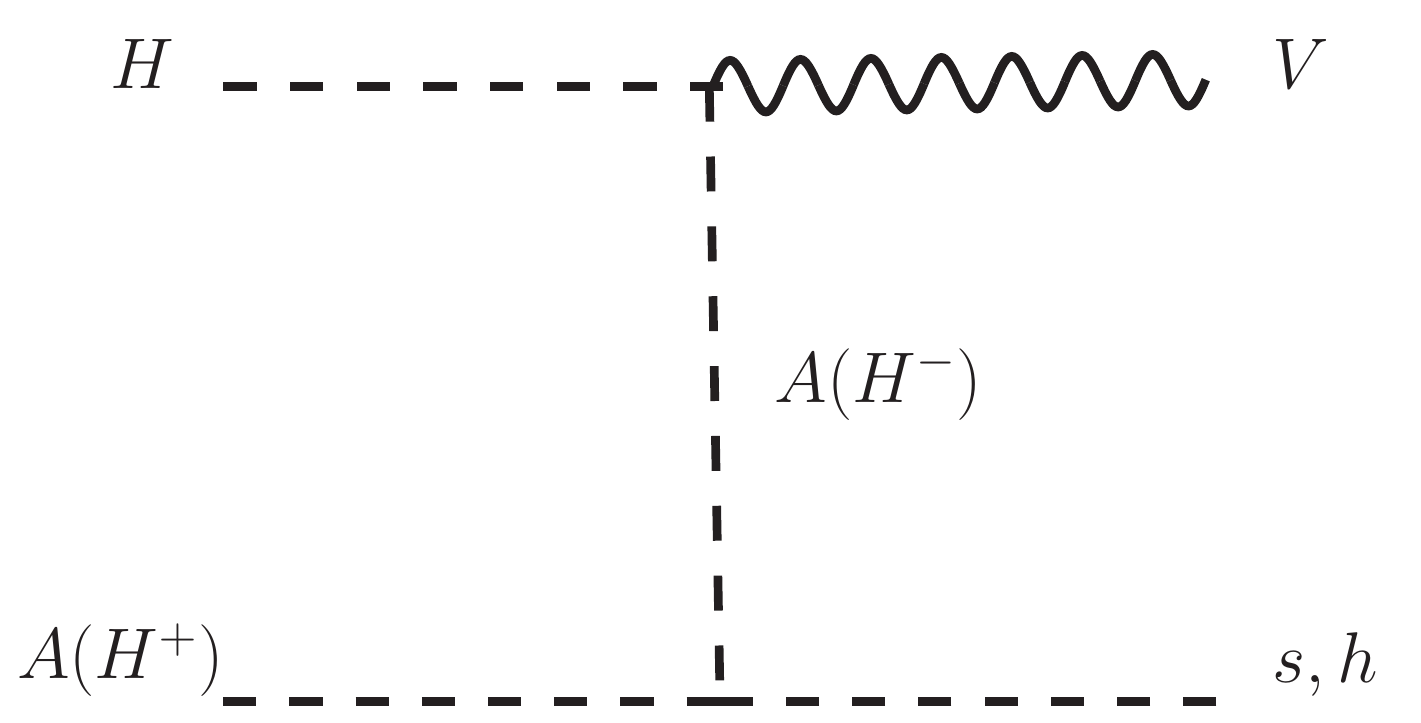}
$$
\caption{Co-annihilation processes for $H$.}
\label{coanni_H}
\end{figure}

\begin{figure}[htb!]
$$
\includegraphics[height=4 cm, width=5 cm,angle=0]{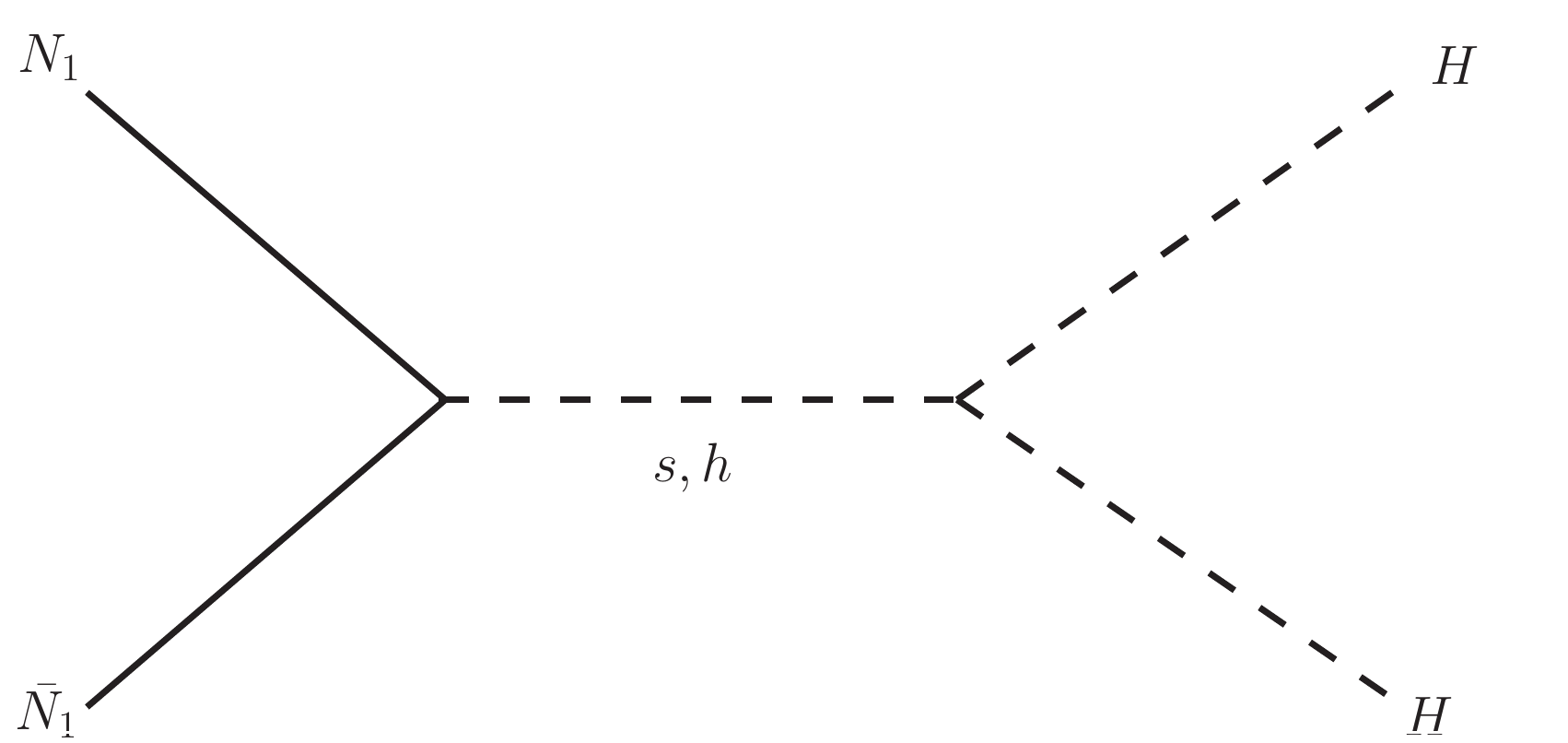}~~~
\includegraphics[height=4 cm, width=5 cm,angle=0]{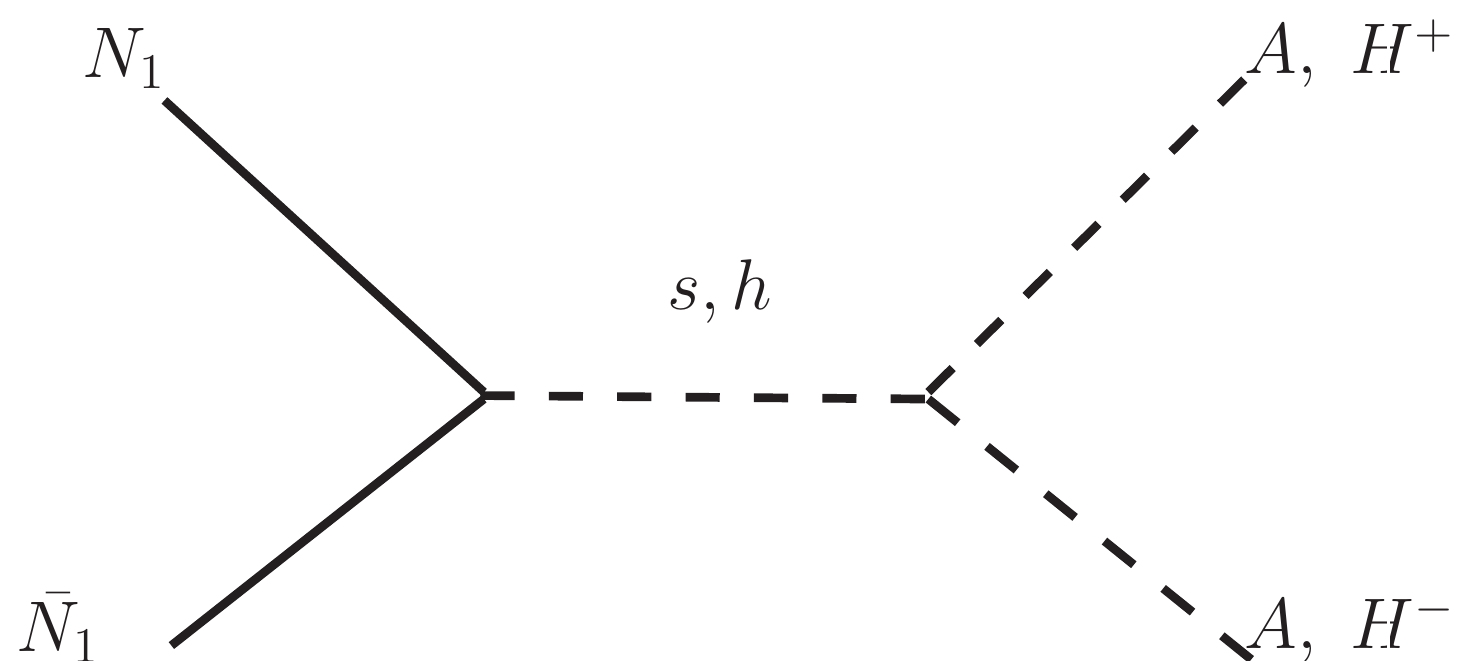}
$$
\caption{$N_1 -H/A/H^+$ conversion processes assuming $M_1 >M_{H,A,H^+} $.}
\label{conversion}
\end{figure}

The two DM candidates of this model are lightest right handed neutrino $N_1$ and the CP-even component $H$ of the inert Higgs 
doublet. $N_1$ talks to SM via Yukawa interaction (recall Eq.(\ref{eq1a})) and $U(1)_{B-L}$ gauge interaction. Relic density of $N_1$ 
(when assumed to be present in equilibrium with SM at early universe) is primarily dictated 
by its annihilations to SM, which are all imperatively s-channel processes mediated by $h, ~s~ \rm{or}~ Z_{BL}$ as shown in Fig.\ref{anni_N}. 
The inert DM $H$ depletes number density via annihilation channels to SM as shown in Fig.~\ref{anni_H}. 
Main contributions come from (i) exchanging $h$ and $s$ in the s-channel (ii) exchanging $A$ and $H^+$ in the t-channel, and (iii) through the four-point like $HH-hh$, 
$HH-ss$,~$HH-sh$ and $HH-VV$ (gauge interactions). Co-annihilation of $H$ with the heavier components of the doublet add to the 
number changing process of inert DM and plays a crucial role as shown in Fig.\ref{coanni_H}. In a two component DM set up, a key role is played by 
DM-DM conversion as we have here. The Feynman graph for such conversion is shown in Fig.~\ref{conversion}. 
Through this, the heavier DM component annihilates into the lighter, 
for example, with $M_{1}>M_H$, $N_1N_1 \to HH$ annihilation occurs via s-channel Higgs and $s$ 
mediation and the contribution directly adds to annihilation cross-section of the heavier component to SM. 
The lighter component being produced from the heavier one, faces milder changes in thermal decoupling, and relic density gets altered if its annihilation to SM is comparable or smaller than the conversion production.

The expressions for $N_1 N_1 \longrightarrow H H, A A, H^+ H^-$ (when $M_1 > M_H , M_A , M_{H^+}$ ) annihilation cross-section is mentioned below for the sake of completion.
\besub
\bea
\sigma_{N_1 N_1 \to H H} &=& \frac{1}{16 \pi s} \sqrt{\frac{s - 4 M_1^2}{s - 4 M_H^2}}|\frac{y_{h N_1 N_1} \l_{H H h}}{s - M_h^2 + i M_h \Gamma_h} + \frac{y_{s N_1 N_1} \l_{H H s}}{s - M_s^2 + i M_s \Gamma_s}|^2 (s - 4 M_1^2), \\
\sigma_{N_1 N_1 \to A A} &=& \frac{1}{16 \pi s} \sqrt{\frac{s - 4 M_1^2}{s - 4 M_A^2}}|\frac{y_{h N_1 N_1} \l_{A A h}}{s - M_h^2 + i M_h \Gamma_h} + \frac{y_{s N_1 N_1} \l_{A A s}}{s - M_s^2 + i M_s \Gamma_s}|^2 (s - 4 M_1^2), \\
\sigma_{N_1 N_1 \to H^+ H^-} &=& \frac{1}{16 \pi s} \sqrt{\frac{s - 4 M_1^2}{s - 4 M_{H^+}^2}}|\frac{y_{h N_1 N_1} \l_{H^+ H^- h}}{s - M_h^2 + i M_h \Gamma_h} + \frac{y_{s N_1 N_1} \l_{H^+ H^- s}}{s - M_s^2 + i M_s \Gamma_s}|^2 (s - 4 M_1^2).
\eea
\label{sigma_conv}
\eesub
The expressions for the various scalar and Yukawa couplings are to be read in the appendix. The comoving number densities of $N_1$ and $H$ are obtained by 
solving the coupled Boltzmann equations below. The parameter $x$ is however redefined to $x = \mu/T$, where $\mu$ is the reduced mass defined through: $~\mu= \frac{M_{1}M_{H_2}}{M_{1}+M_{H}}$\footnote{We adopt the notation from a recent article on two component DM \cite{Bhattacharya:2018cgx}}.

\besub
\bea
\frac{dy_{N_1}}{dx} &=& \frac{-1}{x^2}\bigg{[}\langle \sigma v_{N_{1}N_{1}\rightarrow XX}\rangle \left(y_{N_1}^{2}-(y_{N_1}^{EQ})^2\right)~+~\langle \sigma v_{N_{1}N_{1}\rightarrow HH}\rangle \left ( y_{N_1}^{2}-\frac{(y_{N_1}^{EQ})^2}{(y_{H}^{EQ})^2} y_{H}^{2}\right)\Theta(M_{1}-M_{H}) \nonumber \\
&&
-~\langle\sigma v_{HH \rightarrow N_{1}N_{1}}\rangle \left( y_{H}^{2}-\frac{(y_{H}^{EQ})^2}{(y_{N_1}^{EQ})^2}y_{N_1}^{2}\right)~\Theta(M_{H}-M_{1})\bigg{]} \\
\frac{dy_{H}}{dx} &=& \frac{-1}{x^2}\bigg{[} \langle \sigma v_{HH\rightarrow XX} \rangle \left (y_{H}^{2}-(y_{H}^{EQ})^2\right )~+~\langle \sigma v_{HH\rightarrow N_{1}N_{1}}\rangle \left (y_{H}^{2}-\frac{(y_{H}^{EQ})^2}{(y_{N_1}^{EQ})^2}y_{N_1}^{2}\right )\Theta(M_{H}-M_{1}) \nonumber \\
&&
-~\langle \sigma v_{N_{1}N_{1} \rightarrow HH}\rangle \left (y_{N_1}^{2}-\frac{(y_{N_1}^{EQ})^2}{(y_{H}^{EQ})^2}y_{H}^{2}\right )\Theta(M_{1}-M_{H})\bigg{]}.
\eea
\label{eq:cBEQ}
\eesub

Here $y_{i}$ ($i = N,H$) is related to yield $Y_i=\frac{n_i}{s}$ (where $n_i$ refers to DM density and $s$ is entropy density) 
by $y_i=0.264M_{\text{Pl}}\sqrt{g_*}\mu Y_{i}$; similarly for equilibrium density, $y_i^{EQ}= 0.264M_{\text{Pl}}\sqrt{g_*}\mu Y_{i}^{EQ}$, with equilibrium distributions ($Y_{i}^{EQ}$) in terms of $\mu$ take the form
\bea
Y_{i}^{EQ}(x) = 0.145\frac{g}{g_{*}}x^{3/2}\bigg{(}\frac{m_{i}}{\mu}\bigg{)}^{3/2}e^{-x\big{(}\frac{m_{i}}{\mu}\big{)}}.
\eea 
Here $M_{\rm Pl} = 1.22\times10^{19} ~{\rm GeV}$, $g_{*} = 106.7$\footnote{One is supposed to use $g_{*s}$ in the above equations. However, $g_{*s} \simeq g_{*}$ holds for temperatures $\sim \mathcal{O}$ (GeV) or above\cite{Kolb:1990vq}.} and $m_i$ stands for $M_1$ and $M_H$. 
In Eqn.~\ref{eq:cBEQ}, $X$ represents SM particles, $H^{\pm}$ and $A$. This is because $H^{\pm}$ is expected to be in equilibrium with the thermal plasma by  electromagnetic interactions whereas $A$ being heavier than $H$, can also decay to $H$ and SM fermions ($f$) via off shell $Z \to f\bar{f}$ to be in equlibrium with the thermal bath. The thermally averaged annihilation cross section, given by 
\bea
\langle \sigma v\rangle = \frac{1}{8m^{4}_{i}T K_2^2(\frac{m_{i}}{T})}\int\limits^{\infty}_{4m_{i}^2}\sigma(s-4m_{i}^2)\sqrt{s}K_1\bigg{(}\frac{\sqrt{s}}{T}\bigg{)}ds
\label{eq:sigmav}
\eea
is evaluated at $T_f$ and denoted by $\langle \sigma v \rangle_f$. The freeze-out temperature $T_f$ is derived from 
the equality condition of DM interaction rate $\Gamma = n_{\rm DM} \langle \sigma v \rangle$ with the rate of expansion of the universe $\bar{H}(T) \simeq \sqrt{\frac{\pi^2 g_*}{90}}\frac{T^2}{M_{\rm Pl}}$. In the above expression of Eq.(\ref{eq:sigmav}), $K_{1,2}(x)$ are the modified Bessel functions.
\\One should note that the contribution to the Boltzmann equations coming from the DM-DM conversion (corresponding to Fig.\ref{conversion}) will depend on the mass hierarchy of DM particles. This is described by the use of $\Theta$ function in the above equations. These coupled equations can be solved numerically to find the asymptotic abundance of the DM particles, $y_{i} \left (\frac{\mu}{m_{i}}x_{\infty} \right)$, which can be further used to calculate the relic:
\besub
\bea
\Omega_{i}h^2 &=& \frac{854.45\times 10^{-13}}{\sqrt{g_{*}}}\frac{m_{i}}{\mu}y_{i}\left ( \frac{\mu}{m_{i}}x_{\infty}\right ),
\eea
\eesub
where $x_{\infty}$ indicates an asymptotic value of $x$ after the freeze-out. The index $i$ stands for DM components in our scenario: $N_1, H$. The total relic abundance is a sum of the individual components.
\bea
\Omega_{T} h^2 &=& \Omega_{N_1}h^2+\Omega_{H}h^2
\eea

\noindent However, we use numerical techniques to solve for relic density of this two component model. The model was first implemented in \texttt{LanHEP}~\cite{Semenov:2008jy}. A compatible output was then fed into 
the publicly available tool \texttt{micrOMEGAs}4.1 ({capable of handling multipartite DM scenarios)\cite{Belanger:2014vza} to compute the relic densities of $N_1$ and $H$.

\subsection{Direct detection}

Direct detection experiments like LUX~\cite{Akerib:2016vxi}, PandaX-II~\cite{Zhang:2018xdp} and Xenon-1T~\cite{Aprile:2018dbl} search
for the evidence of dark matter via dark matter-nucleon scattering producing nuclear recoil signature. Unfortunately no events of such kind have been confirmed so far, which evidently provide 
bounds on the dark matter-nucleon scattering cross-section. In this section, we will illustrate the processes through which the DM components in our model interact with detector, and compute direct search 
cross-section. This will be required to obtain the limit on relevant DM parameters from non-observation in direct search to be compatible with correct relic abundance.

The elastic scattering processes for $N_1$ and $H$ with detector nucleon are shown in Fig. \ref{DD}. While both DM components can interact via t-channel Higgs and $s$ portal interactions 
(the latter suppressed by mixing angle), $N_1$ having $U(1)_{B-L}$ charge can also interact to nucleon via gauge interaction mediated by $Z_{BL}$. 

\begin{figure}[H]
\centering
\includegraphics[height=4 cm, width=5 cm,angle=0]{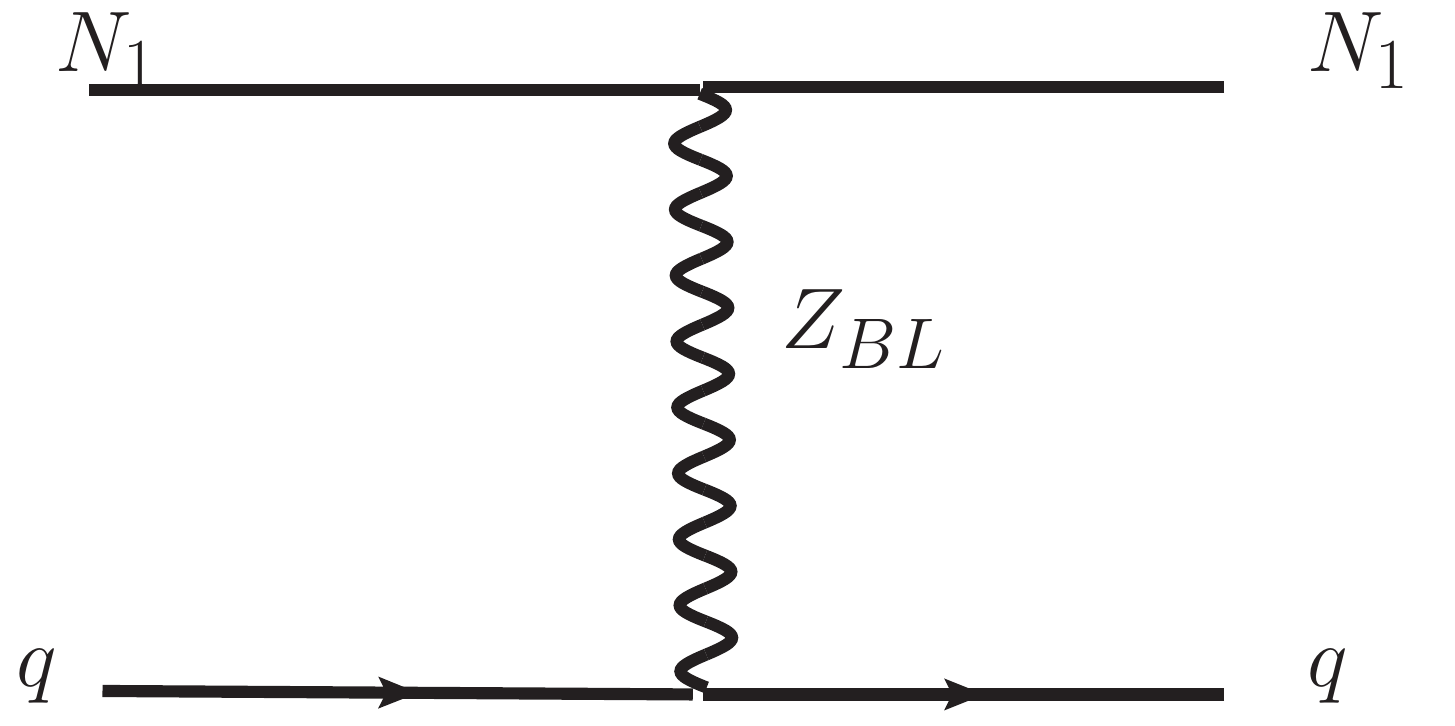}~~~
\includegraphics[height=4 cm, width=5 cm,angle=0]{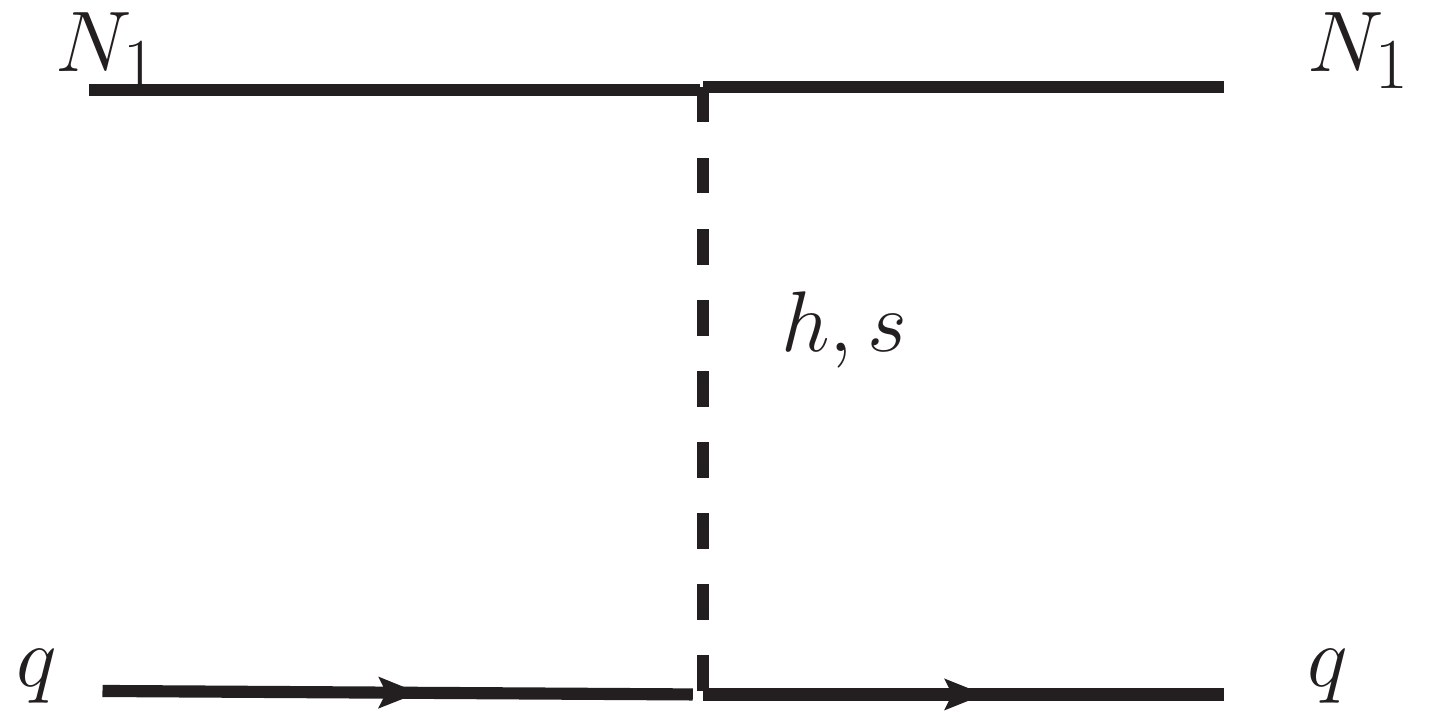} \\
\includegraphics[height=4 cm, width=5 cm,angle=0]{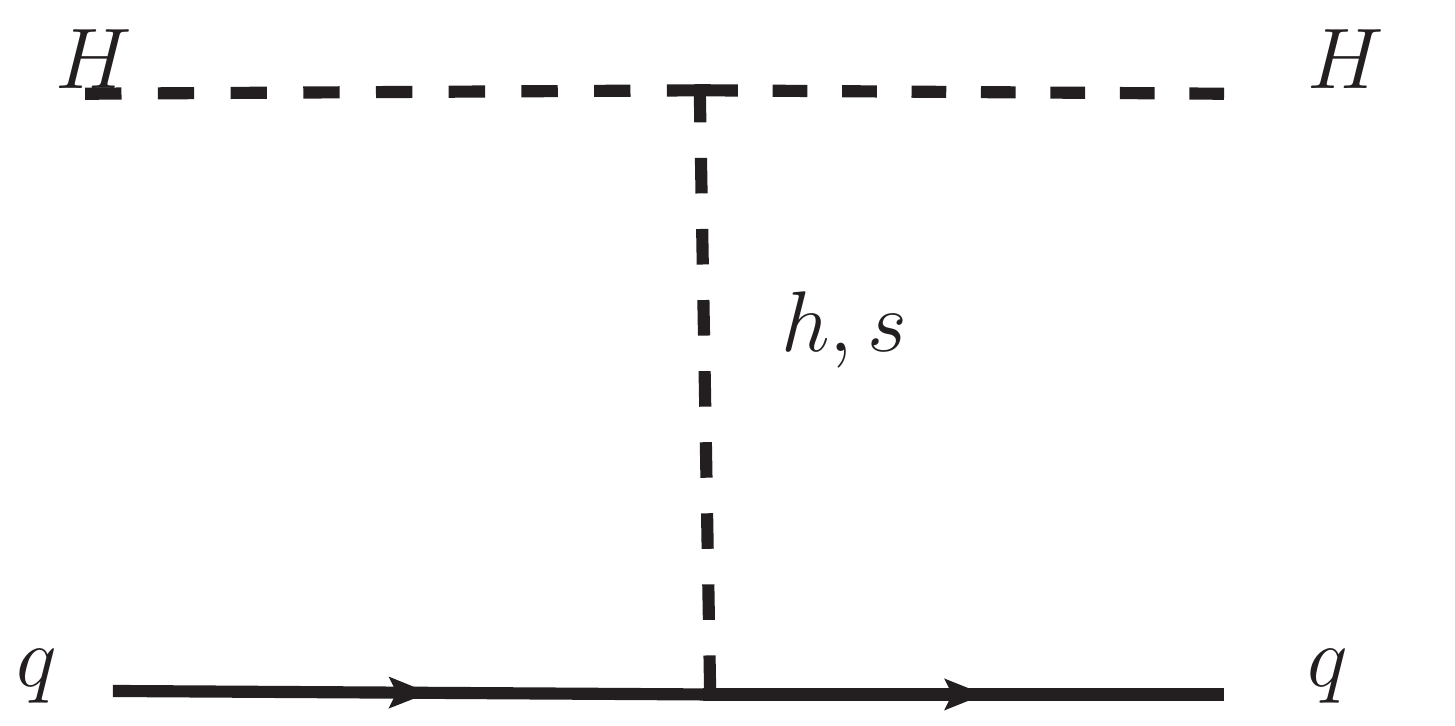}
\caption{$N_1, H$ scattering off nucleons elastically }
\label{DD}
\end{figure}

The spin-independent direct detection (SI-DD) cross section for $H$ and $N_1$ reads respectively
\besub
\bea
\sigma^{SI}_{H} = \frac{\mu_{H,n}^2}{4\pi}\bigg{[}\frac{m_n~f_n}{M_{H}~v}\bigg{(}\frac{\l_{HHh}}{M_h^2}+\frac{\l_{HHs}}{M_s^2}\bigg{)}\bigg{]}^2.\\
\sigma^{SI}_{N_1} = \sin{2\theta}\frac{\mu_{N_1,n}^2}{4\pi}\bigg{[}\frac{y_{11}~m_n~f_n}{v}\bigg{(}\frac{1}{M_s^2}-\frac{1}{M_h^2}\bigg{)}\bigg{]}^2
\eea
\eesub
where $\mu_{H,n} = m_n M_H /(m_n + M_H )$, $\mu_{N_1,n} = m_n M_1 /(m_n + M_1 )$ are the DM-nucleon reduced masses, $\l_{HHh}$ and $\l_{HHs}$ are the quartic coupling, $y_{11}$ is the Yukawa coupling involved 
in DM-Higgs interaction and $f_n$ = 0.2837 is the nucleon form factor~\cite{Alarcon:2011zs,Alarcon:2012nr} and $v$ is the SM Higgs VEV. In this two-component DM framework, the 
\emph{effective} SI-DD cross sections relevant for each of the candidates can be expressed by the individual DM-nucleon cross-section multiplied by the relative abundance of that particular component ($\Omega_i$) in total DM relic density ($\Omega_T$):
\bea
\sigma^{SI}_{i,eff} = \frac{\Omega_i}{\Omega_T}\sigma_{i}^{SI}.
\eea
A more careful analysis for multiparticle DM direct search cross section can be performed by computing total recoil rate (see for example, \cite{Bhattacharya:2016ysw,Herrero-Garcia:2017vrl}), however above procedure 
provides a correct order of magnitude estimate for individual components. Also note here, that direct search prospect and therefore constraint from non-observation of DM in direct detection only appears because the 
DM components are assumed to be present in early universe and thereafter freezes out via thermal decoupling. On the contrary, if DM is produced via `freeze-in', then the DM-SM coupling turns insignificant for correct relic, 
to produce no direct search signal. For example, we consider later a possibility of $N_1$ freezing in, where 
the direct search prospect of that component will simply die. 

\subsection{Role of DM-DM conversion}

In this section, we will illustrate the role of DM-DM conversion to alter relic density outcome of the individual DM components.
We first demonstrate the differences between the minimal $U(1)_{B-L}$ model and the present scenario at the quantitative level. In the former, we first recall that the lightest right-handed neutrino ($N_1$) DM, annihilates to SM 
particles by s-channel mediations of $Z_{BL}$, $h$ and $s$. In our case it also does the same, while additionally, it may annihilate to other DM component, if allowed kinematically. Before proceeding further, let us remind the parameters 
relevant for DM analysis of this model, as we will treat some of them as variables, keeping others at some fixed values in the analysis hereafter
\bea
\nonumber
\{M_1,M_H,M_A,M_{H^+},M_s,\sin\theta,v_{BL},g_{BL},\l_7,\l_L\}
\eea 
 }
\begin{figure}[H]
\centering
\includegraphics[scale=0.45]{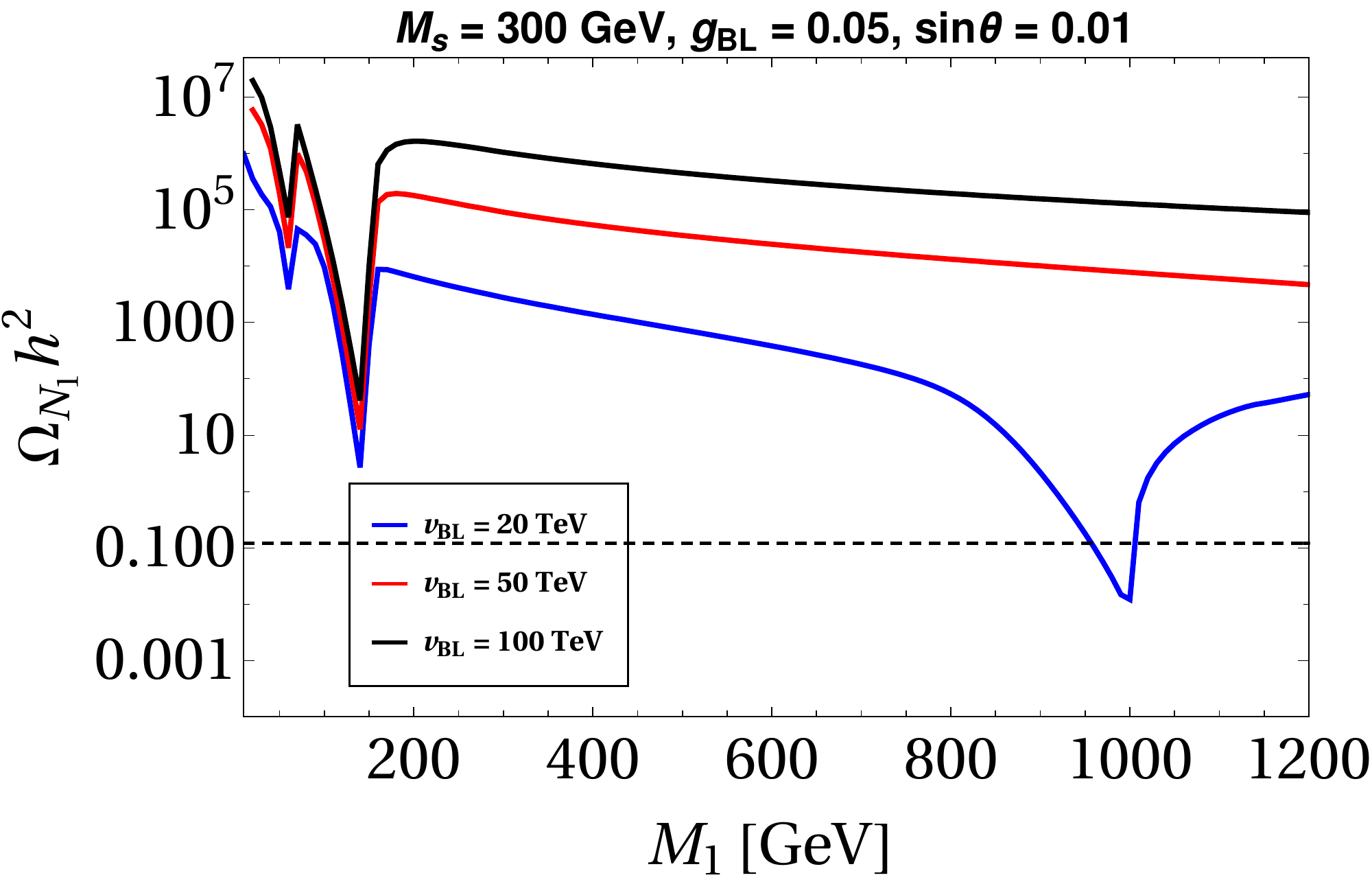}
\caption{The variation $\Omega_{N_1}$ versus $M_1$ in the minimal $U(1)_{B-L}$ case. The colour coding is explained in the legends.}
\label{pure_B-L}
\end{figure}

Fig.\ref{pure_B-L} depicts the variation of the relic density in the minimal $U(1)_{B-L}$ case for a particular choice of the parameters as shown in Fig.~\ref{pure_B-L} inset.
The annihilations for such a choice are mostly gauge-driven thereby making the corresponding amplitude $\propto \frac{g_{BL}^2}{M_{Z_{BL}}^2}\propto \frac{1}{v^2_{BL}}$. 
This explains the increase in the thermal relic with increasing $v_{BL}$ (from 20 TeV to 100 TeV). Annihilations through the scalars also turn important near the resonance regions. In fact, for $v_{BL} = $ 20 TeV ($M_{Z_{BL}} = 2g_{BL}v_{BL}=2$ TeV), all three resonance dips around $M_1 = \frac{M_h}{2} ,~ \frac{M_s}{2} ~\rm{and}~ 
\frac{M_{Z_{BL}}}{2}$ are visible as opposed to 
$v_{BL} = 50~ \rm{TeV} ~\rm{and}~ 100~ \rm{TeV}$ when the $M_1 =\frac{M_{Z_{BL}}}{2}$ 
dips no longer fit in the shown range. In all, the key feature identified here is that the minimal $U(1)_{B-L}$ model satisfies the requisite relic in the vicinity of the resonance dips only. As mentioned above, $v_{BL}=20$ TeV ($M_{Z_{BL}} = 2g_{BL}v_{BL}=2$ TeV) is altough disfavoured from the dilepton searches at LHC, we keep it for demostration purpose that the resonance dip is good enough to satisfy relic density given the choices of other parameters. However, we do our further analysis with $v_{BL}=50$ TeV.\\

\begin{figure}[H]
\centering
\includegraphics[scale=0.35]{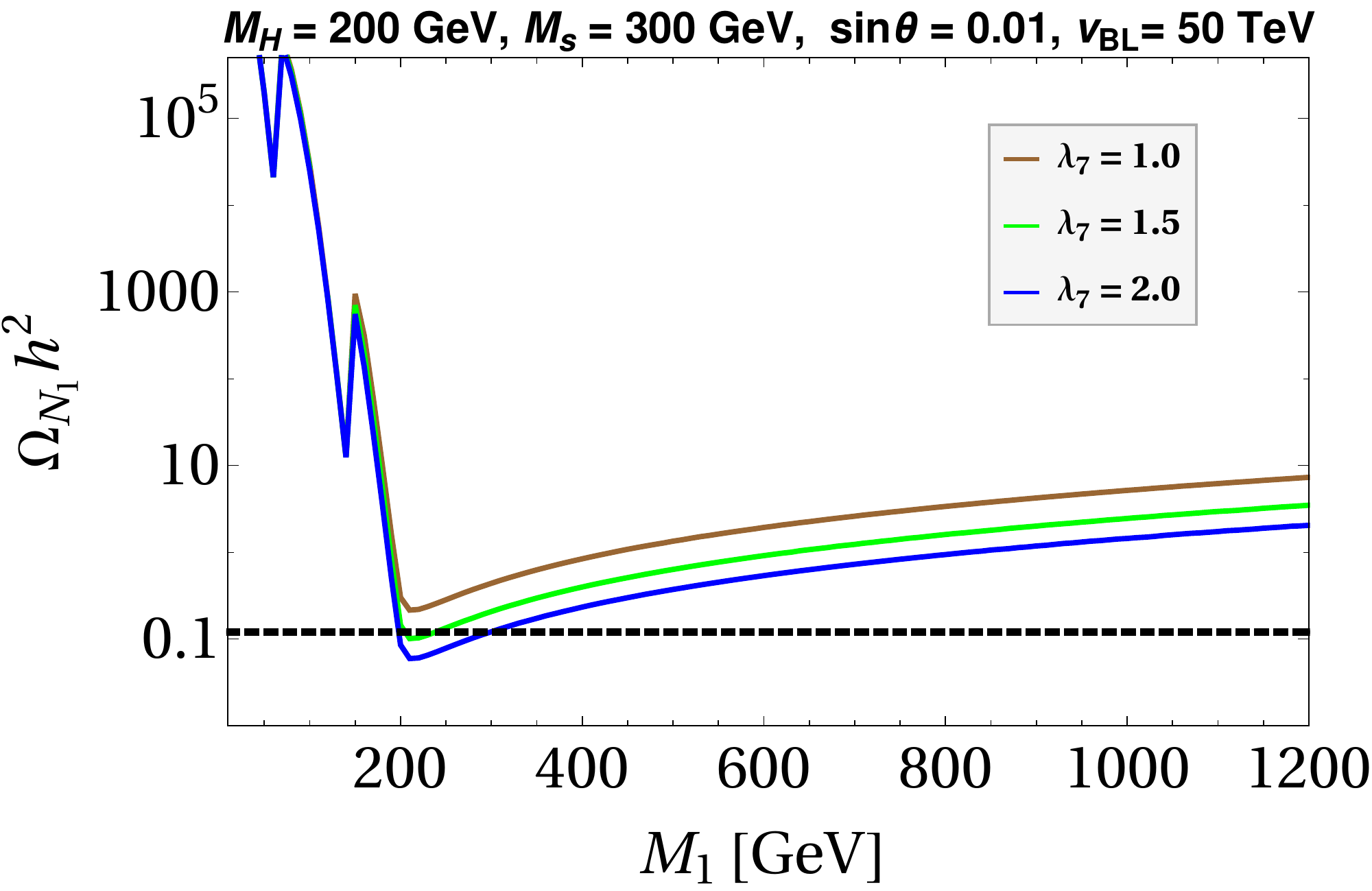}~~~~
\includegraphics[scale=0.35]{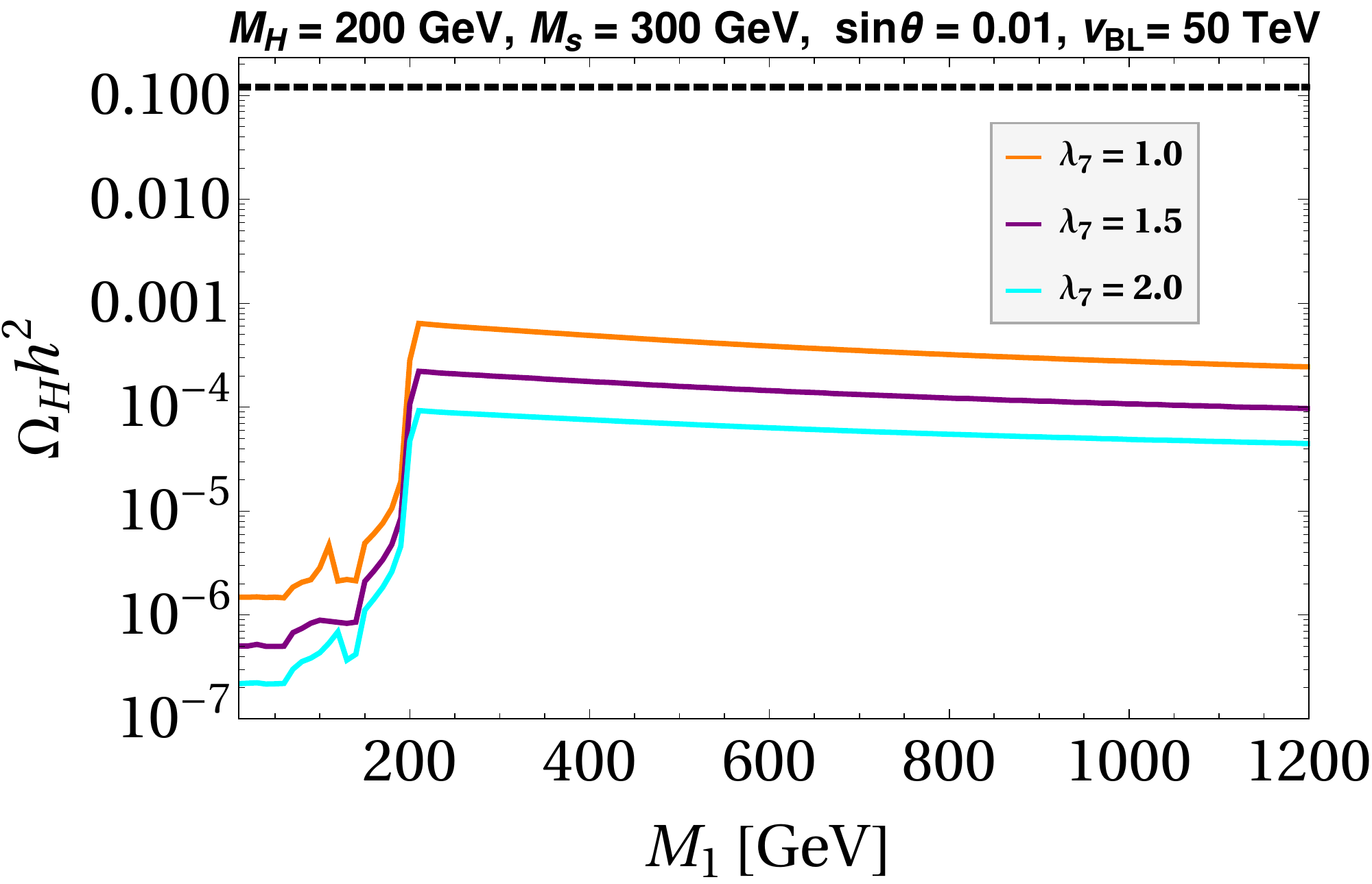}
\caption{The variation $\Omega_{N_1}$ versus $M_1$ (left) and $\Omega_{H}$ versus $M_1$ (right) for $g_{BL}=0.05$, $v_{BL}$ = 50 TeV. The colour coding is explained in the legends.}
\label{relic_20_50}
\end{figure}


Turning to the behaviour of the relic density in the present case, we first take $M_H = 200$ GeV (along with $M_A - M_H = M_{H^+} - M_A$ = 1 GeV throughout the section), a choice motivated from the fact that $M_H$ lies in the well known \emph{intermediate} mass region ($M_W < M_H <$  500 GeV) of the inert Higgs doublet where $H H \to V V$ annihilations are turned on leading to an under-abundant relic. We also choose $\l_L = 10^{-4}$. The presence of an additional scalar $s$ implies that the additional annihilation channels $H H \to s s, s h$ are liable to open up thereby causing further under-abundance. We take $M_s$ = 300 GeV intending to 
kinematically close the aforementioned channels. However, since $\l_{H H s} \simeq \l_7 v_{BL}$ and $\l_{hhs} \simeq \l_6 v_{BL}$ {\footnote{$\l_6$ is determined from Eq.(~\ref{quartic} f)} for small $s_\theta$, the process $H H \longrightarrow h h$ will have copious rates for $v_{BL} \sim \mathcal{O}(10)$ TeV and a sizeable $\l_{7}$. This causes the relic of $H$ to further decrease compared to the pure IDM value. The contribution remains 
$\sim \mathcal{O}(10^{-3})$ at best.    


As for the relic of $N_1$, an inspection of Fig.\ref{relic_20_50} (left panel) also reveals that an $\Omega_{N_1}h^2 \simeq 0.1$ also occurs for $M_1 \simeq  200$ GeV, a mass value distinctly away from any of the resonance dips. 
This is due to onset of the $N_1 N_1 \longrightarrow H H, A A, H^+ H^-$ (collectively written $N_1 N_1 \longrightarrow \phi_2 \phi_2$) conversion processes near the $M_1 \simeq M_H$ threshold 
(the small difference can be attributed to a small DM velocity). And the higher the value of $\l_7$ taken, the higher are the $H-H-s$, $A-A-s$ and $H^+-H^--s$ interaction strengths, the higher are the 
$N_1 N_1 \longrightarrow H H, A A, H^+ H^-$ cross sections (see eqn.(\ref{sigma_conv})), and ultimately, the higher is the attrition in the abundance of 
$N_1$. 
One can estimate the relic density for $N_1$ including conversion to $\phi_2$ as $\Omega_{N_1}h^2 \sim (\langle \sigma v\rangle_{N_1N_1 \to SM~SM}+\langle \sigma v\rangle_{N_1N_1 \to \phi_2\phi_2})^{-1}$.
For example, in case of $v_{BL}$ = 50 TeV and $\l_7 = 1.5$, the relic curve hits the $\simeq$ 0.1 mark for $M_1 \sim$ 200 GeV.
One also notes that $\l_7 = 1.0$ does not suffice to bring down $\Omega_{N_1}h^2$ to the requisite ball-park. The dynamics of the $N_1 N_1 \longrightarrow \phi_2 \phi_2$ remains qualitatively the same for each $v_{BL}$ however with pronounced differences in the relic. The different choices of $v_{BL}$ though spell pronouncedly different $\Omega_{N_1}h^2$ very much due to the same reason as in the pure $U(1)_{B-L}$ case, the conversion region witnesses only small differences, an observation elucidated at the end of section \ref{dm+hsv}. Overall, $\Omega_{N_1} h^2 > > \Omega_H h^2$ and therefore $\Omega h^2 \simeq \Omega_{N_1} h^2$. This can be clearly visible from Fig. \ref{relic_20_50} (right panel), where we plot $\Omega_Hh^2$ against $M_1$ with fixed $M_H=200~\rm{GeV}$ for $v_{BL}=50~\rm{TeV}~\rm{(right~ panel)}$. This behaviour remains qualitatively the same for a different $(M_H,M_s)$ but a similar mass hierarchy as in this case. In a word, one cannot emphasize more the role of the DM-DM conversion processes in the generation of relic density, and, the parameter $\l_7$ here, the former being inextricably linked to the latter. 
Of course, the model survives beyond resonance regions for $N_1$, only with $M_1>M_H$, thanks to DM-DM conversion as described above. Also, one may note, that the effect of conversion of $N_1$ to $H$ affects the latter mildly, 
 and therefore the relic of $H$ do not undergo a sea change from its single component status. The following remark is in order. For a fixed $M_1$ and $\l_7$, since $y_{11} \propto \frac{1}{v_{BL}}$ and $\l_{HHs} \propto v_{BL}$ for small $s_\theta$, the $N_1 N_1 \to \phi_2 \phi_2$ amplitude has a very weak dependence on $v_{BL}$. The same is therefore expected for the relic density in the conversion region. This has been checked for $v_{BL}$ = 100 TeV.
 
Thus, despite the present model having a particle content same as in \cite{Kanemura:2011vm} and \cite{Borah:2018smz}, 
the difference in the assignment of the discrete charges makes it phenomenologically distinct. All $\phi_2,~N_{1,2,3}$ are charged negatively under a common $\mathbb{Z}_2$ in the quoted studies. 
Since the inert doublet is in the same dark sector with the RH neutrinos,
$N N \to \phi_2 \phi_2$ conversion absent in such cases.  
In \cite{Kanemura:2011vm}, the inert doublet does not participate in the DM phenomenology and its role is seemingly restricted to neutrino mass-generation only.  
The requisite relic is observed to be satisfied around scalar resonance dips only. On the other hand, \cite{Borah:2018smz} considers the possibility of $N_1$ coannihilating with $N_{2,3}$ or with the inert scalars. Therefore, in either case, the DM phenomenology is qualitatively different from ours. 
In addition, since all $N_{1,2,3}$ couple to the inert doublet in the aforementioned studies, all three SM neutrinos can acquire non-zero masses, as opposed to our case where $N_1$ does not participate in that interaction leading to a massless SM neutrino.

\begin{figure}[H]
\centering
\includegraphics[scale=0.52]{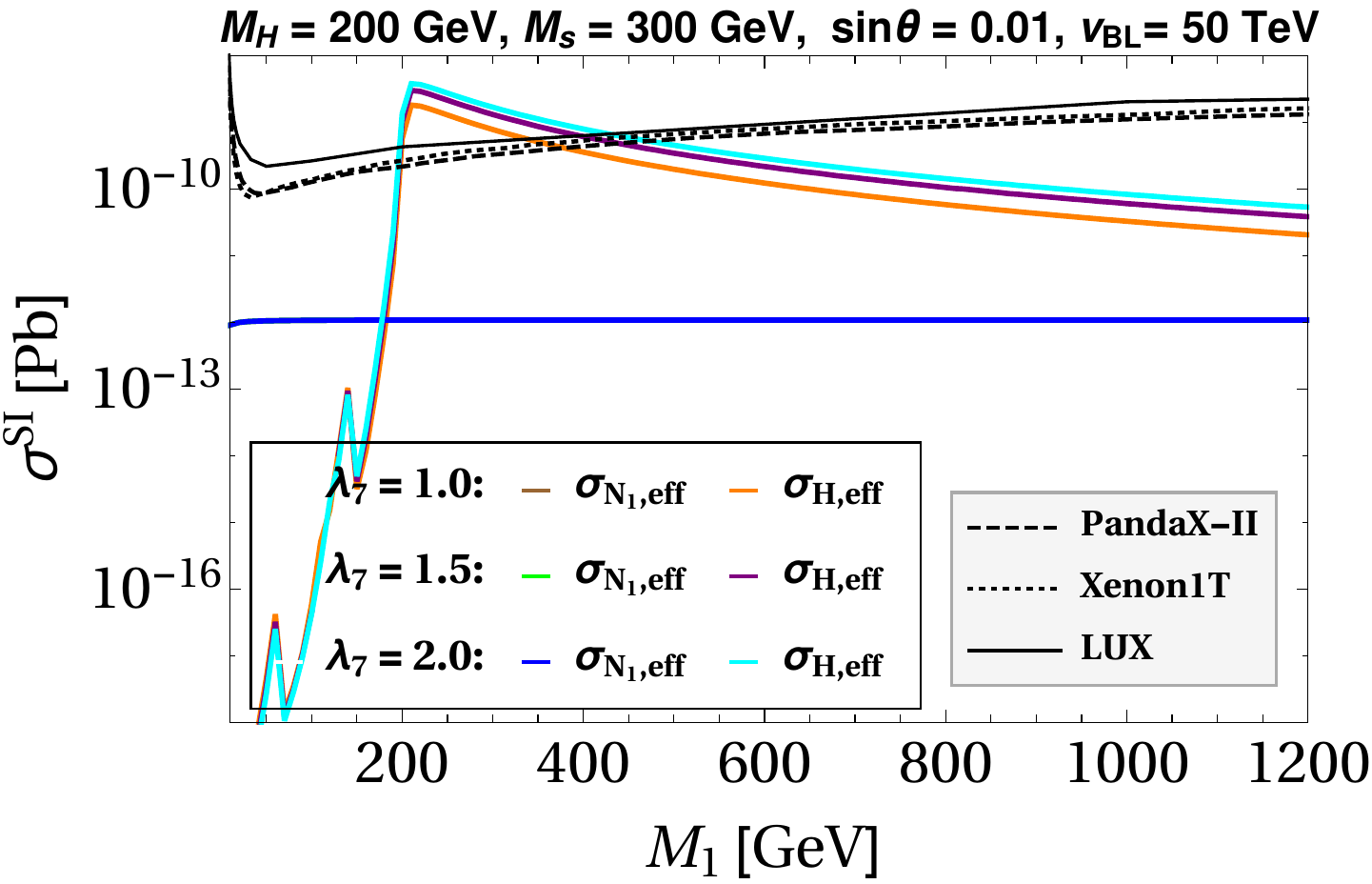}
\caption{The variation of the effective SI-DD cross sections versus $M_1$ for $g_{BL}=0.05$, $v_{BL}$ = 50 TeV. The colour coding is explained in the legends.}
\label{DD_20_50}
\end{figure}

We summarise the finding from direct detection next. Plots in Fig.~\ref{DD_20_50}
  depicts the effective SI-DD cross sections of both $N_1$ and $H$ ($\l_7=1.0$: brown for $N_1$, orange for $H$; $\l_7=1.5$: green for $N_1$, purple for $H$; $\l_7=2.0$: blue for $N_1$, cyan for $H$; ) versus mass of $N_1$. Note that $\sigma_{N_1}^{SI}$ is also dominated by $v_{BL}$ which was instrumental for $\Omega_{N_1}$ contribution as stated before. For a fixed $v_{BL}$, we observe only a mild variation of  $\sigma_{N_1,eff}^{SI}$ with $M_1$ (overlapping of brown, green and blue lines in Fig.~\ref{DD_20_50}). Also 
$\Omega_T h^2 \simeq \Omega_{N_1} h^2$ which implies $\sigma_{N_1,eff}^{SI} \sim$ $\sigma_{N_1}^{SI}$. In all, the effective SI-DD rate for $N_1$ always remains below the XENON-1T bound for the $v_{BL}$ chosen. On the other hand, the presence of $\frac{\Omega_H}{\Omega_T}(=1-\frac{\Omega_{N_1}}{\Omega_T})$  in $\sigma_{H,eff}^{SI}$ implies that the dips in $\Omega_{N_1}$ translate to spikes in $\sigma_{H,eff}^{SI}$. One should note here that the $s$-mediated DD amplitude for $H$ is proportional to $\l_7 v_{BL} s_\theta$, and consequently, almost five orders of magnitude higher than the $h$-mediated DD amplitude for the parameters shown in Fig.~\ref{DD_20_50}. Therefore, despite relic density suppression for $H$ by the scaling factor 
$\frac{\Omega_H}{\Omega_T}(=1-\frac{\Omega_{N_1}}{\Omega_T})$, the effective DD cross section in this setup remains higher than pure IDM by almost two orders of magnitude. One however does not have to commit to $M_s = 300$ GeV. The lower value $M_s = 210$ GeV opens up the $H H \longrightarrow h s$ mode thereby causing the $H$ yield to drop further. The direct detection rate of $H$ also diminishes accordingly. A choice $\l_7 = 2$ here (see Fig. \ref{DD_allowed}) maintains both the thermal relic and the direct detection rates within their respective permissible values \footnote{In case of the IDM with $\l_L = 10^{-4}$, \cite{PhysRevD.87.075025} reports a one-loop enhancement of the DD by a factor $\sim 100$. For $\l_L = 10^{-4}$, the tree level amplitude becomes small compared to the one-loop amplitude that is dominantly borne out of  the gauge interactions. However, this is not the case with the present model, where, in case of $v_{BL} \sim \mathcal{O}$(10) TeV and $s_\theta = 0.01$, the DD amplitude for $H$ at the tree level (dominantly driven by $s$-mediation) itself is expected to yield the leading contribution.}. 
\begin{figure}[H]
\centering
\includegraphics[scale=0.34]{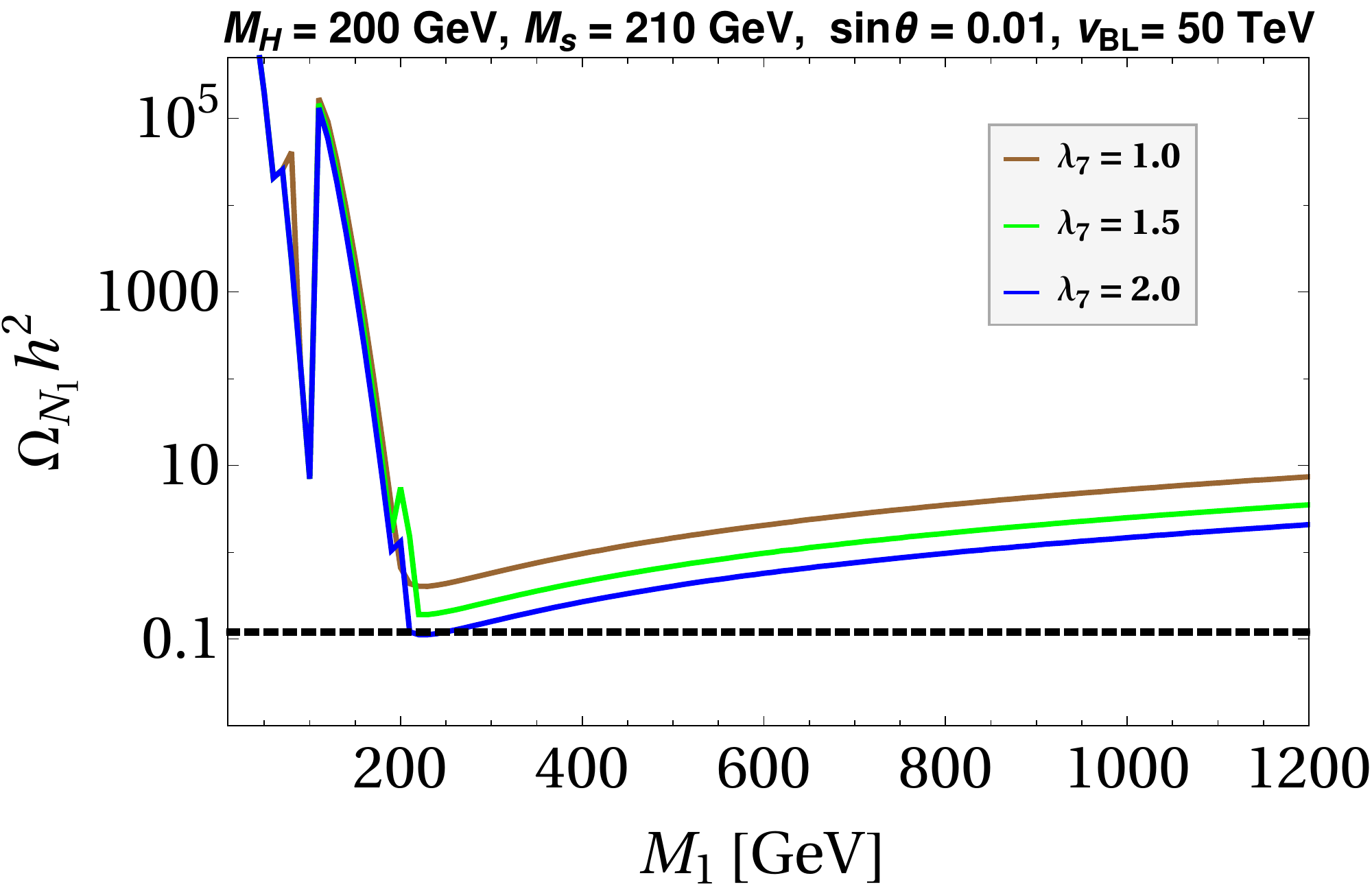}
\includegraphics[scale=0.48]{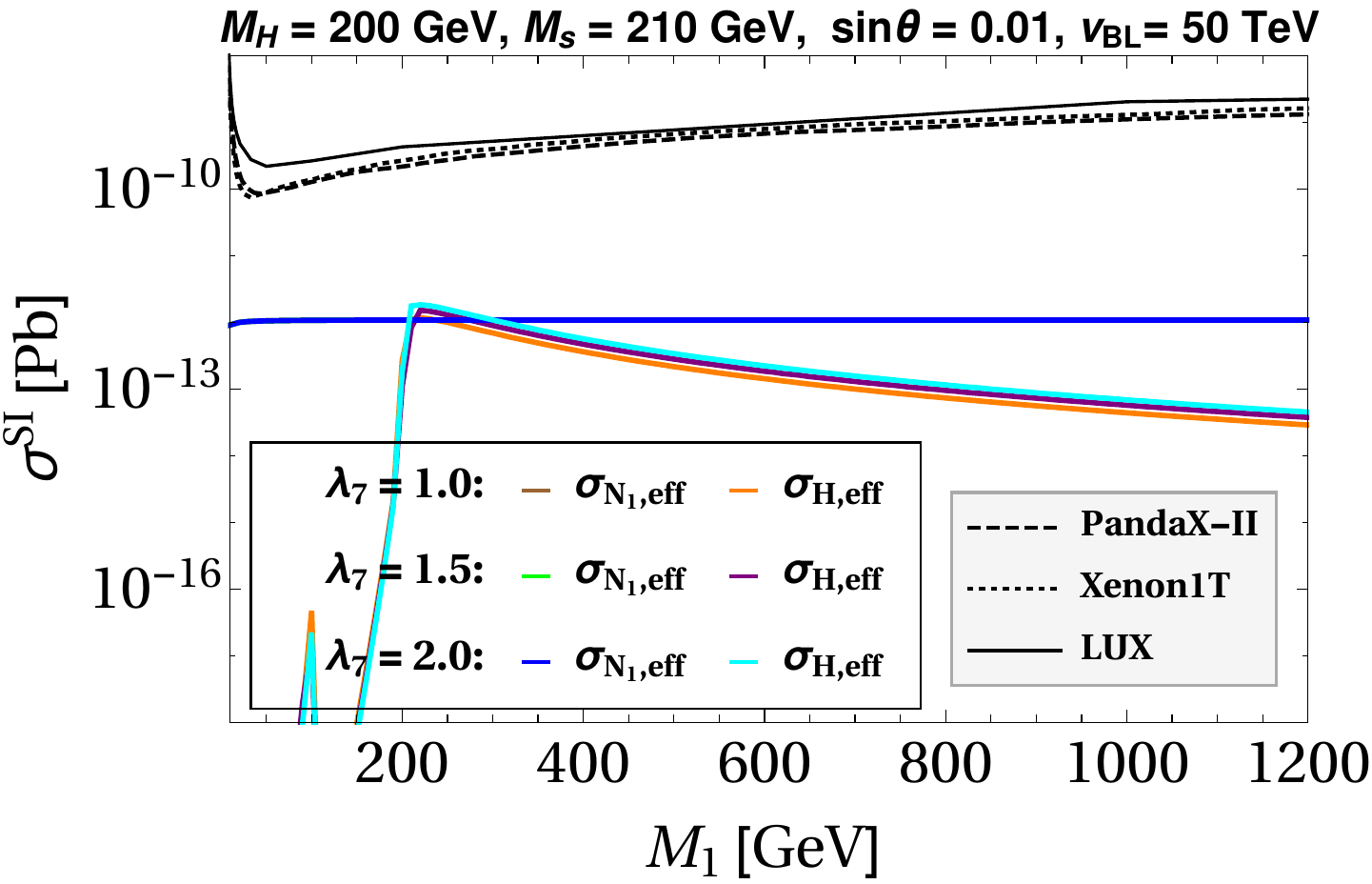}
\caption{The variation of the $\Omega_{N_1}h^2$ and effective SI-DD cross sections versus $M_1$ for $v_{BL}$ = 50 TeV. 
The colour coding is explained in the legends.}
\label{DD_allowed}
\end{figure}

A scan of the model parameter space therefore becomes necessary to find a parameter region that meets both the relic and DD requirements, a task we take up in section~\ref{dm+hsv}. Nonetheless, the results present in this section are demonstrative of the main aspects of DM phenomenology for the present scenario.

\section{High-scale validity}\label{vacstab}

The fate of the model at high energy scales can be understood by studying the RG evolution of its couplings. Particularly interesting is the evolution of the quartic couplings where the presence of additional bosonic degrees 
of freedom in the model can potentially introduce an interesting interplay between high-scale perturbativity and vacuum stability. The vacuum is stable 
up to a cut-off if eqn.(\ref{vsc}) are satisfied at each intermediate scale till that cut-off. Likewise $|\l_i(\mu)| < 4\pi$ must also hold all along up to the cut-off. 
Some explorations of high scale validity (HSV) of TeV-scale neutrinos are~\cite{Bhattacharya:2019fgs,Ghosh:2017fmr,Bambhaniya:2016rbb,Chakrabortty:2013zja,Chakrabortty:2012np,Mohapatra:2014qva,Das:2019pua,Chun:2012jw,Khan:2012zw,Coriano:2014mpa,Ng:2015eia,Bonilla:2015kna,Garg:2017iva,Chen:2012faa,Rose:2015fua,Das:2015nwk,Lindner:2015qva,Chakrabarty:2015yia}.

We choose $\mu = M_t = 173.34$ GeV as the initial scale. 
The $t$-Yukawa and the gauge couplings are evaluated at this scale
incorporating the necessary threshold corrections. 
Besides, $M_{N_{2,3}}$ are taken $\simeq$ 1 TeV.
In principle the effect $N_{i}$ must be turned on in the RG equations only when $\mu > M_{N_i}$. However for RH neutrino masses not exceeding 1 TeV, the gap between $M_t$ 
and the RH neutrino mass scale is not wide and therefore turning on $N_i$ from $\mu = M_t$ itself is a reasonable approximation. In addition, the smallness of $\zeta_{i\a}$ allows to neglect their effects in the 
$\beta$-functions. Below we list the 1-loop beta functions of the model couplings.
\\ \\
\textbf{$\beta$ functions for the gauge couplings~}\cite{Basso:2010jm}:

\besub
\bea
16 \pi^2 \beta_{g_1} &=& 7 g_1^3, \\
16 \pi^2 \beta_{g_2} &=& -3 g_2^3, \\
16 \pi^2 \beta_{g_3} &=& -7 g_3^3, \\
16 \pi^2 \beta_{g_{B-L}} &=& 12 g_{B-L}^3.
\eea
\eesub

\textbf{$\beta$ functions for the quartic couplings}~\cite{Branco:2011iw,Basso:2010jm}:

\besub
\bea
16 \pi^2 \beta_{\l_1} &=& 12 \l_1^2 + 4 \l_3^2 + 4 \l_3 \l_4 + 2 \l_4^2
 + 2 \l_5^2 + 2 \l_6^2 + 12 \l_1 y^2_t - 12 y^4_t \nonumber \\
&&
+ \frac{3}{4} g_1^4 + \frac{9}{4} g_2^4
+ \frac{3}{2} g_1^2 g_2^2  - \l_1 (3 g_1^2 + 9 g_2^2), \\
16 \pi^2 \beta_{\l_2} &=& 12 \l_2^2 + 4 \l_3^2 + 4 \l_3 \l_4 + 2 \l_4^2
 + 2 \l_5^2 + 2 \l_7^2 
  + \frac{3}{4} g_1^4 + \frac{9}{4} g_2^4
  + \frac{3}{2} g_1^2 g_2^2 \nonumber \\
&& 
  - \l_2 (3 g_1^2 + 9 g_2^2), \\
16 \pi^2 \beta_{\l_3} &=& 6 \l_1 \l_3 + 2 \l_2 \l_3 + 4 \l_3^2 + 2 \l_1 \l_4
+ 2 \l_2 \l_4 + 2 \l_4^2 + 2 \l_5^2 + 2 \l_6 \l_7 + 6 \l_3 y^2_t \nonumber \\ 
&&
+ \frac{3}{4} g_1^4 + \frac{9}{4} g_2^4
- \frac{3}{2} g_1^2 g_2^2  - \l_3 (3 g_1^2 + 9 g_2^2), \\
16 \pi^2 \beta_{\l_4} &=& 2 \l_1 \l_4 + 2 \l_2 \l_4 + 8 \l_3 \l_4 + 4 \l_4^2 + 8 \l_5^2 + 6 \l_4 y^2_t  + 3 g_1^2 g_2^2  - \l_4 (3 g_1^2 + 9 g_2^2),  \\
16 \pi^2 \beta_{\l_5} &=& 2 \l_1 \l_5 + 2 \l_2 \l_5 + 8 \l_3 \l_5
 + 12 \l_4 \l_5 + 6 \l_5 y^2_t  - \l_5 (3 g_1^2 + 9 g_2^2), \\
16 \pi^2 \beta_{\l_6} &=& 6 \l_1 \l_6 + 4 \l_3 \l_7 + 2 \l_4 \l_7 + 4 \l_6^2 + 8 \l_6 \l_8 + 6 \l_6 y^2_t + 4 \l_6 \text{Tr}[Y^\dagger Y] - 24 \l_6 g^2_{B-L}, \\
16 \pi^2 \beta_{\l_7} &=& 6 \l_2 \l_7 + 4 \l_3 \l_6 + 2 \l_4 \l_6 + 4 \l_7^2 + 8 \l_7 \l_8 + 4 \l_7 \text{Tr}[Y^\dagger Y] - 24 \l_7 g^2_{B-L},\\
16 \pi^2 \beta_{\l_8} &=& 2 \l_6^2 + 2 \l_7^2 + 20 \l_8^2
+ 8 \l_8 \text{Tr}[Y^\dagger Y] - \text{Tr}[Y^\dagger Y Y^\dagger Y] - 48 \l_8 g^2_{B-L} + 96 g^4_{B-L}.  
\eea
\label{RGquartic}
\eesub

\textbf{$\beta$ functions for the Yukawa couplings~}\cite{Basso:2010jm}:
\besub
\bea
16 \pi^2 \beta_{Y} &=& 4 Y Y^{\dagger} Y + 2 Y \text{Tr}[Y^{\dagger} Y] - 6 g_{BL}^2 Y, \\
16 \pi^2 \beta_{y_t} &=& \frac{9}{2}y^3_t
 - y_t \Big(\frac{17}{12} g_1^2 +\frac{9}{4} g_2^2 + 8 g_3^2
  + \frac{2}{3} g^2_{B-L}\Big).
\eea
\eesub
Here $Y$ = diag($y_{11},y_{22},y_{33}$). 
With an aim to understand the high-scale behaviour of the model, we first take
$v_{BL} = 50$ TeV, $M_1 - M_s = 40$ GeV, $\l_L = 10^{-4}, \l_2 = 0.01$, $s_{\theta} = 0.01$ and propose the following benchmark values for the rest of the parameters as listed in Table~\ref{BP}. 
It is important to mention that the choice $s_{\theta}$ = 0.01 is compatible with the proposed benchmarks as can be read from 
Fig.~\ref{mugaga}. The corresponding $\mu_{\gamma \gamma}$ and $\Delta T$-values corresponding to the BPs are given in Table~\ref{mugagaT}.

\begin{figure}[htbp]
\centering
\includegraphics[scale=0.4]{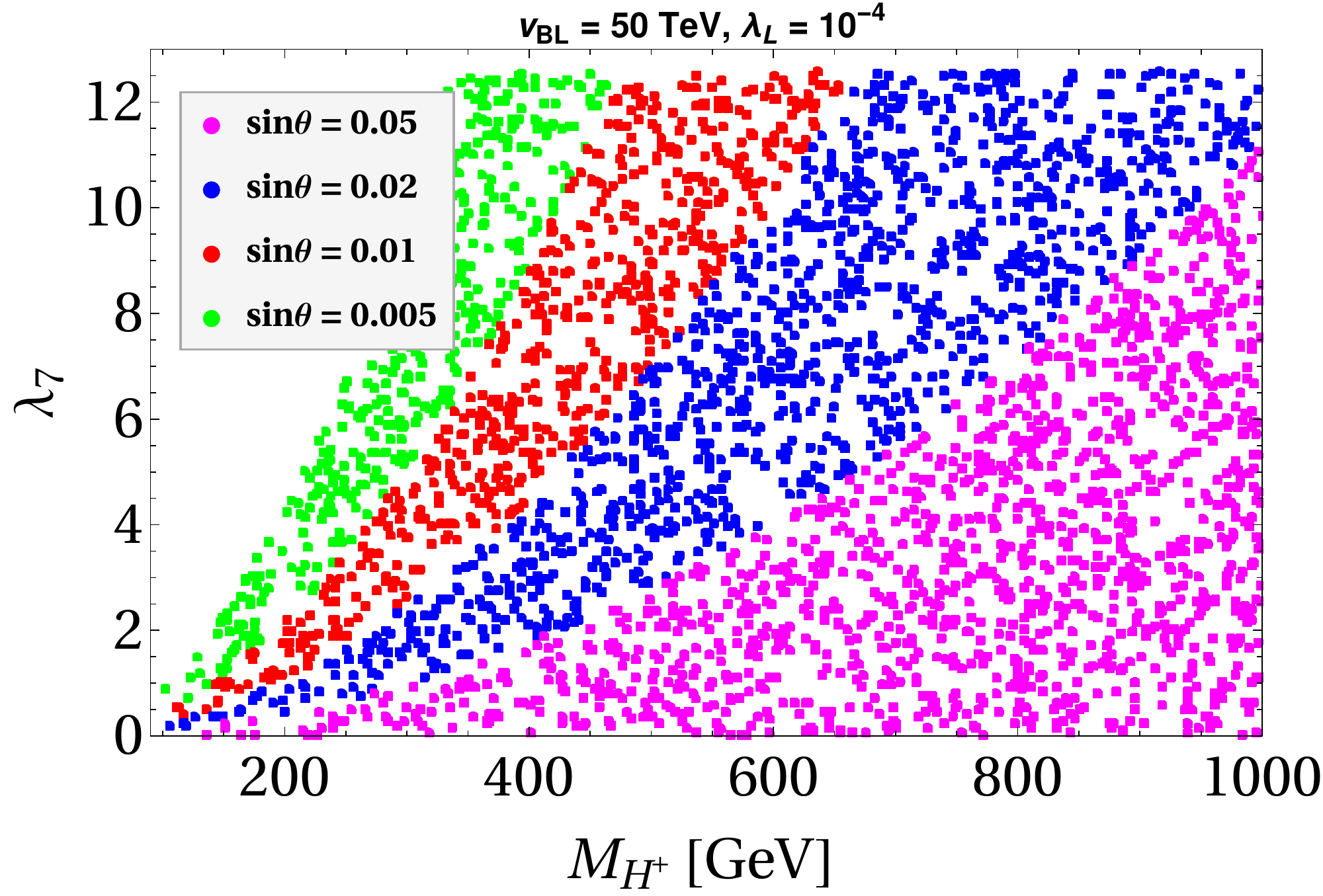}
\caption{Parameter space in the $\l_7 - M_{H^+}$ plane allowed by the 
$\mu_{\gamma \gamma}$ constraint}
\label{mugaga}
\end{figure}

\begin{table}[h]
\centering
\begin{tabular}{|c c c c c c c c c c|}
\hline
BP & $M_1$ & $M_H$ & $M_A - M_H$ & $s_\theta$ & $\l_7$ & $\Omega_{H} h^2$ & $\Omega_{N_1} h^2$ & $\sigma^{SI}_H$ (cm$^2$) & $\sigma^{SI}_{N_1}$ (cm$^2$) \\ \hline
BP1 & 250 & 200 & 10 & 0.01 & 2.1  & $1.10 \times 10^{-7}$ & 0.121 & $1.5 \times 10^{-48}$ & $1.7 \times 10^{-48}$ \\ \hline
BP2 & 185 & 135 & 10 & 0.005 & 1.5 & $1.07 \times 10^{-5}$ & 0.120 & $1.5\times10^{-49}$ &  $4.2 \times 10^{-47}$ \\
\hline
\end{tabular}
\caption{Benchmark parameters to demonstrate high-scale validity. All masses and mass-splittings are in GeV.}
\label{BP}
\end{table}

\begin{table}[h]
\centering
\begin{tabular}{|c c c|}
\hline
BP & $\mu_{\gamma \gamma}$ & $\Delta T$ \\
\hline
BP1 & 1.28 & 0.0035\\
BP2 & 1.19 & 0.0035 \\ 
\hline
\end{tabular}
\caption{$h \to \gamma \gamma$ signal strength and $T$-parameter for the BPs}
\label{mugagaT}
\end{table}

\noindent Fig.~\ref{BP1_p05_lam} displays the RG running of $\l_i$  for BP1 with $M_{2} = 1,10$ TeV, $M_{3} = 1.1,11$ TeV and $g_{BL} = 0.05$. This parameter point offers a bounded-from-below potential and perturbative couplings up to $\simeq 
4 \times 10^{6}$ GeV. The largest Yukawa coupling strength is by far that of the $t$-quark. Therefore, $\l_1$ experiences the strongest fermionic downward pull in course of evolution amongst other quartic couplings. This can be countered by adjusting $\l_3, \l_4, \l_5$ and $\l_6$ appropriately ({as seen from Eq.(\ref{RGquartic}a)}). Now $\l_6 \sim 10^{-5}$ for the aforementioned benchmarks and therefore it is too small to counter the fermionic effect. The size of $\l_3, \l_4, \l_5$ is controlled by the mass splitting amongst $H,A$ and $H^+$. We find that a splitting of $\simeq 10 - 20$ GeV prevents $\l_1(\mu) < 0$ throughout.  

On a similar note, the presence of a $4 \l_7^2$ term in $\beta_{\l_7}$ implies that $\l_7 > 1$ at the EW scale in this case causes the coupling to grow rapidly and become non-perturbative around the said cut-off. One the other hand, according to the left plot in Fig.~\ref{BP1_p05_yuk}, the variation of $y_{ii}$ however remains negligible due to the smallness of their initial values. 
\begin{figure}[H]
\centering
\includegraphics[scale=0.35]{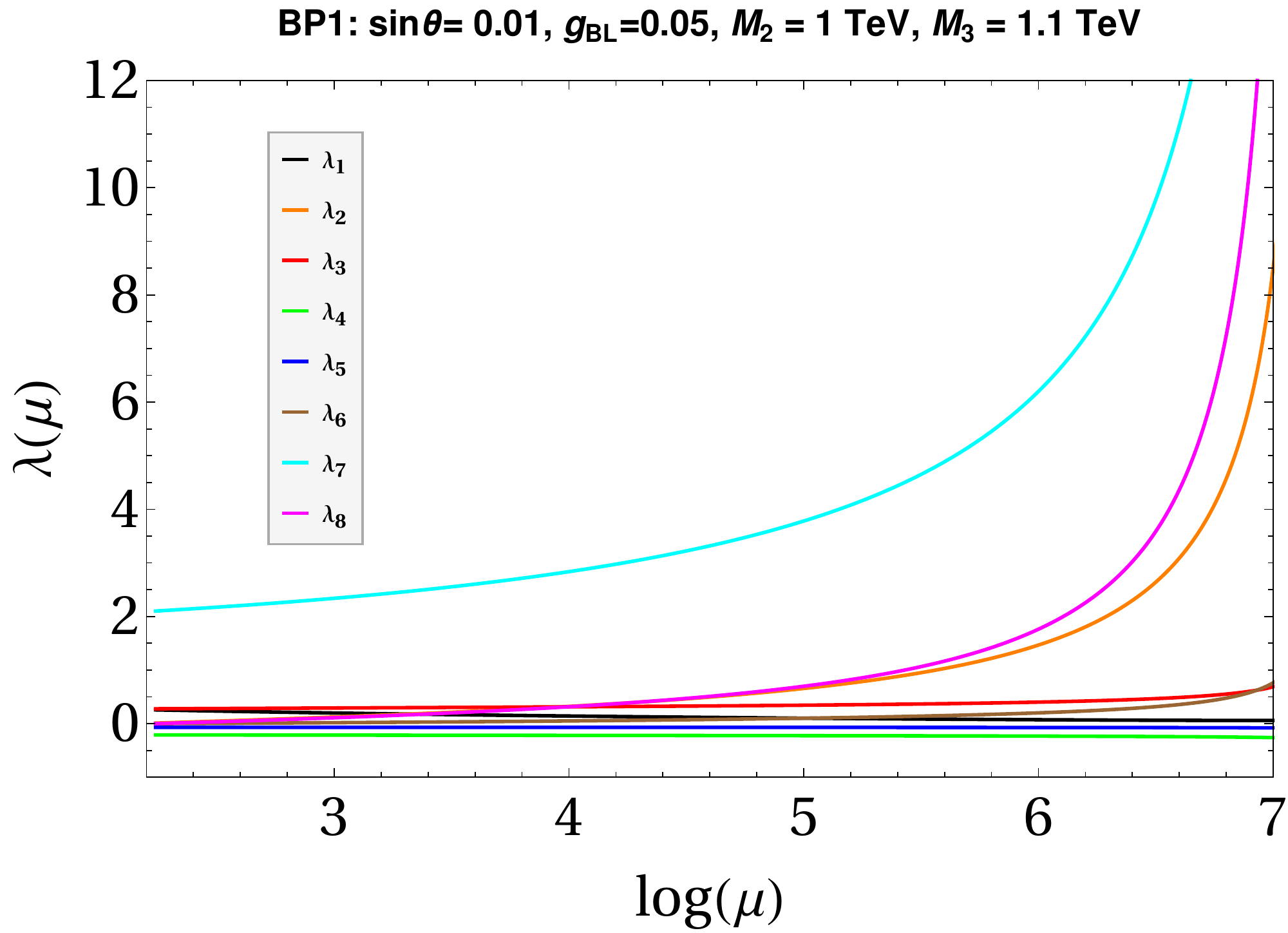}~~~
\includegraphics[scale=0.35]{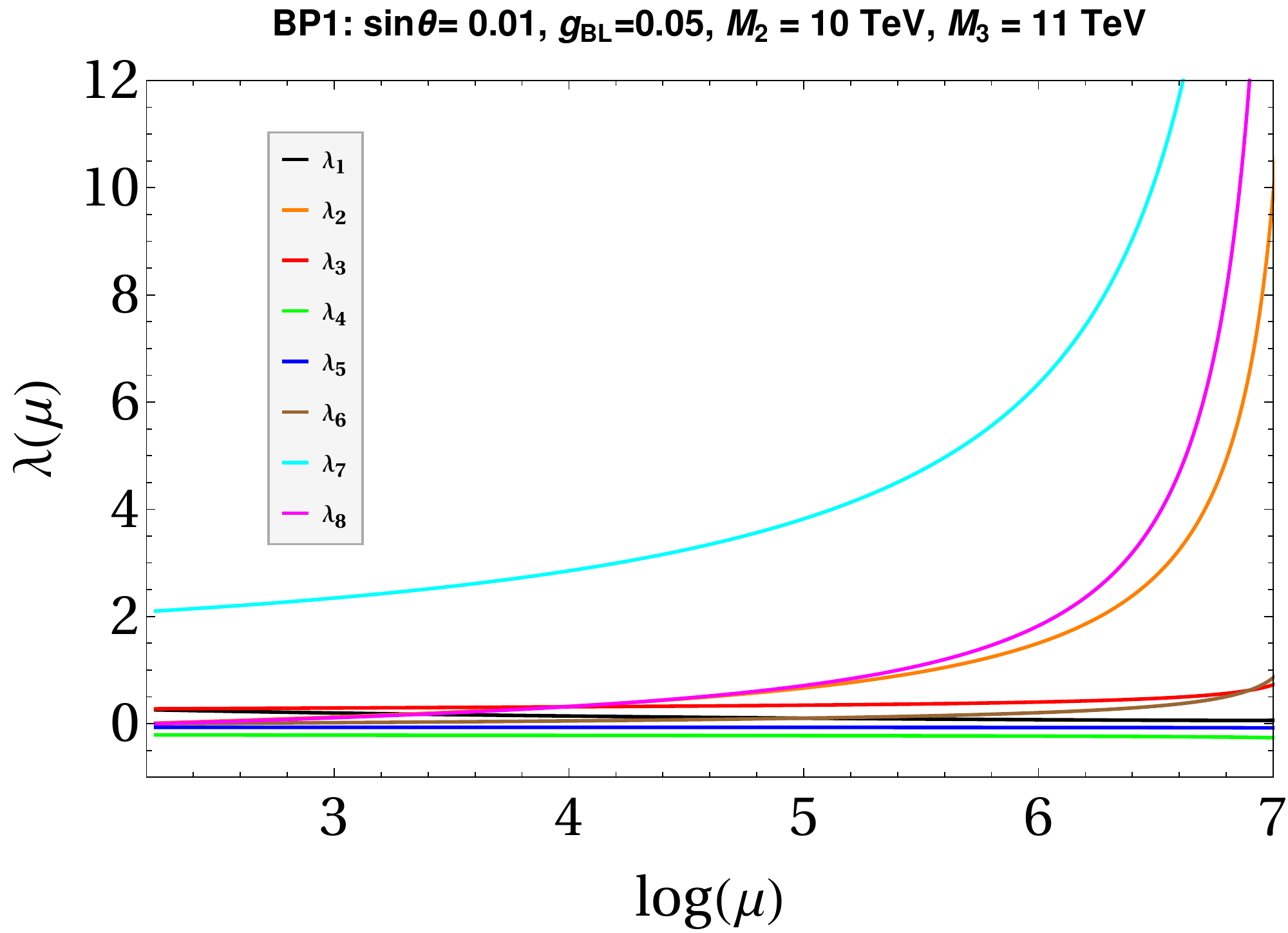}~
\caption{The evolution of the quartic couplings for BP1. The left (right) plot corresponds to $M_2 = 1(10)$ TeV and $M_3 = 1.1(11)$ TeV. The color coding is explained in the legends.}
\label{BP1_p05_lam}
\end{figure}
The same BP1 evolves as shown in the right plot of Fig.~\ref{BP1_p05_lam} when taken along with $M_2 = 10$ TeV, $M_3 = 11$ TeV. The Yukawa couplings $y_{22}$ and $y_{33}$ register a gentle rise in this case owing to larger initial values. This in turn causes $\l_7$ to grow slightly faster compared to the previous case. 

\begin{figure}[H]
\centering
\includegraphics[scale=0.35]{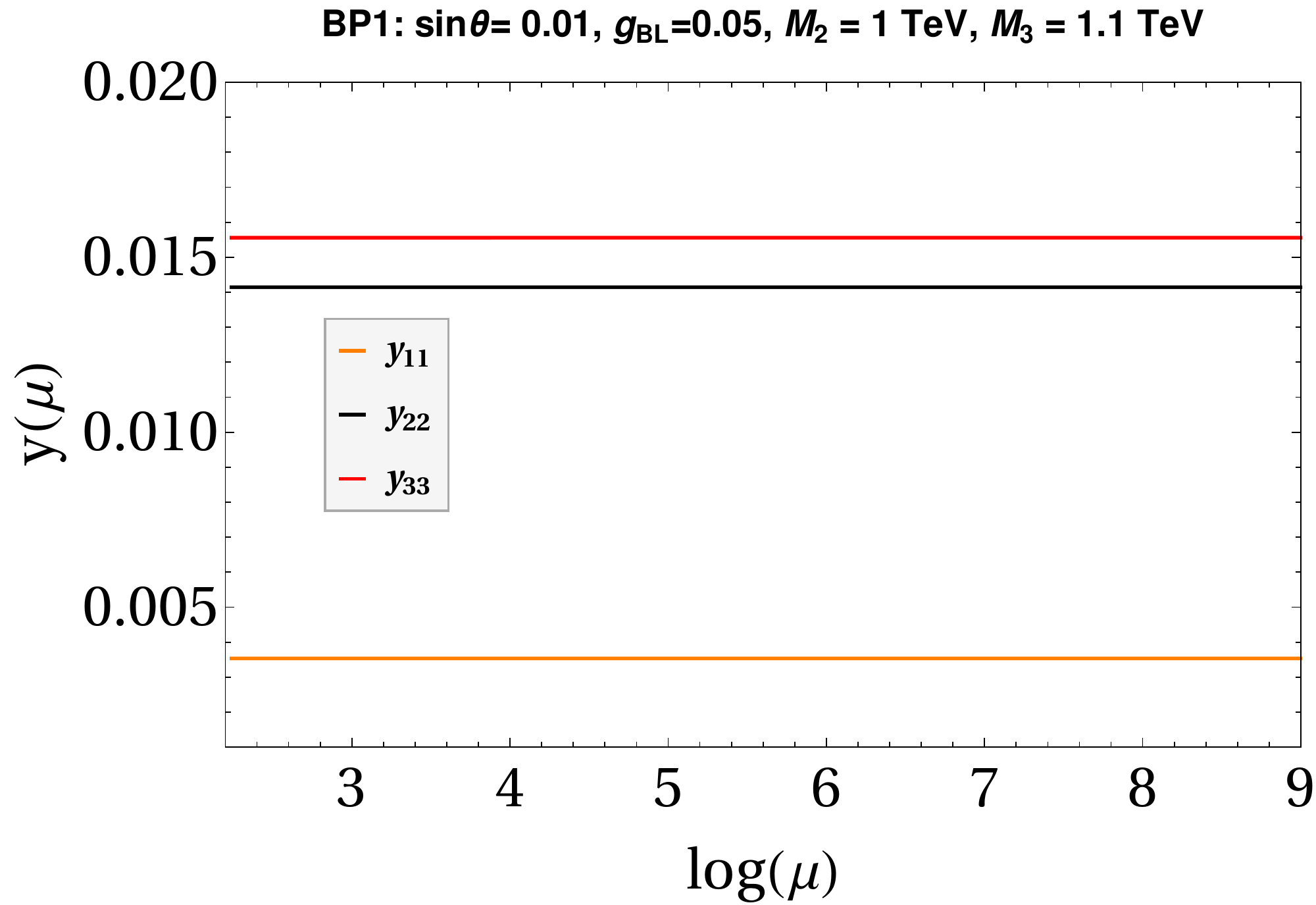}~~~
\includegraphics[scale=0.35]{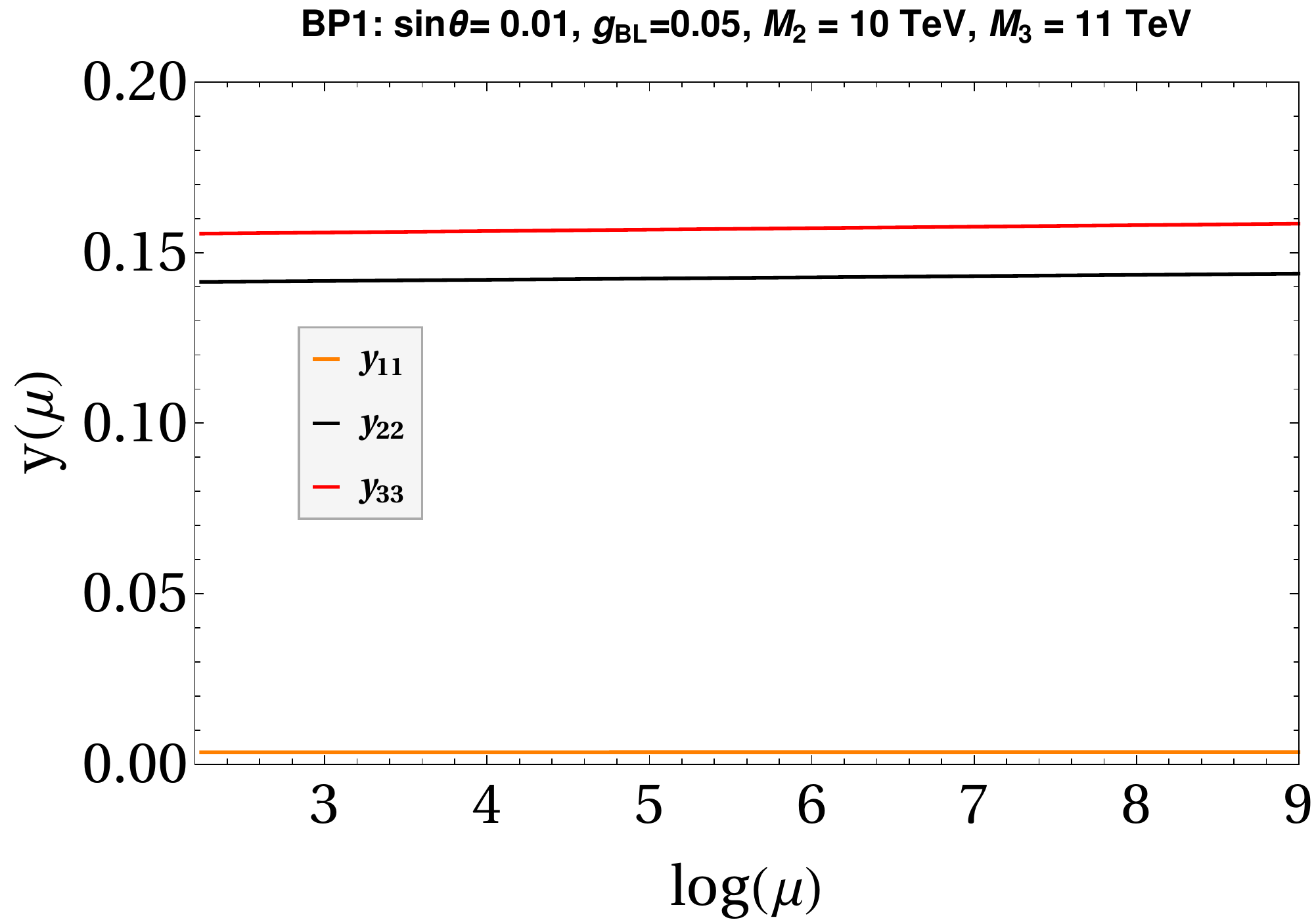}~
\caption{The evolution of the Yukawa couplings $y_{ii}$ for BP1. The left (right) plot corresponds to $M_2 = 1(10)$ TeV and $M_3 = 1.1(11)$ TeV. The color coding is explained in the legends.}
\label{BP1_p05_yuk}
\end{figure}
Compared to BP1, the lighter $N_1$ and $H$ featuring in BP2 tend to generate the requisite $N_1 - \phi_2$ conversion rate for a smaller value of $\l_7 = 1.5$ as shown in table~\ref{BP}. And this smaller $\l_7$ when used as an initial condition in the RG equations ensures perturbativity up to a higher scale ($\sim 10^{9}$ GeV) compared to BP1. We further state the qualitative features of the RG evolution of the two benchmarks remain unchanged \emph{w.r.t} a 
 $0.05 < g_{BL}(M_t) < 0.3$ variation. Elevating $N_{2,3}$ to $\simeq 10$ TeV masses lowers the perturbative cut-off of the model negligibly.
 
\begin{figure}[H]
\centering
\includegraphics[scale=0.35]{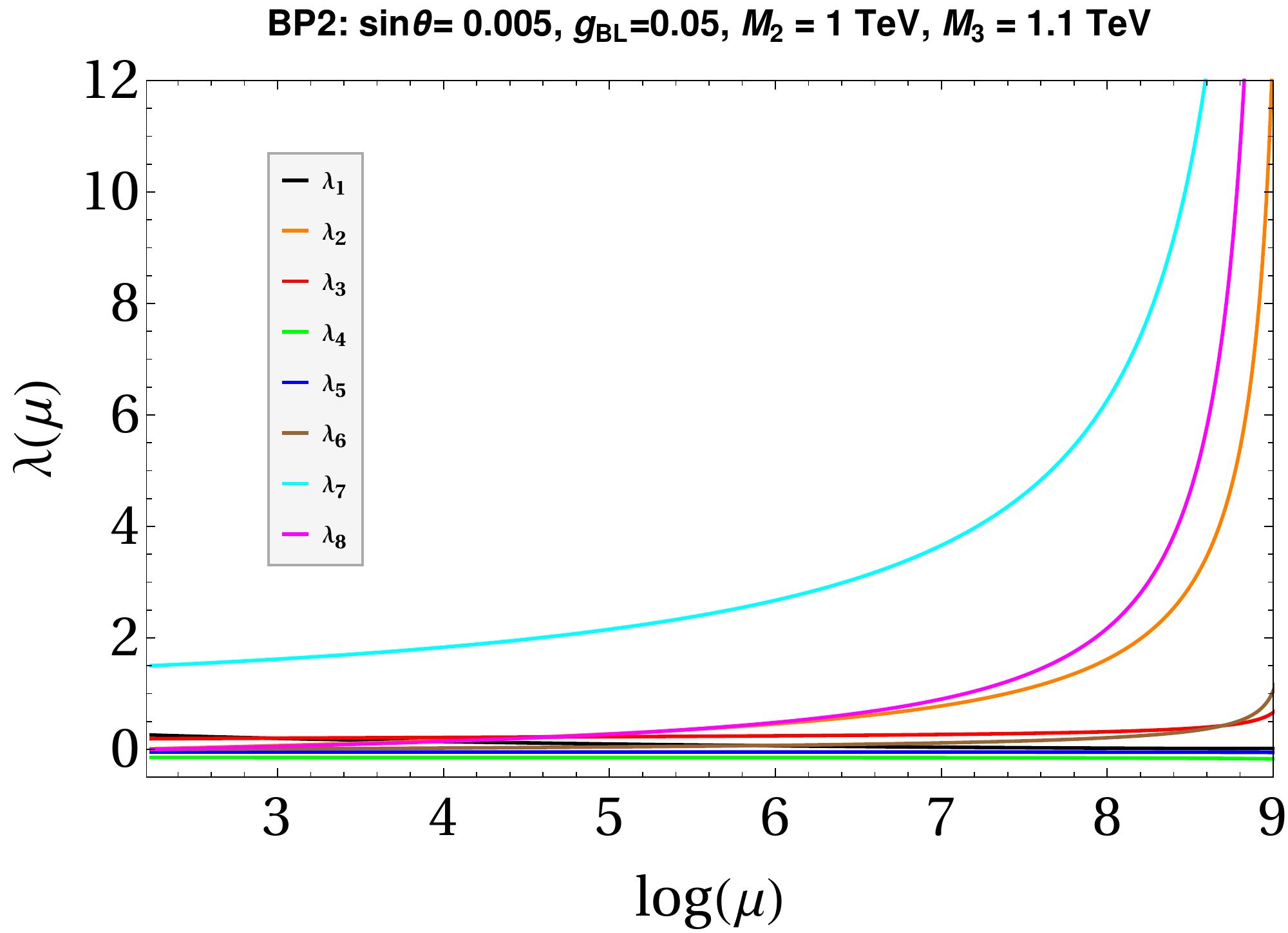}~
\includegraphics[scale=0.35]{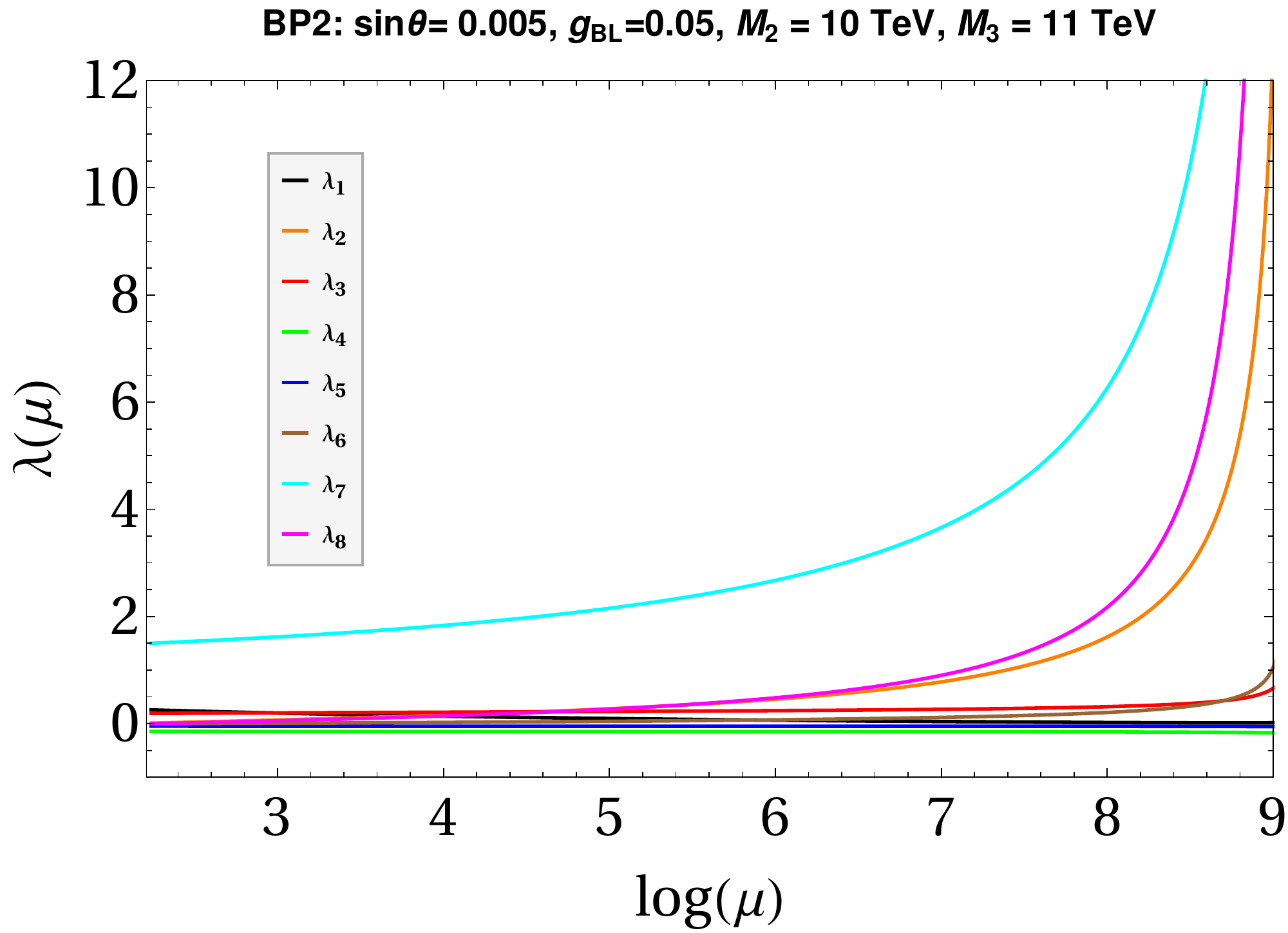}~
\caption{The evolution of the quartic couplings for BP2. The left (right) plot corresponds to $M_2 = 1(10)$ TeV and $M_3 = 1.1(11)$ TeV. The color coding is explained in the legends.}
\label{BP2_p05_lam}
\end{figure} 
We conclude this section by reiterating the most important finding. The parameter $\l_7$ turns out to be crucial in (a) generating the observed thermal relic via triggering $N_1 - \phi_2$ conversions, and, 
(b) determining the highest energy scale up to which the model can be deemed perturbative. This only goes to show that adding an inert scalar doublet to the minimal $U(1)_{B-L}$ model bears interesting effects 
both from experimental as well as theoretical perspectives. Secondly, we also find that the choice $M_{2,3} \simeq 1$ TeV is seemingly more favourable from a high-scale validity perspective compared to higher values of the same.   

\section{Combined constraints from DM and high scale validity}\label{dm+hsv}

This section is aimed towards combining the constraints coming from relic density and direct detection with those coming from high scale vacuum stability and perturbativity. We take the approach of fixing some of the model parameters so that (a) the computational time is reduced, and, (b) the analysis is not unwieldy and the scan results bring out the dominant effects that go into this interplay of dark matter and RG evolution. Therefore
\begin{itemize}

\item We take $v_{BL}$ = 50 TeV.

\item $M_{2(3)}$ is fixed to 1(1.1) TeV. This choice is motivated from the finding from the previous section that smaller $y_{22}(M_t)$ and $y_{33}(M_t)$ aid towards high scale perturbativity.

\item $M_A - M_H = M_{H^+} - M_A$ are taken to be 10 GeV and 20 GeV.

\item We choose $M_{1} - M_s$ = 30 GeV, 40 GeV.

\item sin$\theta$ is fixed to 0.005 \footnote{The value of $\mu_{\g\g}$ for BP1 lies at the boundary of the $2\sigma$ and $3\sigma$ whereas the corresponding value for BP2 lies well within the $2\sigma$ range. In the parameter scan, we therefore have imposed a more conservative 2$\sigma$ condition (which requires a smaller $\sin{\theta}\simeq0.005$ in our set-up).}.

\item We also fix $(\l_L,\l_2) = (10^{-4},10^{-2})$.

\end{itemize} 
The reason for choosing $M_A - M_H = M_{H^+} - M_A$ to 10, 20 GeV is to aid coannihilation and to give appropriately sizeable values to $\l_{3,4,5}$.
The variation $\l_7 \in [0,4\pi], M_1 \in [200~\text{GeV},1~\text{TeV}]$ subject to the constraints yields scatter plots of the allowed parameter points in the $\l_7 - M_1$ plane. Fig.~\ref{scan1} displays the corresponding parameter points for $\Delta M_{NH}=~M_1 - M_H$ = 50 GeV, $\Delta M_{NS}=~M_1 - M_s$ = 40 GeV, $v_{BL} = 50$  TeV. The ensuing observations based on Fig.~\ref{scan1} are detailed below.

\begin{figure}[htbp]
\includegraphics[scale=0.35]{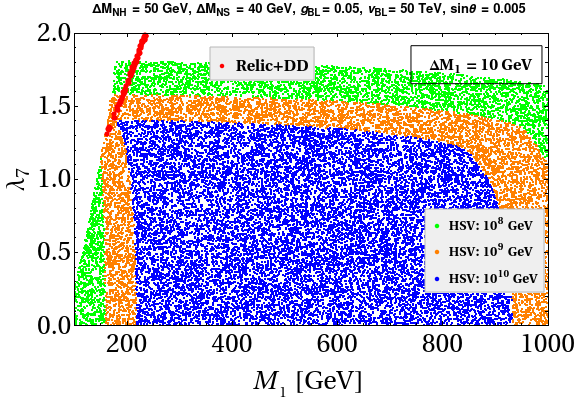}~
\includegraphics[scale=0.35]{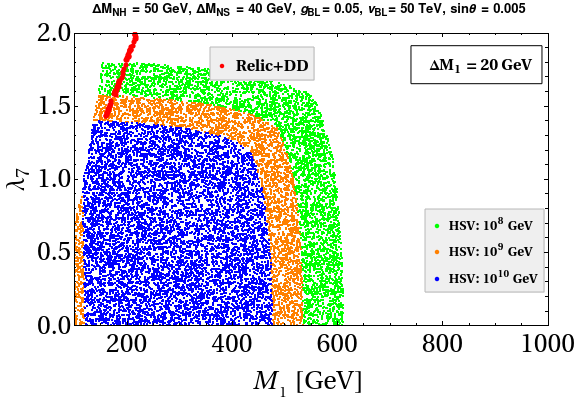}~
\caption{Allowed points in the $\l_7 - M_1$ plane surviving the dark matter (red) and high-scale validity constraints up to $10^{8}$ GeV (green) and $10^{9}$ GeV (yellow). Here $\Delta M_1$ denotes $M_A - M_H$. 
The left (right) plot corresponds to $\Delta M_1 = 10(20)$ GeV. The values for the other parameters can be read at the top.}
\label{scan1}
\end{figure}

Firstly, the parameter points allowed by the DM constraints as shown are entirely generated by the $N_1 - \phi_2$ conversion. This is easy to understand since the $h,s$ and $Z_{BL}$ resonance regions can only show themselves up in 
the $\l_7 - M_1$ plane as vertical dips around $M_1 = \frac{M_h}{2}, \frac{M_s}{2}$ and $\frac{M_{Z_{BL}}}{2}$. Of these, the smallness of the $h-N_1-N_1$ Yukawa coupling (for $s_{\theta}$ = 0.005) causes the $\frac{M_h}{2}$ 
to lose prominence. Besides, $M_1 = M_s + 40$ GeV in the aforementioned scan range forbids the possibility of $M_1 \simeq \frac{M_s}{2}$. In addition, the dip at $M_1 \simeq \frac{M_{Z_{BL}}}{2}$ = 2.5 TeV 
would also not be seen in the plot where the mass of $N_1$ does not exceed 1 TeV. The conversion region of the model therefore has been segregated in the $\l_7 - M_1$ plane and its interplay with high scale validity can be commented upon. 
  
Fig.~\ref{scan1} shows that for $M_A - M_H$ = 10 GeV, the highest scale up to which the conversion region is
extrapolatable is some intermediate scale lying between $10^{9}$ GeV- $10^{10}$ GeV. RG constraints alone lead to $\l_7 \lesssim 1.8$ for validity till $10^{8}$ GeV. This obviously tightens to $\l_7 \lesssim 1.5$ in case of $10^{9}$ GeV since we expect the parameter space to shrink when the cut-off scale is raised. An upper bound on $M_1$ (for example, $\simeq$  930 GeV for $10^{10}$ GeV) is understood as follows. Demanding perturbativity up to a given scale restricts $|\l_3|, |\l_4|$ and $|\l_5|$. For a fixed mass splitting amongst the inert scalars, this restriction translates to an upper bound on the individual masses (see eqn.(~\ref{quartic})). And for a fixed $M_1 - M_H$, this in turn puts an upper limit on the mass of the RH neutrino DM.

\begin{figure}[htbp]
\includegraphics[scale=0.350]{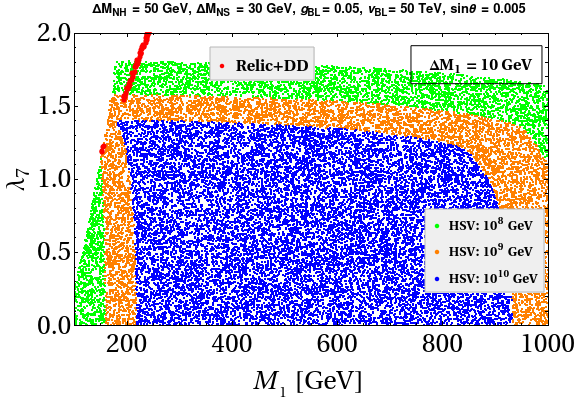}~
\includegraphics[scale=0.350]{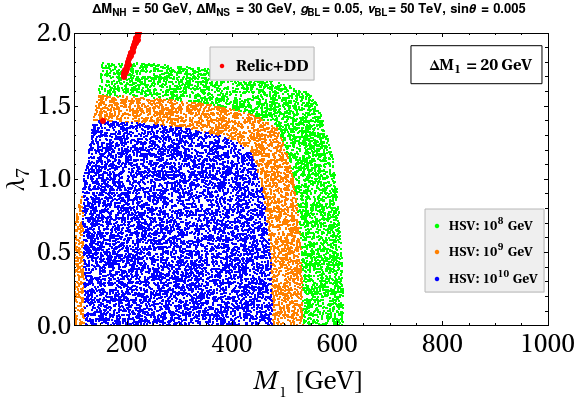}~
\caption{Same as Fig.\ref{scan1} but with $\Delta M_{NS}$ = 30 GeV}
\label{scan2}
\end{figure}

When $M_A - M_H$ = 10 GeV ensures high scale validity up to a cut-off, say $\Lambda$, increasing the mass splitting to 20 GeV implies that the individual masses of the inert scalars have to be appropriately smaller 
so as to give to $\l_3,\l_4$ and $\l_5$ the requisite values that ensure validity up to the quoted $\Lambda$. And thus the upper bound on $M_1$ will also get tighter. This is ascertained by an inspection the left plot in 
Fig.~\ref{scan1} where the $M_1 \lesssim 600(530)$ GeV for $10^{8}(10^{9})$ GeV bound is more stringent than the corresponding bounds in Fig.~\ref{scan1}. As a result, the entire RG-allowed region shifts towards left. 
This stands as an important finding in this regard. The conversion region is slightly displaced \emph{w.r.t.} the $M_A - M_H$ = 10 GeV case and this is traced back to the slight reduction in the 
$N_1 N_1 \longrightarrow A A, H^+ H^-$ for fixed values of the other parameters.  

Fig.~\ref{scan2} corresponds to $\Delta M_{NS} = 30$ GeV, other parameters being the same as in Fig.~\ref{scan1}. The parameter region allowed by the DM constraints undergoes a minute change \emph{w.r.t.} the 
$\Delta M_{NS} = 40$ GeV case. Other important features remain unchanged. In fact, such is also the case with a higher $v_{BL}$ (say 100 TeV). Extracting an UV extrapolatable scale 
$\sim 10^{9}$ GeV out of the conversion dynamics seen in this model is a clear upshot of this analysis.

%
%

\section{\emph{Freeze-in} production of $N_1$}

Here we briefly comment on the possibility of including FIMP type DM within the present setup. The limit where $N_1$ couples \emph{feebly} to other particles paves the way for the former's non-thermal production. The initial abundance of the DM candidate is taken to be zero and, as the Universe cools, the DM is expected to be dominantly produced by the decay or scattering of other particles. By virtue of the tiny strengths of the couplings at play here, the interaction rate(s) is always smaller than the Hubble expansion rate ($\Gamma < \bar{H}$, where 
$\Gamma$ and $\bar{H}$ respectively denote the relevant decay rate and the Hubble parameter). \\
In the present model, $N_1$ can be produced through the decays\footnote{The present FIMP scenario qualitatively resembles the one elaborated in \cite{Biswas:2016bfo} where it is shown that the scattering processes contribute negligibly to $N_1$ production. The scattering contribution is therefore throughout omitted in the present study.}:
$Z_{BL} \to N_1 N_1,~h \to N_1 N_1,~S \to N_1 N_1$. The expressions for the decay widths are to be seen in the Appendix. Since the detailed analysis of FIMP is beyond the scope of this  work, we present our result for the specific parameter set: $M_{Z_{BL}} = 150$ GeV, 
$M_s = 500$ GeV and $M_1 =$ 20 MeV and $g_{BL} = 10^{-10}$. Interestingly, this tiny gauge coupling $g_{BL}$ implies that, similar to $N_1$, $Z_{BL}$ too will not be in thermal equilibrium with the thermal soup. The comoving number densities of $Z_{BL}$ and $N_1$ are then dictated by the following set of coupled Boltzmann equations:
\besub
\bea
\frac{d Y_{Z_{BL}}}{d x} &=& \frac{2 M_{Pl}}{1.66 M^2_h} \frac{x \sqrt{g_*(x)}}{g_s(x)}
\Bigg(\Gamma_{h \to Z_{BL} Z_{BL}}Y_h^{EQ} + \Gamma_{s \to Z_{BL} Z_{BL}}Y_s^{EQ} 
 - \Gamma_{Z_{BL} \to f \bar{f}} Y_{Z_{BL}}\Bigg), \\
\frac{d Y_{N_1}}{d x} &=& \frac{2 M_{Pl}}{1.66 M^2_h} \frac{x \sqrt{g_*(x)}}{g_s(x)} 
\Bigg(\Gamma_{Z_{BL} \to N_1 N_1}(Y_{Z_{BL}} - Y_{N_1})
 + \Gamma_{h \to N_1 N_1}(Y_h^{EQ} - Y_{N_1}) \nonumber \\
&& 
 + \Gamma_{s \to N_1 N_1}(Y_s^{EQ} - Y_{N_1}) \Bigg). 
\eea
\label{FIMP}
\eesub
\noindent where $x=\frac{M_{\text{ref}}}{T}$ and T is the temperature of the Universe. For simplicity we
have taken $M_{\text{ref}}\sim M_h$, the mass of SM Higgs boson. As stated above, we refrain from performing a detailed scan of the parameter space
for a non-thermally produced RH neutrino since the same for a similar scenario has already been carried out in \cite{Biswas:2016bfo}. We rather take up to demonstrate the high-scale validity of the parameter region consistent with a \emph{frozen-in} $N_1$.\\

We first plot the $Z_{BL}$ and $N_1$ yields using the Eqns.~(\ref{FIMP}) and, the relic abundance of $N_1$ as shown in Fig.~\ref{relic_fimp}. In the left panel, the raising segment of the $Z_{BL}$ yield indicates the production of $Z_{BL}$ from the decay of heavy scalar $s$ and the plateau corresponds to the region where the production and decay rates of $Z_{BL}$ are equal. Then the decreasing portion is for the region where decay of the $Z_{BL}$ field dominates and it explains the increasing yields of $N_1$. In the right panel, relic contribution of $N_1$ is indicated for which the frozen-in contribution of $N_1$ satifies the required relic.
\begin{figure}[H]
\centering
\includegraphics[scale=0.35]{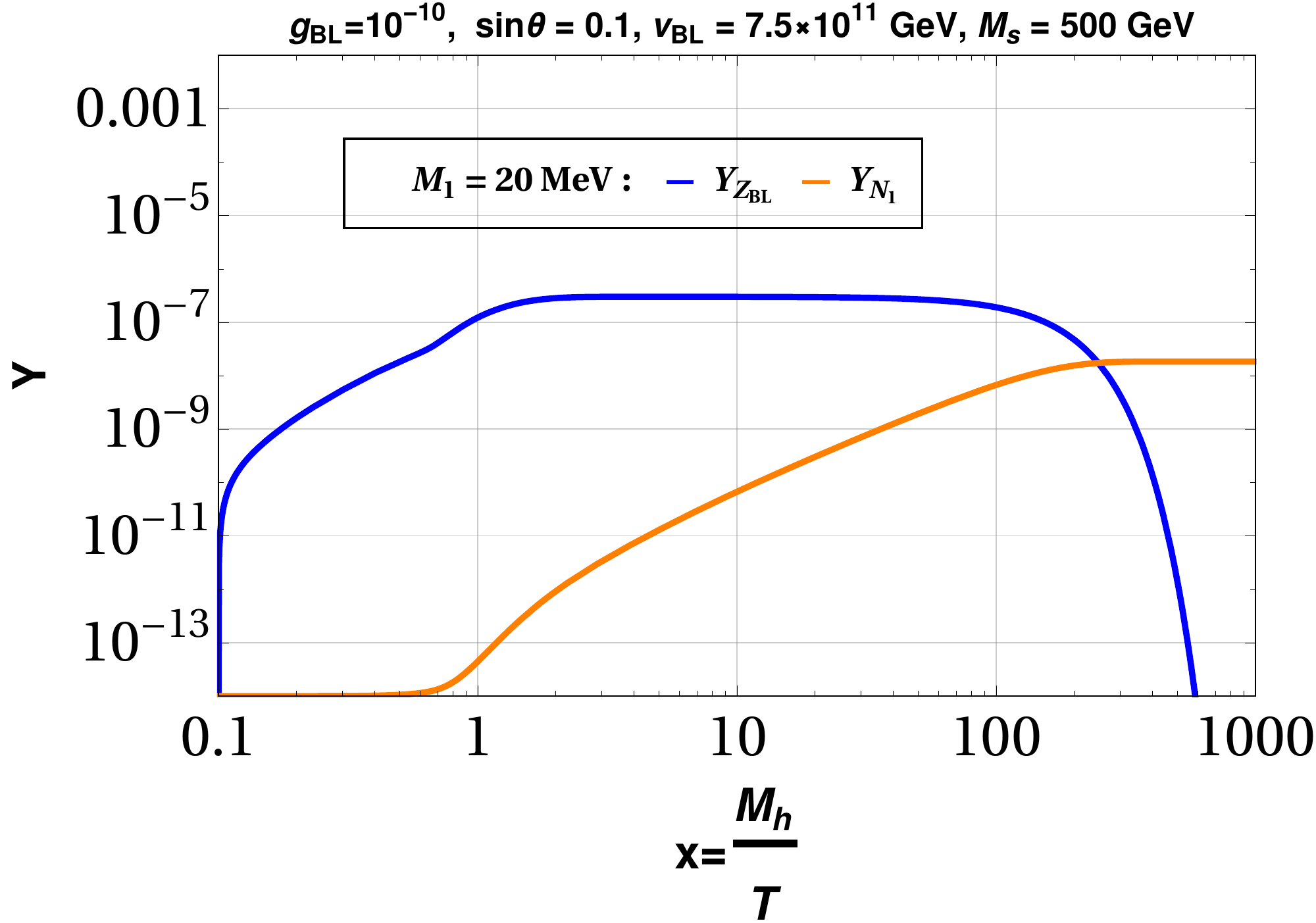}~~~
\includegraphics[scale=0.35]{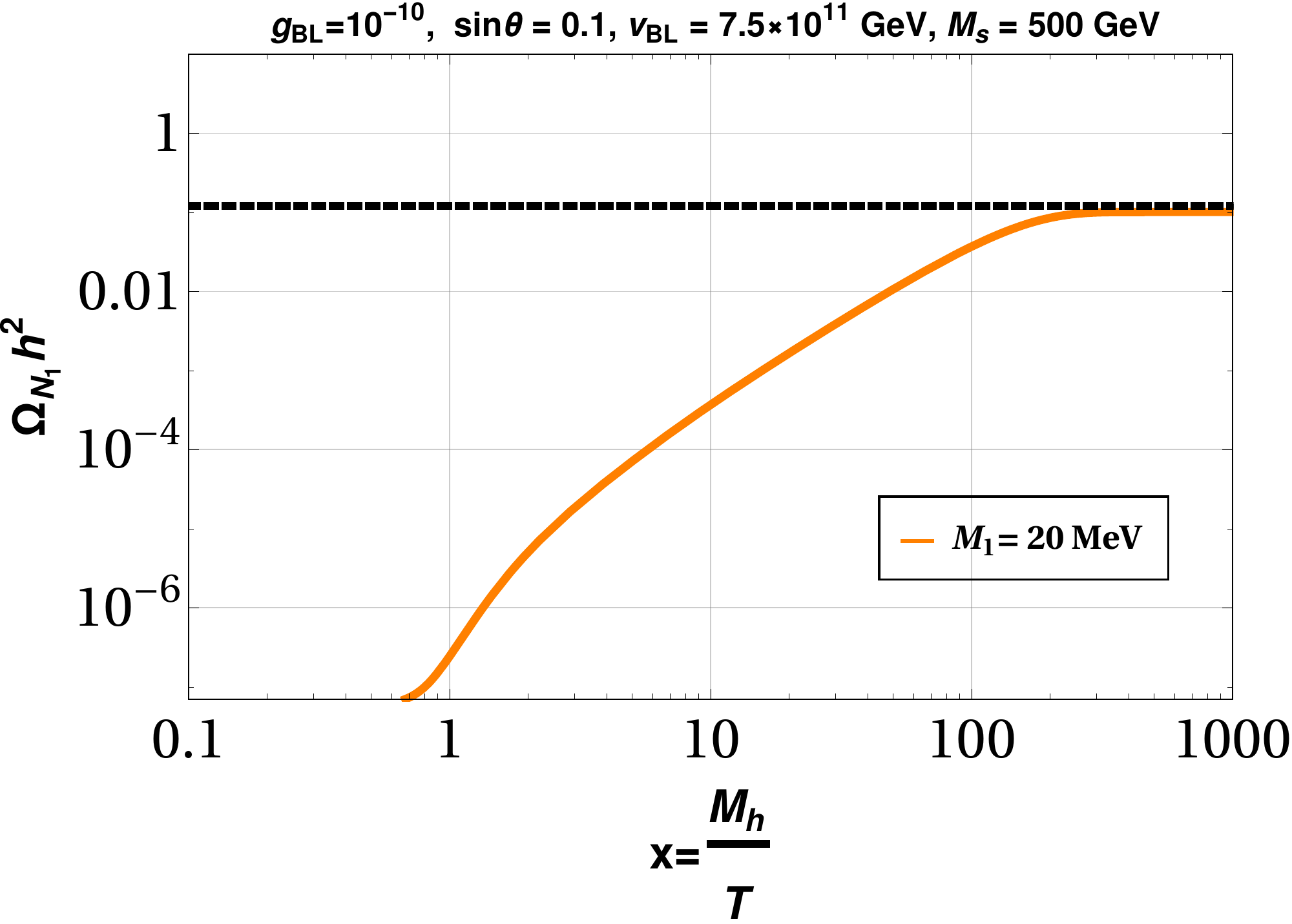}~
\caption{The solution to the coupled Boltzmann equations. The left plot depicts
 evolution of the (comoving) number densities of $Z_{BL}$ and $N_1$. The right plot displays the relic of $N_1$.}
\label{relic_fimp}
\end{figure}
Note that such a tiny value for $g_{BL}$ (= $10^{-10}$) corresponding to a 150 GeV $Z_{BL} ~(M_{Z_{BL}}=2g_{BL}v_{BL})$ implies a value for $v_{BL} (= 7.5\times10^{11} ~\rm{GeV})$ that is several orders of magnitude higher than the TeV scale. Such a large $v_{BL}$ forces some of the other model parameters to take extremely small values. Eqns.~(\ref{quartic}) show that $\l_6$ and $\l_8$ are accordingly small. The Yukawa coupling
$y_{11} = 1.9 \times 10^{-14}$ for this parameter point implies $\Gamma_{h,s \to N_1 N_1} < \bar{H}$ is obeyed. Accordingly such a small $y_{11}$ implies that a possible $H H \longrightarrow N_1 N_1$ conversion is too small to play any role in the generation of relic.

Comments on possible constraints on $\l_7$ are in order here as it was crucial in determining the parameter space consistant with high scale validity in the WIMP case. Since the $H-H-s$ coupling is $\simeq \l_7 v_{BL} s_\theta$, an $\mathcal{O}$(1) value for $\l_7$ would lead to a hopelessly tiny $\Omega_H$ through $s$ mediated annihilations. Although the DD cross-section   $\sigma_{H n \to H n}$ is then expected to be accordingly large, the effective direct detection cross-section $\sigma^{\text{eff}}_{H n \to H n}
 = (\frac{\Omega_H}{\Omega_T})\sigma_{H n \to H n}$ will be suppressed (as $H$ will contribute a miniscule fraction of the total relic) and hence within the permissible limit. Such a value for $\l_7$ will then be allowed. Turning to a case with significantly small $\l_7$, for instance, $\l_7 = 10^{-7}$ would lead to $\l_7 v_{BL} \simeq 7.5 \times 10^4$ GeV here, a value similar to the corresponding number for the thermal case. For this case also, we obtain approximately similar value for $\sigma^{\text{eff}}_{H n \to H n}$. 
Hence it turns out that role of $\l_7$ is insignificant in the non-thermal scenario (at least for $\mathcal{O}$(100 GeV) mass of the decaying particle) contrary to what we have found in the thermal case.

In other words, the stringent constraint on $\l_7$ that we encountered in the thermal case no longer applies to the non-thermal case and this emerges as an important takeaway in this section. There is therefore the lucrative possibility of choosing a small value for $\l_7$ and maintaining the perturbativity of the scalar potential up to energy scales higher than what was obtained for the thermal case. The aforementioned parameters augmented with a $\l_7 = 10^{-7}$ gives rise to the following RG evolution of $\l_{1-5}$ (Fig.~\ref{lambda_fimp}).
\begin{figure}[H]
\centering
\includegraphics[scale=0.40]{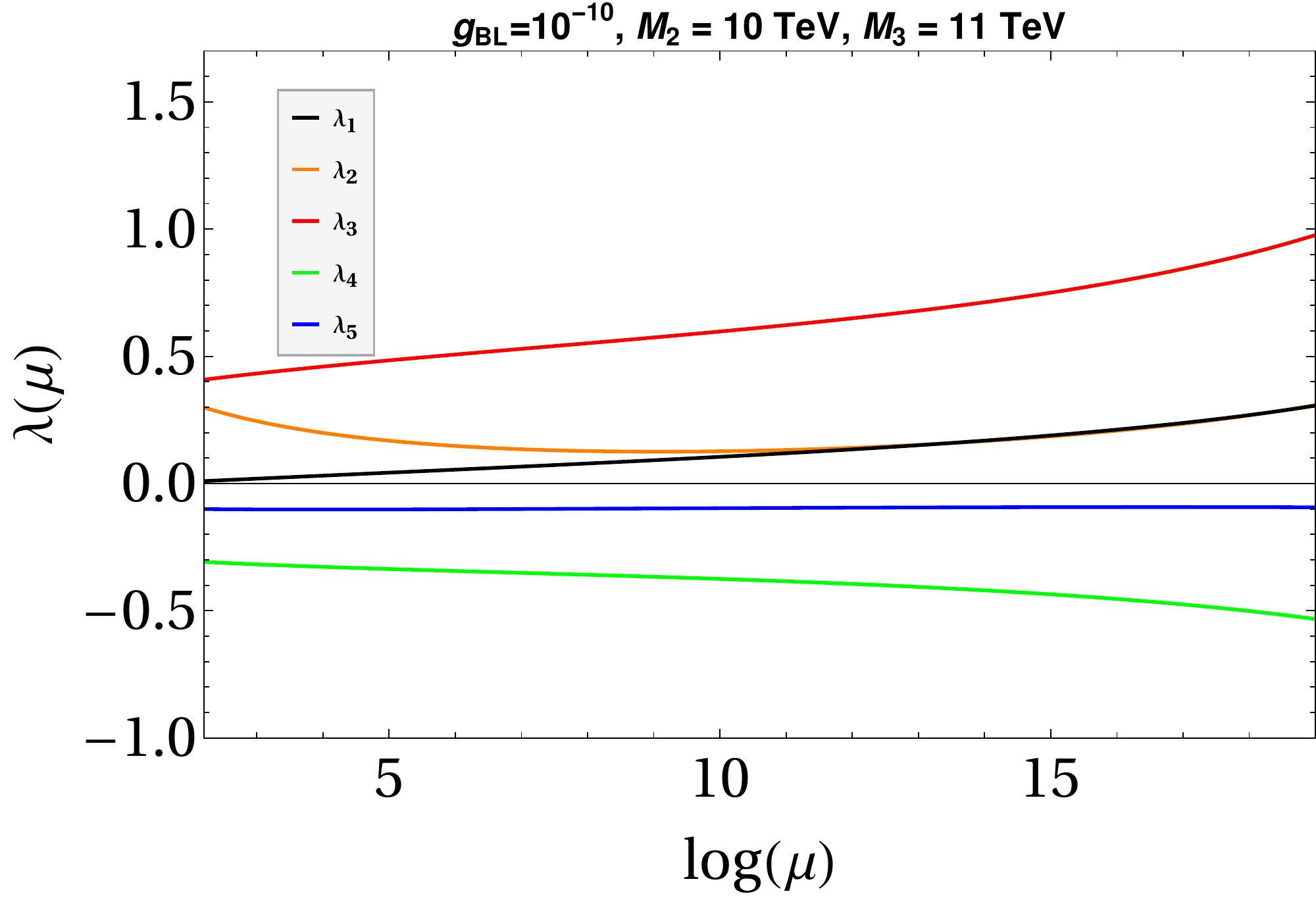}~~~
\caption{RG running of $\l_{j}$ for $j$ = 1,2,3,4,5. The color coding can be read in the legends.}
\label{lambda_fimp}
\end{figure}
\noindent In the above, $\l_6,\l_7$ and $\l_8$ are too small to cast an impact on the evolution of 
$\l_{1-5}$. It is precisely due to choosing a small $\l_7$ that we get a stable vacuum and perturbative couplings all the way up to the Planck scale. We have $\l_6 = 1.26 \times 10^{-10},\l_{7} = 10^{-7},\l_{8} =2.2\times 10^{-19} $ at $\mu = M_t$. Each term in 
$\beta_{\l_{6,7,8}}$ contain at least one power of either $\l_6,\l_7$ or $\l_8$. Therefore, small $\l_6,\l_7$ or $\l_8$ automatically imply small first derivatives and hence a mild evolution rate. We reiterate what we consider the most important finding in this case: the high-scale validity of the model for a \emph{frozen-in} $N_1$ can be extended all the way till the Planck scale as opposed to the much lower scale of $10^{9}$ GeV for a 
\emph{frozen-out} $N_1$. It is obvious, but, still worth mentioning once that, one can not have a {\em freeze-in} of inert scalar DM, simply due to its known $SU(2)_L$ coupling.


\section{Summary and conclusions}\label{summary}

In this work, we extend the minimal $U(1)_{B-L}$ model by an inert scalar doublet. The lightest RH neutrino and the $CP$-even inert scalar emerge as DM candidates and masses for the light neutrinos is generated 
radiatively following the scotogenic mechanism. The proposed scenario opens up the attractive possibility of Higgs-mediated DM-DM conversion, a phenomenon that goes on to become the main theme of the study. 
The parameter region leading to the optimal conversion rates is subjected to renormalisation group evolution up to high energy scales. The following conclusions are derived.

\begin{itemize}

\item Conversion processes of the $N_1 \to \phi_2$ form can lead to the desired relic density for $N_1$ in a mass region of $N_1$ that would give an overabundant relic in absence of the inert doublet $\phi_2$. 
The relic contributed by the inert doublet alone although becomes negligible in the process owing to enhanced annihilations.
The requisite conversion amplitudes are found to be triggered whenever the quartic coupling $\l_7 \gtrsim$ 1 and the $U(1)_{B-L}$ breaking VEV $v_{BL} \sim \mathcal{O}$(10) TeV. 
These observations of course comply with the constraints coming from the direct detection and collider experiments.

\item A sizeable $\l_7$, as necessitated by the conversion dynamics, tends to grow under renormalisation group evolution and eventually become non perturbative at some high energy scale below the Planck scale. 
While this behaviour is qualitatively robust, the exact cut-off scale is determined by a choice of the other parameters. Taking all of that into account, the conversion region is found to be extrapolatable up to a maximum of 
$\sim 10^{9}$ GeV. 

\item Nonthermal production of $N_1$ from the decays of $Z_{BL}, h, s$ are also possible in this model. In that case, the \emph{frozen in} $N_1$ can explain the observed relic for a feeble $g_{BL}\sim 10^{-10}$. 
In such a case, however, having the $Z_{BL}$ and scalar masses in the $\mathcal{O}$(100) GeV ball-park relaxes the stringent constraint on $\l_7$. The model then becomes extrapolable all the way till the Planck scale.

\item It must be noted therefore, that this model serves as the simplest multipartite DM framework in  $U(1)_{B-L}$ scenario, compatible with relic density, direct search and high scale validity constraints 
to have a viable parameter space beyond resonance regions. For example, a similar analysis of $U(1)_{B-L}$ model in presence of a scalar singlet DM component ($\phi$) would be disfavoured from both the facts that DM-DM interaction 
would have failed to keep the model on-board in regions beyond $N_1$ resonance, as it would be first extremely difficult to get under abundance of such a DM ($\phi$), compatible with direct search constraint absent coannihilation channels, 
secondly it would pose even a stronger bound on DM-DM conversion coupling from EW vacuum stability.

\end{itemize}

Possible collider signals to test the proposed scenario at the LHC is to look for hadronically quiet dilepton signatures arising from  production of the heavier components of the inert doublet ($H^\pm, A$) through Drell-Yan process 
and its further decay to DM ($H$) associated with off-shell $W^\pm \to \ell^\pm+\nu_\ell$ yielding
\besub
\bea
p p \longrightarrow H^+ H^- \longrightarrow \ell^+ \ell^- + \fsl{E_T}, \\
p p \longrightarrow H A \longrightarrow \ell^+ \ell^- + \fsl{E_T}.
\eea
\eesub

For $M_H > M_W$, the conversion dynamics in the present setup extracts a correct relic even in the $M_W < M_H < 500$ GeV mass range, 
as opposed to the pure inert doublet model where the corresponding range is $M_H < M_W \cup M_H > 500$ GeV. And when it comes to probing the two cosmologically motivated mass ranges, the former is kinematically more prospective. 
The proposed model thus clearly offers better observability at the energy frontier than the pure inert doublet case. However, we should also note that the preferred mass difference between the charged and neutral (DM) component 
of the inert doublet is on the smaller side, 10, 20 GeVs, so that we can effectively use co-annihilation channels to yield under abundance. In terms of segregating the dilepton signal arising from the inert doublet as mentioned above, from 
SM background, one often needs to use missing energy and effective mass cuts judiciously. Having a smaller mass difference between the parent ($H^\pm$) and daughter ($H$) yields a  signal distribution almost identical to that of SM 
background and becomes difficult to distinguish. The international linear collider (ILC) may be able to probe such a scenario.\\

\acknowledgments
NC acknowledges financial assistance from National Center for Theoretical Sciences and Centre for High Energy Physics, Indian Institute of Science. He also thanks Indian Institute of Technology Guwahati for hospitality during the formative stages of the project. RR thanks  Amit Dutta Banik, Purusottam Ghosh, Basabendu Barman and Dibyendu Nanda for various useful discussions during the course of this work.

\section{Appendix}\label{appendix}

We list below expressions for the relevant annihilation cross sections and decay widths.

\subsection{Couplings}

\textbf{Yukawa interations:}
\besub
\bea
y_{h N_1 N_1} &=& -\frac{1}{\sqrt{2}}y_{11} s_{\theta}, \\
y_{s N_1 N_1} &=& \frac{1}{\sqrt{2}}y_{11} c_{\theta}, \\
y_{h f f} &=& \frac{M_f}{v} c_{\theta}, \\
y_{s f f} &=& \frac{M_f}{v} s_{\theta} ~\text{where $f$ is a SM fermion.}
\eea
\eesub

\textbf{Gauge interations:}
\besub
\bea
g_{h V V} &=& \frac{2 M_V^2}{v} c_{\theta},\\
g_{s V V} &=& \frac{2 M_V^2}{v} s_{\theta}
~\text{where} ~V = W^+, Z \\
g_{h Z_{BL} Z_{BL}} &=& -\frac{2 M_V^2}{v_{BL}} s_{\theta},\\
g_{s Z_{BL} Z_{BL}} &=& \frac{2 M_V^2}{v_{BL}} c_{\theta}.
\eea
\eesub

\textbf{Scalar interations:}
\besub
\bea
\l_{H H h} &=& (\l_3 + \l_4 + \l_5) v c_{\theta} - \l_7 v_{BL}s_{\theta},\\
\l_{H H s} &=& (\l_3 + \l_4 + \l_5) v s_{\theta} + \l_7 v_{BL}c_{\theta},\\
\l_{A A h} &=& (\l_3 + \l_4 - \l_5) v c_{\theta} - \l_7 v_{BL}s_{\theta},\\
\l_{A A s} &=& (\l_3 + \l_4 - \l_5) v s_{\theta} + \l_7 v_{BL}c_{\theta},\\
\l_{H^+ H^- h} &=& \l_3 v c_{\theta} - \l_7 v_{BL}s_{\theta},\\
\l_{H^+ H^- s} &=& \l_3 v s_{\theta} + \l_7 v_{BL}c_{\theta}.
\eea
\eesub

\subsection{Decay widths}

The scalar $\phi = h, s$ and $Z_{BL}$ have the following decay widths to the $N_1 N_1$ final state:
\begin{align}
\Gamma_{\phi \longrightarrow N_1 N_1} = \frac{M_{\phi}}{16 \pi} 
~y^2_{\phi N_1 N_1} ~\Big(1 - \frac{4 M^2_1}{M^2_\phi}\Big)^{3/2}, \\
\Gamma_{Z_{BL} \longrightarrow f \bar{f}} = \frac{M_{Z_{BL}}}{12 \pi} 
~g^2_{BL} ~\Big(1 + \frac{2 M^2_f}{M^2_{Z_{BL}}}\Big)
\Big(1 - \frac{4 M^2_f}{M^2_{Z_{BL}}}\Big)^{1/2}, \\
\Gamma_{Z_{BL} \longrightarrow N_1 N_1} = \frac{M_{Z_{BL}}}{24 \pi} 
~g^2_{BL} ~\Big(1 - \frac{4 M^2_1}{M^2_{Z_{BL}}}\Big)^{3/2}.
\end{align}


\bibliographystyle{JHEP}
\bibliography{ref}


\end{document}